\documentclass[12pt,a4paper]{article}
\usepackage{type1cm,amsmath}
\usepackage[psamsfonts]{amssymb}
\usepackage{cite}
\usepackage{graphicx}
\numberwithin{equation}{section}

\newcommand{\om}{\omega}
\newcommand{\al}{\alpha}
\newcommand{\ep}{\epsilon}

\newcommand{\la}{\lambda}

\newcommand{\lb}{\lbrack}
\newcommand{\rb}{\rbrack}

\newcommand{\msc}[1]{\mbox{\scriptsize #1}}
\newcommand{\dsp}{\displaystyle}

\newcommand{\br}{\mathbb R}
\newcommand{\bz}{\mathbb Z}

\newcommand{\bone}{\mbox{{\bf 1}}}
\newcommand{\bsz}{\msc{{\bf Z}}}
\newcommand{\da}{\dot{a}}
\newcommand{\dal}{\dot{\alpha}}
\newcommand{\db}{\dot{b}}

\newcommand{\cA}{{\cal A}}
\newcommand{\cB}{{\cal B}}

\newcommand{\cK}{{\cal K}}

\newcommand{\cJ}{{\cal J}}

\newcommand{\cO}{{\cal O}}

\newcommand{\cN}{{\cal N}}

\newcommand{\cP}{{\cal P}}
\newcommand{\cF}{{\cal F}}
\newcommand{\cQ}{{\cal Q}}
\newcommand{\cH}{{\cal H}}

\newcommand{\ket}[1]{{|#1\rangle}}
\newcommand{\bra}[1]{{\langle#1|}}

\newcommand{\tr}{\mbox{Tr}}

\DeclareMathOperator*{\im}{{\rm Im}}

\newcommand{\nn}{\nonumber\\}

\newcommand {\eqn}[1]{(\ref{#1})}

\makeatletter
\@addtoreset{equation}{section}

\makeatother

\begin{document}
\vskip 7mm

\begin{titlepage}

 \renewcommand{\thefootnote}{\fnsymbol{footnote}}
 \font\csc=cmcsc10 scaled\magstep1
 {\baselineskip=14pt
 \rightline{
 \vbox{
       }}}

 \vfill
 \baselineskip=20pt
 \begin{center}
 \font\titlerm=cmr10 scaled\magstep4
 \font\titlei=cmmi10 scaled\magstep4
 \font\titleis=cmmi7 scaled\magstep4
 \centerline{\titlerm Superstrings on NSNS PP-Waves} 
 \vskip 3mm 
 \centerline{\titlerm and Their CFT Duals}
 \vskip 2 truecm

{ \large Yasuaki Hikida}\\
{\sf hikida@hep-th.phys.s.u-tokyo.ac.jp}
\bigskip

 \vskip .6 truecm
 {\baselineskip=15pt
 {\it Department of Physics,  Faculty of Science, University of Tokyo \\
  Hongo 7-3-1, Bunkyo-ku, Tokyo 113-0033, Japan} 
 }
 \vskip 2 truecm
 A dissertation presented to the faculty of University of Tokyo\\
 in candidacy for the degree of Doctor of Science\\
 \vskip .6 truecm
 December 20, 2002
 \vskip .4 truecm

 \end{center}

 \vfill
 \vskip 0.5 truecm

\begin{abstract}

We investigate the correspondence between superstring theory on pp-wave
background with NSNS-flux and superconformal field theory on a symmetric
orbifold.
This correspondence can be regarded as the ``Penrose limit''
of $AdS_3/CFT_2$ correspondence. 
Superstring theory on the Penrose limit of 
$AdS_3 \times S^3 (\times M^4)$ ($M^4 = T^4$ or $K3$) 
with NSNS-flux can be described by a generalization of Nappi-Witten
model.
We quantize this system in the covariant gauge and obtain the spectrum
of superstring theory.
In the dual CFT point of view, the Penrose limit means
concentrating on the subsector of almost BPS states with large R-charges.
We show that stringy states can be embedded in the single-particle
Hilbert space of symmetric orbifold theory. 
 
\end{abstract}
 \vfill
 \vskip 0.5 truecm

\setcounter{footnote}{0}
\renewcommand{\thefootnote}{\arabic{footnote}}
\end{titlepage}

\newpage

\tableofcontents

\newpage

\section{Introduction}
\label{Introduction}

Superstring theory\footnote{There are good text books on superstring
theory \cite{GSW,Polchinski}. 
Please refer to them for the background of superstring theory.
} has been investigated intensively since it is the
only known consistent theory describing quantum gravity.
In recent years, important aspects of superstring theory,
which are closely related to string duality, have been revealed
and D-branes play important roles in these developments.
While D-branes can be treated as black-brane solutions in
supergravity theory, effective field theory on D-branes is given
by super Yang-Mills theory.
The fact that we can deal with D-branes in two ways implies
a duality between supergravity theory (or superstring theory) and
supersymmetric gauge theory.

This duality is called as $AdS/CFT$ correspondence 
\cite{Maldacena} (for a review, see \cite{AdSCFT}) and
the most famous example is $AdS_5/CFT_4$ correspondence.  
Let us consider large $N$ D3-branes.
If we take the near horizon limit, corresponding supergravity solution
becomes $AdS_5 \times S^5$.
On the other hand, effective field theory on D-branes is given by
$\cN=4$ $SU(N)$ super Yang-Mills theory.
Therefore, supergravity on $AdS_5 \times S^5$ is supposed to be dual
to $\cN=4$ $SU(N)$ super Yang-Mills theory.
Because the region where supergravity limit is valid corresponds to
strong coupling region of super Yang-Mills theory, 
it is expected to understand some non-perturbative aspects of super
Yang-Mills theory by using classical supergravity theory.
For this reason, $AdS_5/CFT_4$ correspondence has been investigated 
by many authors.

It is natural to expect that there is a correspondence
beyond supergravity level. 
However, it is difficult to quantize superstrings on 
$AdS_5 \times S^5$ because of the existence of RR-charges.
Recently, an important progress has been made on this aspect.
It was found in \cite{BFHP} that there is a maximally supersymmetric
background with RR-flux in type IIB supergravity in addition to the
flat Minkowski space and the $AdS_5 \times S^5$ background.
This background is called as maximally supersymmetric pp-wave
background and it was shown in \cite{BFHP2,BFP} that this background is
obtained by so-called Penrose limit \cite{Penrose,Gueven} of 
$AdS_5 \times S^5$ background.
Moreover, it was realized that superstring theory on the pp-wave
background with RR-flux can be quantized by using Green-Schwarz formalism
in the light-cone gauge \cite{Metsaev,Metsaev2}. 
Motivated by these facts, the authors of \cite{BMN} applied
the ``Penrose limit'' to the $AdS_5/CFT_4$ correspondence. 
They have shown that the Penrose limit of superstring
theory on $AdS_5 \times S^5$ corresponds to the subsector of 
``almost BPS states'' with large R-charges in $\cN=4$ super Yang-Mills
theory  and identified stringy excitations with single trace operators.

The other famous example is $AdS_3/CFT_2$ correspondence.
This is the correspondence between superstring theory on 
$AdS_3 \times S^3 \times M^4$ ($M^4 = T^4$ or $K3$) and two
dimensional superconformal field theory on a symmetric orbifold
\cite{Maldacena,MS}. 
The correspondence is related to D1/D5 system,
which is the configuration of $Q_5$ D5-branes wrapped on a small $M^4$ 
and $Q_1$ D1-branes put parallel to the extra $1+1$ dimensions.
Since D1/D5 system was used for the microscopic description of 
black holes \cite{SV}, the investigation of $AdS_3/CFT_2$ correspondence
may give insights to black hole physics.
In the IR limit, effective field theory on the two dimensional D-brane 
configuration is believed to be given by 
$\cN=(4,4)$ supersymmetric non-linear sigma model on the symmetric orbifold 
\begin{equation}
Sym^{Q_1Q_5}(M^4) = (M^4)^{Q_1Q_5}/S_{Q_1Q_5} ~,
\end{equation}
where $S_N$ represents the symmetric group of $N$ elements. 
The near horizon geometry of D1/D5 system becomes 
$AdS_3 \times S^3 \times M^4$, and hence superstring theory on 
$AdS_3 \times S^3 \times M^4$ has been conjectured to be dual to
superconformal field theory on $Sym^{Q_1Q_5}(M^4)$.

There are many configurations related to D1/D5 system by U-duality and
in this thesis we concentrate on F1/NS5 system (the system with
$Q_1$ fundamental strings and $Q_5$ NS5-branes), 
which is S-dual to D1/D5 system. 
The near horizon geometry of F1/NS5 system can be described by
superstring theory on $AdS_3 \times S^3 \times M^4$ with NSNS-flux.
This superstring theory can be written in terms of $SL(2;\br) \times
SU(2) $ WZW models, and hence we can quantize this system. 
The superstring theory is dual to
superconformal field theory on $Sym^{Q_1Q_5}(M^4)$ and 
the correspondence has attracted much attention since it can be examined
beyond supergravity approximation (see, e.g., \cite{GKS,DORT,KS}).

One of the important tests of this duality is the comparison of spectrum.
There is a moduli space in the dual CFT
and it is difficult to analyze spectrum at
general moduli point.
Since BPS states do not depend on the deformation of moduli
parameter, 
we can use BPS spectrum at the orbifold point, where we can obtain 
general spectrum by using the orbifold theory technique 
\cite{orbifoldCFT,orbifoldCFT2,orbifoldCFT3}.   
Therefore, we can obtain BPS spectrum and compare the 
supergravity modes of superstring theory.
BPS states with small R-charge are explicitly identified 
in \cite{KLL,HS} and
higher R-charged BPS states are also reproduced successfully 
in \cite{HHS,AGS} by making use of the spectral flow symmetry on the
$AdS_3$ string theory \cite{HHRS,MO}. 
However, it is still difficult to investigate general spectrum
other than BPS states.

In order to compare more general spectrum, we
apply a Penrose limit to $AdS_3/CFT_2$ correspondence \cite{HS'}.
The Penrose limit of $AdS_3 \times S^3 (\times M^4)$ with NSNS-flux 
becomes 6 dimensional pp-wave background with NSNS-flux
and superstring theory on this background can be described by
a generalization of Nappi-Witten model \cite{NW}. 
This model was investigated in a different context by several authors
\cite{KK,Sfetsos,Sfetsos2,NW2,NW2-2,NW2-3,NW2-4,NW2-5,NW2-6,Sfetsos3,KK2,RT2}
and it can be quantized by the current algebra approach 
\cite{KK,KK2,KP}
and by the sigma model approach \cite{RT2-2,RT2-3,RT}. 
Taking a Penrose limit corresponds to concentrating on
almost BPS states with very large R-charges in the dual CFT, 
where we can use spectrum at the orbifold point in this limit.
We show that stringy states can be embedded in the
Hilbert space of the symmetric orbifold theory. 
The other works on the Penrose limit of $AdS_3/CFT_2$ correspondence are
given in \cite{Parnachev,Das,Lunin,GMS,HS'2,Gava}.

The organization of this thesis is given as follows.
In section \ref{AdSBMN}, we first review the $AdS_5/CFT_4$ correspondence.
Then, we see what the Penrose limit is and how we apply the limit to the
$AdS_5/CFT_4$ correspondence.
In section \ref{NSNSPenrose},
we begin with the $AdS_3/CFT_2$ correspondence and then apply the
Penrose limit to it. 
After explaining that a generalization of Nappi-Witten model 
appears in the Penrose limit of $AdS_3 \times S^3 \times M^4$,
we review the Nappi-Witten group (and Lie algebra) and quantize this model 
by the sigma model approach in the light-cone gauge. 
In section \ref{NSNSppwave}, 
we quantize superstring theory on this pp-wave background
by using the current algebra approach in the covariant gauge and obtain
Hilbert space by explicitly constructing DDF operators.
In section \ref{compare}, we first review superconformal field theory on the 
symmetric orbifold $Sym^{Q_1Q_5}(T^4)$ and then compare 
(almost) BPS spectrum with string spectrum on the pp-wave background.
Then we comment on the case with RR-flux in subsection \ref{RR}.
We also discuss a extension to the case of $M^4 = T^4/\bz_2$\footnote{
The K3 surface is obtained by resolving the singularity of $T^4/\bz_2$
and we use $T^4/\bz_2$ as a solvable case of $K3$.}
in section \ref{T^4/Z_2} 
and we conclude this thesis in section \ref{conclusion}. 
The notation of the Gamma matrices is summarized in appendix \ref{gamma}.
The space-time super pp-wave algebra is constructed by using the 
contraction of the super $AdS_3 \times S^3$ algebra in appendix \ref{STSS}.
In appendix \ref{KS}, we also investigate space-time supersymmetry on
pp-waves with NSNS-flux by examining Killing spinors.
In particular, we show that there are 24 supersymmetries in the 
$AdS_3 \times S^3 \times T^4$ background.

\newpage

\section{$AdS_5/CFT_4$ Correspondence and BMN Conjecture}
\label{AdSBMN}

Before examining $AdS_3/CFT_2$ correspondence,
we first review $AdS_5/CFT_4$ correspondence in next subsection in order to 
show the general concept of $AdS/CFT$ correspondence.
After explaining Penrose limit in subsection \ref{Gueven},
we apply the Penrose limit to the both sides of $AdS_5/CFT_4$ correspondence 
and we explicitly construct single trace operators in super Yang-Mills
theory corresponding to states of superstrings in subsection \ref{BMN}.

\subsection{$AdS_5/CFT_4$ Correspondence}
\label{AdS_5/CFT_4}

As we mentioned in introduction, $AdS/CFT$ correspondence was
deduced from the fact that there are two ways to describe D-branes.
One is the supergravity or superstring description and the other is
the super Yang-Mills description.
Therefore, it is natural to expect that there is a correspondence between
a supergravity or superstring theory and a super Yang-Mills theory.
The concrete example is given by Maldacena \cite{Maldacena} as 
$AdS/CFT$ correspondence.
Although we concentrate on $AdS_3/CFT_2$ correspondence in this
thesis, we first overview $AdS_5/CFT_4$ correspondence in order to
see the general concept of $AdS/CFT$ correspondence.

\begin{figure}
\begin{center}
\includegraphics{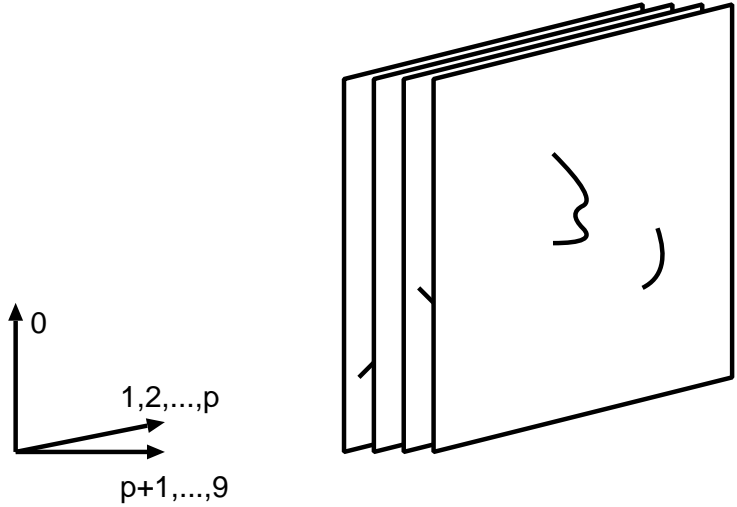}
\end{center}
\caption{D$p$-branes are $p+1$ dimensional
 hypersurfaces and open strings are stretched between the branes.}
\label{FIG:Dp-branes}
\end{figure}

Let us begin with D-branes.
A D-brane can be defined as a hypersurface which open strings are
attached to as in figure \ref{FIG:Dp-branes} \cite{D-brane}.
Open strings have their ends and we have to assign
some boundary conditions to the fields at the ends. 
The conditions consistent with the equations of motion are given by
Neumann or Dirichlet boundary conditions.
Let us assign the Neumann boundary condition for $p+1$ coordinates and
the Dirichlet boundary conditions for $9-p$ coordinates.
More precisely, we assign
\begin{align}
 &\partial_{\sigma} X^{\mu} (\tau,\sigma)|_{\sigma=0, \pi}
      = 0  \qquad (\mu = 0,1,\cdots , p) ~, \nn
 &X^{\mu} (\tau,\sigma) |_{\sigma=0,\pi}
      = {\rm const. } \qquad (\mu = p+1, \cdots , 9) ~.
\end{align}
We have assumed the flat background and we use $(\tau, \sigma)$ as
the coordinates of world-sheet. 
The open strings can move only along $p+1$ dimensional
hypersurface and the hypersurface is called as D$p$-brane.
$N$ D-branes can be described by open strings with $U(N) \times U(N)$
Chan-Paton factors at their ends.
In the limit that the string length $l_s = \sqrt{\alpha'} \to 0$,
only the massless excitations of open strings remain
and low energy effective theory on $N$ D$3$-branes
becomes $U(N)$ super Yang-Mills theory.
In $p=3$ case, the effective theory is given by 4 dimensional 
$\cN=4$ super Yang-Mills theory.

We can describe D-branes also in terms of supergravity.
The supergravity solution corresponding to multiple D$p$-branes
can be constructed \cite{Horowitz}
and for $N$ D$3$-branes it is given by 
\begin{align}
 ds^2 &= f^{-\frac{1}{2}} (-dx^2_0 + dx^2_1 + dx_2^2 + dx_3^2) 
   + f^{\frac{1}{2}} (dr^2 + r^2 d \Omega_5^2)~, \nn
 f &= 1 + \frac{R^4}{r^4} ~, \qquad R^4 = 4 \pi g_s {\alpha'}^{2} N~,
\label{D3-brane}
\end{align}
supported by nontrivial 5-form RR-field strength. 
The string coupling is denoted as $g_s$.
In the super Yang-Mills description, there are two decoupling sectors;
One is the super Yang-Mills theory on D-branes and the other is the
supergravity theory on flat background.
In the supergravity description, we have also two decoupling regions;
One is the region near D-branes and the other is the region away from
D-branes.
In order to extract the information at the region near D-branes, we take
the near horizon limit of D3-brane solution \eqn{D3-brane}.
By taking the limit $r \ll R$, the geometry becomes $AdS_5 \times S^5$
\begin{align}
 ds^2 &= \frac{r^2}{R^2} (-dx^2_0 + dx^2_1 + dx_2^2 + dx_3^2) 
   + \frac{R^2} {r^2} dr^2 + R^2 d \Omega_5^2~.
\label{AdS5}
\end{align}
Since the supergravity theory on the flat background away from
the D-branes are the same in the both descriptions, the super Yang-Mills theory
on D3-branes are conjectured to be dual to superstring theory on the 
$AdS_5 \times S^5$ background \cite{Maldacena}.
The $AdS_5$ space has a codimension one boundary and 
the dual CFT is believed to live in the 4 dimensional boundary.

One evidence of the conjecture is the correspondence of symmetry.
4 dimensional $\cN =4$ super Yang-Mills theory is known to be
a superconformal field theory.
The combination of $3+1$ dimensional Poincar\'{e} symmetry, 
dilatation symmetry and special conformal symmetry is described by 
$SO(4,2)$, and there is $SU(4)$ R-symmetry in $\cN=4$ supersymmetric
theory. 
The isometry of $AdS_5 \times S^5$ space is also given by
$SO(4,2) \times SO(6) (\backsimeq SU(4))$ and precisely reproduce
the symmetry of $\cN=4$ superconformal field theory.

In the large $N$ $U(N)$ Yang-Mills theory, effective coupling is given
by 't Hooft coupling $\lambda = g_{\rm YM}^2 N$.
By using the relation $g_{\rm YM}^2 = 4 \pi g_s$ and \eqn{D3-brane},
the perturbative region is given by
\begin{equation}
 g_{\rm YM}^2 N \sim g_s N \sim \frac{R^4}{l_s^4} \ll 1 ~.
\end{equation}
On the other hand, the supergravity approximation is valid when
\begin{equation}
 \frac{R^4}{l_s^4} \sim  g_s N  \sim  g_{\rm YM}^2 N \gg 1 ~,
\end{equation}
thus we may investigate super Yang-Mills theory at strong coupling
region $\lambda \gg 1$ by using the supergravity on $AdS_5 \times S^5$.

\begin{figure}
\begin{center}
\includegraphics{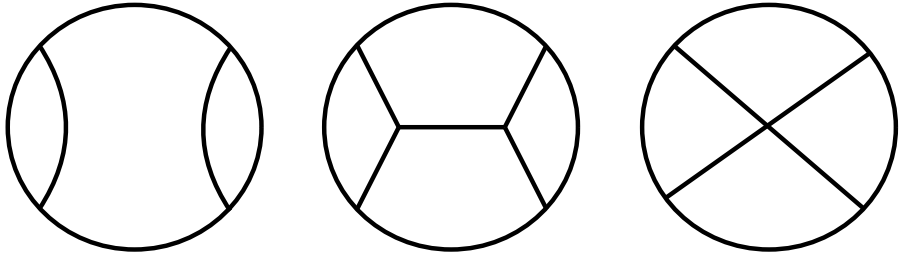}
\end{center}
\caption{Diagrams of 4-point function.}
\label{FIG:Benz}
\end{figure}

In a conformal field theory, basic objects are operators.
In the context of $AdS/CFT$ correspondence, it was proposed in 
\cite{GKP,Witten} that the correlation functions of operators 
are obtained from the following relations as
\begin{align}
 \left \langle e^{\int dx^4 \phi_0 (x) {\cal O}(x) } \right \rangle_{\rm CFT}
  = e^{-S_{\rm string}[\phi |_{\rm boundary} = \phi_0 (x)]} ~. 
\label{Benz}
\end{align}
In the CFT side, $\phi_0$ is a source coupling to a
operator and in the AdS side $\phi_0$ is the value of a field $\phi$ at
the boundary.
The left hand side is the generating function of correlation function
and the right hand side is the partition function of superstring
theory on $AdS_5 \times S^5$.
This equation implies that the operator ${\cal O}$ in the conformal
field theory corresponds to the field $\phi$ in the superstring theory.
In fact, it is shown that there is a one-to-one correspondence between
Kaluza-Klein modes compactified on $S^5$ in the supergravity 
and single trace BPS operators in the super Yang-Mills theory. 
In the CFT side, we obtain a correlation function by differentiating
source currents $\phi_0$. In the AdS side, the correlation function
corresponds to S-matrix for the fields $\phi$. 
For example, 4-point functions of operators $\cO$ in the super Yang-Mills
theory can be calculated by Feynman diagrams with the 4 external fields
$\phi$ as in figure \ref{FIG:Benz}.

\subsection{PP-Waves as Penrose Limits}
\label{Gueven}

In \cite{Penrose} Penrose pointed out that pp-wave 
(plane fronted wave with parallel rays) geometry is realized as a
limit of any geometry.
Furthermore, this limiting procedure is generalized in \cite{Gueven}
to supergravity on ten or eleven dimensional Lorentzian space-time $M$.

Let us denote a null geodesic of $M$ without conjugate points by $\gamma$.
In a neighborhood of $\gamma$, the metric of any geometry can be written
by introducing local coordinates $U$, $V$, $Y^i$ as
\begin{equation}
 g = dV \left( dU + \alpha (U,V,Y^k) dV + \sum_i \beta_i(U,V,Y^k) d Y^i \right)
   + \sum_{i,j} C_{ij} (U,V,Y^k) dY^i dY^j ~.
\end{equation}
We use $\alpha$ and $\beta_i$ as real numbers and $C_{ij}$ as a
symmetric positive-definite matrix.
In this coordinate, the null geodesic is given by the surface with
$V$, $Y^i = {\rm const.}$
and affine parameter is $U$.

In the Penrose limit, we see the neighborhood of the null
geodesic very closely. 
In other words, it is obtained by rescaling the coordinates as
\begin{equation}
 U=u ~, \qquad V=\Omega^2 v ~,\qquad Y^i = \Omega y^i ~,
\label{rescale}
\end{equation}
and taking the limit of $\Omega \to 0$.
Under this limiting procedure, the metric become
\begin{equation}
 \bar g = \lim_{\Omega \to 0} \Omega^{-2} g
        = du dv + \sum_{i,j} \bar C_{ij} (u) dy^i dy^j ~.
\label{scalemetric}
\end{equation}
Notice that the matrix $\bar C_{ij}$ only depends on $u$ after taking 
the limit.  This metric is of the form of pp-waves written by Rosen
coordinates.

In supergravity theory, there are a dilaton $\Phi$ and gauge potentials 
$A_p$. By using the gauge freedom, we choose a gauge such as
\begin{equation}
 A_{Ui_1 \cdots i_{p-1}} = A_{UVi_1 \cdots i_{p-2}} = 0 ~.
\end{equation} 
We use the dilaton and the gauge potentials after taking the Penrose limit as
\begin{equation}
 \bar \Phi = \lim_{\Omega \to 0} \Phi (\Omega)~, \qquad 
 \bar A_p = \lim_{\Omega \to 0} \Omega^{-p} A_p (\Omega)~.
\label{scalegauge}
\end{equation}
In this limit, the non-trivial components of gauge potentials are only 
transverse ones and field strengths are of the forms
\begin{equation}
 \bar F_{p+1} = du \wedge \frac{d}{du} \bar A_p (u) ~.
\label{FS}
\end{equation}
Because supergravity Lagrangians transform homogeneously under the
scaling of metric \eqn{scalemetric} and dilaton and gauge
potentials \eqn{scalegauge}, the Penrose limits of supergravity
solutions are still solutions to the supergravity equations of motion.

In the following section, we will use the expression of pp-wave metric
different from \eqn{scalemetric}, which is written by Brinkman
coordinates as 
\begin{equation}
 \bar g = -4 d x^+ d x^- + \sum_{i,j} A_{ij} (x^+) x^i x^j dx^+ dx^+ 
  + \sum_i d x^i dx^i ~.
\end{equation}
In eleven dimensional supergravity, there are three types of the
maximally supersymmetric solutions, i.e., flat, 
$AdS_{4,7} \times S^{7,4}$ and pp-wave background.
The maximally supersymmetric solution of pp-wave type was found in
\cite{KG} and its metric is written as
\begin{equation}
 g = -4 d x^+ d x^- + \left(\sum_{i=1}^3 \left(\frac{\mu}{3}\right)^2x^i x^i 
  +  \sum_{i=4}^9 \left(\frac{\mu}{6}\right)^2 x^i x^i \right)dx^+ dx^+ 
  + \sum_{i=1}^9 d x^i dx^i ~.
\end{equation}
The normalization factor $\mu$ can be changed by using the redefinition
of light-cone coordinates $x^{\pm}$.
Remarkably, this pp-wave background can be obtained as a Penrose limit of
$AdS_{4,7} \times S^{7,4}$ \cite{BFHP2}.  
In type IIB supergravity, flat and $AdS_5 \times S^5$
backgrounds are know to be maximally supersymmetric solutions.
Recently, it was found in \cite{BFHP} that there is another maximally
supersymmetric solution given by pp-wave geometry 
\begin{equation}
 g = -4 d x^+ d x^- + \sum_{i=1}^8 \mu^2 x^i x^i dx^+ dx^+ 
  + \sum_{i=1}^8 d x^i dx^i ~.
\label{RRpp-wave}
\end{equation}
This metric can be also obtained as a Penrose limit of $AdS_5 \times S^5$.
In next subsection, we apply this fact to the $AdS_5/CFT_4$
correspondence.

The above two coordinates (Rosen and Brinkman coordinates)
can be exchanged by performing the following transformation
\begin{align}
 u = -4 x^+ ~, \quad
 v = x^- + \sum_{i,j} \frac{1}{4} M_{ij} (x^+) x^i x^j ~, \quad
 y^i = \sum_{j} Q^i_j (x^+) x^j ~,
\label{R2B}
\end{align}
where $Q^i_j$ is a sort of inverse of $C_{ij}$ and $M_{ij}$ is chosen as
\begin{equation}
 C_{ij} Q^i_k Q^j_l = \delta_{kl} ~, \qquad M_{ij} = C_{kl} {Q'}^k_i Q^l_j ~.
\end{equation}
We use $'$ as the derivative $d/d x^+$.
The relation between $C_{ij}$ and $A_{ij}$ can be calculated as
\begin{equation}
 A_{ij} = - {C'}_{kl}{Q'}^l_j Q^k_i - C_{kl} {Q''}^l_j Q^k_i ~.
\end{equation}
Using the coordinate transformation \eqn{R2B}, we obtain the field
strength \eqn{FS} in terms of Brinkman coordinates as
\begin{equation}
 \bar F_{p+1} = 
 \sum_{i_k, j_k} \frac{d}{dx^+} \bar A_{i_1 \cdots i_p} (-4 x^+)
 Q^{i_1}_{j_1} \cdots  Q^{i_p}_{j_p} 
 dx^+ \wedge dx^{j_1} \wedge \cdots \wedge d x^{j_p} .
\end{equation}

\subsection{BMN Conjecture}
\label{BMN}

\begin{figure}
\begin{center}
\includegraphics{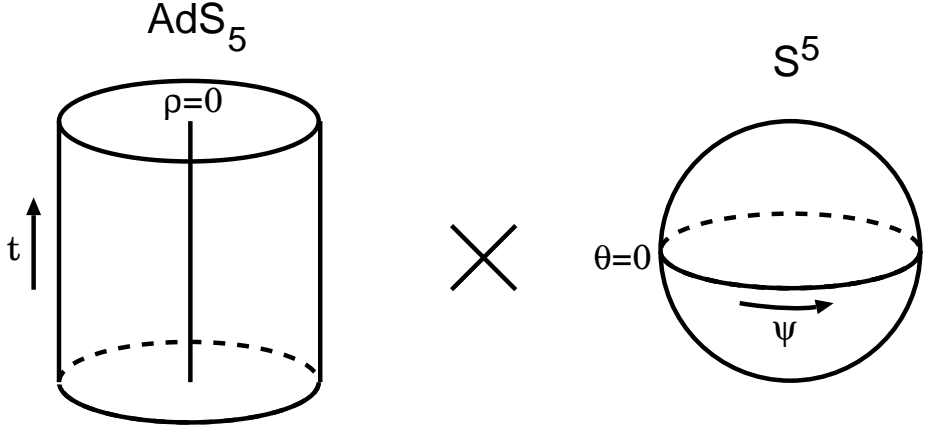}
\end{center}
\caption{We focus on a particle moving along an equator of $S^5$ very
 fast and staying at the center of $AdS_5$. This corresponds to taking a
 Penrose limit of $AdS_5 \times S^5$.}
\label{FIG:Penrose}
\end{figure}

As mentioned in the previous subsection, the maximally supersymmetric 
pp-wave arises as a Penrose limit of $AdS_5 \times S^5$.
To see this, it is convenient to rewrite the metric of  
$AdS_5 \times S^5$ \eqn{AdS5} in terms of global coordinates as
\begin{align}
 ds^2 = R^2 \left[
 - \cosh^2 \rho dt^2 + d \rho^2 + \sinh^2 \rho d \Omega^2_3 + 
 \cos^2 \theta d \psi^2  + d \theta^2 + \sin ^2 \theta d {\Omega'}^2_3
 \right] ~.
\label{global}
\end{align}
We choose a null geodesic as the trajectory of a particle moving very
fast along the $S^5$, more precisely, moving along the $\psi$ direction
and staying at $\theta = 0$ and $\rho = 0$. (See figure \ref{FIG:Penrose}.)
The Penrose limit corresponds to concentrating on this particle.
In terms of the coordinates, we rescale as
\begin{equation}
 x^+ = \frac12 (t+ \psi) ~,\quad
 x^- = R^2 \frac12 (t-\psi) ~,\quad
 r = R \rho ~,\quad y = R \theta ~,
\label{AdS2pp}
\end{equation}
and take the limit of $R = 1/\Omega \to \infty$.
The resultant metric is of the form of the maximally supersymmetric pp-wave 
\eqn{RRpp-wave}
\begin{align}
 &ds^2 = -4 dx^2 dx^- - \mu^2 {\vec{z}}^{\ 2} (dx^+)^2 + 
  d {\vec{z}}^{\ 2} ~,\nn
 &F_{+1234} = F_{+5678} = {\rm const.} \times \mu ~,
\label{RRpp-wave2}
\end{align}
with $\mu = 1$.
The vector $\vec{z}$ represents the points on $R^8$ and the
constant in front of $\mu$ is fixed by choosing a
normalization of field strengths.

Superstrings on the pp-wave \eqn{RRpp-wave2} can be quantized by using
Green-Schwarz formalism in the light-cone gauge \cite{Metsaev}.
In the gauge $x^+ = \tau$, the action becomes
\begin{equation}
 S = \frac{1}{2 \pi \alpha '} \int d \tau 
     \int ^{2 \pi \alpha ' p^+}_0 d \sigma  
     \left[ \frac{1}{2} (\partial_{\tau} \vec{z})^2
        -  \frac{1}{2} (\partial_{\sigma} \vec{z})^2 
        - \frac{1}{2} \mu^2 \vec{z}^{\ 2} 
   + i \bar S (\slash \!\!\! \partial + \mu \Gamma^{1234} S )\right] ~,
\end{equation}
with the world-sheet coordinates $(\tau,\sigma)$.
Green-Schwarz fermion $S$ consists of eight components, 
and they are Majorana spinors on the world-sheet
and space-time spinors with positive chirality of $SO(8)$.

The boundary condition and equation of motion for bosonic field $z^I$ 
are
\begin{align}
 &z^I (\sigma + 2 \pi \alpha ' p^+ , \tau ) = z^I (\sigma, \tau) ~,\nn
 &(\partial_{\tau}^2 - \partial_{\sigma}^2)z^I + \mu^2 z^I = 0~. 
\end{align}
The solution to these equations is given by the following mode
expansion as
\begin{align}
 z^I = z_0^I \cos \mu \tau + p_0^I \cos \mu \tau + i \sum_{n \neq 0}
 \frac{1}{\omega_n} \left( 
  \alpha_n^I e^{-i ( \omega_n \tau - k_n \sigma)}
 + \tilde \alpha_n^I e^{-i ( \omega_n \tau + k_n \sigma)} \right) ~,
\end{align}
where
\begin{align}
 \omega_n = \sqrt{\mu^2 + \frac{n^2}{(\alpha' p^+)^2}} &~~(n > 0) ~, \quad
 \omega_n = - \sqrt{\mu^2 + \frac{n^2}{(\alpha' p^+)^2}} ~~(n < 0) ~, \nn
 k_n &= \frac{n}{\alpha' p^+} ~~(n = \pm 1, \pm 2, \cdots) ~.
\end{align}

By replacing the coefficients $\alpha_n^I$ and $\tilde \alpha_n^I$ with
the corresponding operators, we can quantize the system and obtain the
light-cone Hamiltonian as
\begin{align}
 2 p^- = - p_+ = H_{\rm l.c.} = 
 \sum_{n} N_n \sqrt{\mu^2 + \frac{n^2}{(\alpha' p^+)^2}} ~.
\label{ppspectrum}
\end{align}
The number operator $N_n$ counts one for $\alpha_n$ and 
for $\tilde \alpha_{-n}$.
By virtue of large supersymmetry, the spectrum of fermions are
the same as that of bosons, and the number operator $N_n$ also
counts the fermionic oscillators in the way similar to the bosonic ones.
The level matching  condition is expressed in this notation as
\begin{align}
 \sum_n n N_n = 0 ~.
\label{lmc}
\end{align}

Since the pp-wave geometry can be obtained as a Penrose limit of $AdS_5
\times S^5$, there is a subsector of super Yang-Mills theory dual to
superstrings on the pp-wave.
As we mentioned before, the symmetry on conformal field theory
corresponds to the isometry of $AdS_5$ and the R-symmetry corresponds to
the isometry of $S^5$.
In the superconformal field theory, one of the important quantities is
the conformal weight $\Delta$, which represents how an operator
behaves under the dilatation transformation, and another important
quantity is the $U(1)$ R-charge $Q$ of $SU(4)$ R-symmetry.
The conformal weight $\Delta$ and the $U(1)$ R-charge $Q$ can be
identified as the energy $E=i \partial_t$ and the angular momentum along
$S^5$ $J=-i\partial_{\psi}$ in the coordinates of \eqn{global}.

Using this correspondence and the coordinate transformation
\eqn{AdS2pp}, the light-cone momenta of superstrings on the
pp-wave are written as
\begin{align}
 2p^- &= - p_+ = i (\partial_t + \partial_{\psi}) = \Delta - Q ~, \nn
 2p^+ &= - p_- = \frac{1}{R^2} i (\partial_t - \partial_{\psi})
       = \frac{\Delta + Q}{R^2} ~.
\label{lcmomenta}
\end{align}
Because the light-cone momenta should be finite in the limit of 
$R \to \infty$, we have to consider the states with
\begin{equation}
  \Delta - Q \sim \cO (1) ~, \qquad \Delta + Q \sim \cO (R^2) ~.
\label{ppcondition}
\end{equation}
The BPS bound is given as $\Delta \geq |Q|$, thus these states are 
almost BPS with very large R-charge $Q$.

Using the relation \eqn{lcmomenta}, we can see which operators in the 
super Yang-Mills theory should be identified as states in the 
superstring theory.
Since we have found that the spectrum of superstrings on 
the pp-wave is given by \eqn{ppspectrum}, 
the corresponding operators should have the following quantity
\begin{equation}
 (\Delta -Q)_n = \sqrt{1+ \frac{4 \pi g_s N n^2}{Q^2}} ~.
\label{anomalous}
\end{equation}
We have used the relation \eqn{D3-brane}.
In the remaining part of this section, we construct these operators
explicitly.

$\cN = 4$  $U(N)$ 
super Yang-Mills theory consists of four components of gauge
field $A_i$ $(i=1,2,3,4)$, six scalar fields $\phi^i$ $(i=1,2,3,4)$, 
$Z=\phi^5 +i \phi^6$, $\bar Z = \phi^5 - i \phi^6$ and sixteen
components of gaugino
$\chi^a_{J=1/2}$, $\chi^a_{J=-1/2}$ $(a=1,\cdots,8)$.
We use the adjoint representation of $U(N)$ for these fields and
summarize conformal weighs $\Delta$ and R-charges $Q$ of these
fields in table \ref{delta}.
\begin{table}
\begin{center}
\begin{tabular}{|c|c|c|}
 \hline
 ~ & $\Delta$ & $Q$ \\
 \hline 
 $A_i ~~(i=1,2,3,4) $&$ 1  $&$ 0 $\\ 
 $\phi^i ~~(i=1,2,3,4) $&$ 1  $&$ 0 $\\ 
 $Z=\phi^5 +i \phi^6$  & $1$  & $1$ \\ 
 $\bar Z = \phi^5 - i \phi^6$ & $1$ & $-1$ \\ 
 $\chi^a_{J=1/2} ~~(a=1,\cdots,8)$&$3/2$&$1/2$\\
 $\chi^a_{J=-1/2} ~~(a=1,\cdots,8)$&$3/2$&$-1/2$\\
 \hline
\end{tabular}
\end{center}
\caption{The conformal weights $\Delta$ and the R-charges $Q$ of fields
 in $\cN=4$ super Yang-Mills theory.}
\label{delta}
\end{table}

We are considering large $N$ limit and small fixed $g_{YM}$.
The operators we want to construct have large R-charge $Q \sim \sqrt{N}$
but finite $\Delta - Q$.
The operator with $\Delta -Q = 0$ can be constructed by using many $Z$'s
because this is only the field with $\Delta -Q = 0$.
This operator is identified with the vacuum state in the superstring theory,
thus we have
\begin{equation}
  \tr [Z^Q] \longleftrightarrow \ket{0,p_+} ~.
\end{equation}
We neglect the normalization for simplicity.

Next, we construct the operators with $\Delta - Q =1$.
We can find the fields with $\Delta - Q =1$ from table \ref{delta} as
\begin{align}
 \phi^i ~(i=1,2,3,4) ~,~~D_i Z = \partial_i Z + [A_i,Z] ~,~~
 \chi^a_{J=1/2} ~(a=1,\cdots,8) ~.
\label{D-Q=1}
\end{align}
They corresponds to eight bosonic oscillators and eight fermionic
oscillators 
\begin{equation}
 a_0^I ~ (I=1,\cdots,8) ~,\qquad S_0^a ~ (a=1,\cdots,8)~.
\end{equation}
We obtain the supergravity modes by acting these zero modes
to the vacuum. The corresponding operators can be constructed by
considering the following single trace operators
\begin{equation}
 \frac{1}{Q} \sum_l \tr [Z^l \phi^i Z^{Q-l}] = \tr [\phi^i Z^Q ]
  \longleftrightarrow 
   a^{\dagger   i+4}_0 \ket{0,p_+} ~.
\label{sugrapart}
\end{equation}
We have used the symmetric trace in order to make the operators gauge
invariant.
The operators corresponding to other supergravity modes are of the
form similar to \eqn{sugrapart}.

Let us turn to the stringy modes.
The stringy modes can be created by acting oscillators $a_n^{\dagger I}$ and
$S_n^{\dagger a}$ to the vacuum state.
It was proposed in \cite{BMN} that corresponding operators are given
by single trace operators with appropriate phase factors.
For example, they proposed the correspondence
\begin{equation}
 \sum_{l} \tr [\phi^3 Z^l \phi^4 Z^{Q-l}]e^{2 \pi i n l /Q}
   \longleftrightarrow 
   a^{\dagger  8}_n  a^{\dagger  7 }_{-n} \ket{0,p_+} ~.
\end{equation}
Notice that we need at least two oscillators in order to satisfy the level
matching condition \eqn{lmc}.
In general, the action of oscillators is replaced by the insertion
of operators by using following rules
\begin{align}
 a^{\dagger  i} ~(i=1,2,3,4) &\longleftrightarrow D_i Z ~,\nn
 a^{\dagger  i} ~(i=5,6,7,8) &\longleftrightarrow \phi^{i-4} ~,\nn
 S^a  ~(a=1,\cdots,8) &\longleftrightarrow \chi^a_{J=1/2} ~,
\end{align}
with appropriate phase factors.
It was shown that these operators correctly reproduce the
spectrum \eqn{anomalous} at the first non-trivial order in \cite{BMN},
at the next non-trivial order in \cite{Gross} and at the all order in
\cite{SZ}. 

\newpage
\section{NSNS PP-Wave as a Penrose Limit of $AdS_3 \times S^3$}
\label{NSNSPenrose}

In this thesis, we concentrate on $AdS_3/CFT_2$ correspondence rather
than $AdS_5/CFT_4$ correspondence.
The $AdS_3/CFT_2$ correspondence is related to $D1/D5$ system.
The near horizon geometry of the system is $AdS_3 \times S^3 (\times M^4)$
with $M^4=T^4$ or $K3$ and the effective field theory is given by a two 
dimensional superconformal field theory.
Therefore, it is conjectured that superstrings on $AdS_3 \times S^3$ is
dual to the two dimensional superconformal field theory.
One reason why we analyze this case is that two dimensional superconformal
field theory is easier to deal with than other dimensional one since
symmetry on two dimensional CFT is enhanced to be infinite dimensional.

There is U-duality family of D1/D5 system and we use S-dual one, namely,
F1/NS5 system, which is the system with fundamental strings and NS5-branes.
Superstring theory on the near horizon geometry includes only NSNS-flux and
hence the superstrings can be analyzed.
This is another reason why we investigate on this case.
For these reasons, many authors have worked on this subject as in,
for example, \cite{GKS,DORT,KS}.

In subsequent sections, we will investigate the  ``Penrose limit'' of
this correspondence. 
As we will see below, we can improve the study of correspondence of
spectrum by applying the Penrose limit.
In subsection \ref{PenroseLimit}, we see what the ``Penrose limit'' of
$AdS_3/CFT_2$ correspondence means.
Then, we show that superstrings on the Penrose limit
is described by a generalization of Nappi-Witten model \cite{NW}.
The Nappi-Witten model is the WZW model associated with the Nappi-Witten
group, which is introduced in subsection \ref{NW}.
In subsection \ref{LC}, we obtain the spectrum of superstrings on the
Penrose limit by using the sigma model approach in the
light-cone gauge \cite{RT2-2,RT2-3,RT}.


\subsection{$AdS_3/CFT_2$ Correspondence}
\label{D1D5}

As mentioned in subsection \ref{AdS_5/CFT_4}, the $AdS_5/CFT_4$
correspondence is related to the D3-branes.
For the $AdS_3/CFT_2$ correspondence, the role of D3-branes are played
by D1/D5 system.
\begin{figure}
\begin{center}
\includegraphics{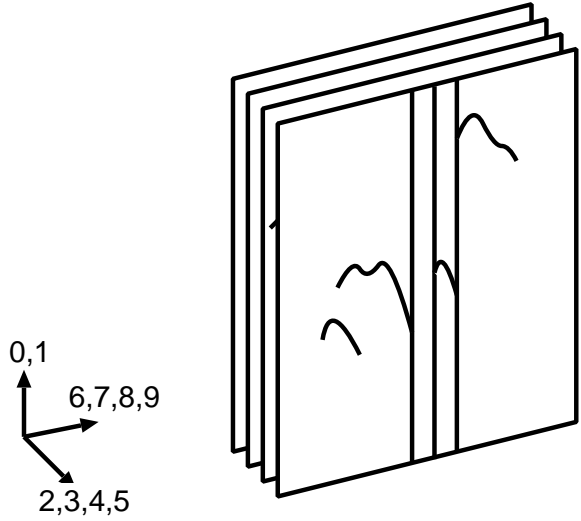}
\end{center}
\caption{The configuration of D1/D5 system. $Q_5$ D5-branes are wrapped on
 a small $T^4$ (6,7,8,9 directions) and $Q_1$ D1-branes are put on
 D5-branes along 0,1 directions.}
\label{FIG:D1D5}
\end{figure}
The D1/D5 system is a configuration with D5-branes and
D1-branes\footnote{See figure \ref{FIG:D1D5}}.
$Q_5$ D5-branes are wrapped on $T^4$ whose volume is fixed small as 
$v l_s^4$.
$Q_1$ D1-strings are set parallel to the extra $1+1$ dimensions.
This configuration can be seen as a one dimensional object in $5+1$
dimensional space-time.
We can replace  $T^4$ with $K3$ and the case of $K3$ will be investigated in
section \ref{T^4/Z_2}.

The supergravity solution corresponding to the D1/D5 system is given by 
\cite{HMS}
\begin{align}
 &ds^2 = f_1^{-\frac12} f_5^{-\frac12} (- d x_0^2 + d x_1^2) 
       +  f_1^{\frac12} f_5^{\frac12} (d r^2 + r^2 d\Omega_3^2) 
       +  f_1^{\frac12}  f_5^{-\frac12}  dx^A dx^A ~, \nn
& \qquad f_1 = \left( 1 + \frac{g \alpha ' Q_1}{v r^2}\right) ~,\quad
   f_5 = \left( 1 + \frac{g \alpha ' Q_5}{ r^2}\right) ~.
\end{align}
There is also non-zero 3-form RR-field strength.
We have set $g$ as the string coupling constant at infinity.
The near horizon geometry of the solution becomes
$AdS_3 \times S^3 \times T^4$
\begin{equation}
 ds^2 = \frac{r^2}{R^2} (- d x_0^2 + d x_1^2) 
       +  \frac{R^2}{r^2} d r^2 + R^2 d\Omega_3^2 
       + \sqrt{\frac{Q_1}{vQ_5}} dx^A dx^A ~,
\end{equation}
where 
\begin{align}
 g_6 = \frac{g}{\sqrt{v}} ~,\qquad R^2 = \alpha' g \sqrt{Q_1 Q_5} ~.
\label{R2QQ}
\end{align}
We should notice that the volume of $T^4$ is fixed as $Q_1/Q_5$ in the
near horizon limit.

According to the conjecture of Maldacena \cite{Maldacena}, 
superstring theory on $AdS_3 \times S^3 \times T^4$ is dual to
the effective theory on this D-brane configurations at the IR limit.
The effective theory is known as $\cN=(4,4)$ superconformal field theory
with central charge $c=6Q_1 Q_5$ and it has two decoupled sectors
called as Coulomb and Higgs branches.

In the D1/D5 system, there are open strings stretched between the same
type of D-branes and the different type of D-branes.
The former ones are called as $(1,1)$ and $(5,5)$ strings.
In the open string spectrum, there are massless scaler fields
corresponding to the positions of D-branes.
When these scalar fields have non-trivial values, the theory is at the
Coulomb branch.
The latter ones are called as $(1,5)$ and $(5,1)$ strings.
The massless scalar fields in the open string spectrum represent how
much the D1-branes are smeared into the D5-branes.
We are interested in the case when these scalar fields have non-trivial
values. 
In other wards, only the Higgs branch is considered.

Among the excitations of a $(1,5)$ string, there are two massless bosons
and two massless fermions.
Taking account of two possible orientations and the $Q_1$ and $Q_5$
degrees of freedom from $U(Q_1) \times U(Q_5)$ Chan-Paton factors, 
there are $4Q_1Q_5$ bosons and $4Q_1Q_5$ bosons.
In fact, it is conjectured in \cite{SV} that the effective theory on the
D1/D5 system is given by
$\cN=(4,4)$ non-linear sigma model on the symmetric orbifold 
\begin{equation}
Sym^{Q_1Q_5}(T^4) = (T^4)^{Q_1Q_5}/S_{Q_1Q_5} ~,
\label{symorb}
\end{equation}
where $S_N$ represents the symmetric group of $N$ elements. 
The degrees of freedom of massless fields are reproduced by
the analysis of massless excitations of open strings and
the symmetry of $S_N$ is a reminiscent of $U(Q_1) \times U(Q_5)$
Chan-Paton factors.

The form of target space \eqn{symorb} is explained as follows.
By applying T-dualities along $x^0$ and $x^1$ directions, we obtain 
D$(-1)$/D3 system, and $Q_1$ D$(-1)$-brane can be interpreted as $Q_1$
instanton solutions in $SU(Q_5)$ Yang-Mills theory.
It is supposed that a D$(-1)$-brane separates into $Q_5$ fractional D$(-1)$ 
branes with $1/Q_5$ D$(-1)$-brane charge.
The position of a fractional D-instanton corresponds to one $T^4$ and
the fact that these fractional D-instantons are indistinguishable 
leads to the $Q_1 Q_5$ symmetric products of $T^4$.
This is a heuristic explanation of the reason why the moduli space of this
instanton solution is given by \eqn{symorb}.

Let us perform S-duality to the D1/D5 system.
The S-duality transforms D1-strings to fundamental strings and D5-branes to
NS5-branes.
Thus, we obtain F1/NS5 system with $Q_5$ NS5-branes wrapped on the
$T^4$ and $Q_1$ fundamental string set parallel to the extra $1+1$
directions.
By using the S-duality, the metric of supergravity solution is obtained as
\begin{align}
 &ds^2 = f_1^{-1} (- d x_0^2 + d x_1^2) 
       + f_5  (d r^2 + r^2 d\Omega_3^2) 
       + dx^A dx^A ~, \nn
& \qquad f_1 = \left( 1 + \frac{g^2 \alpha ' Q_1}{v r^2}\right) ~,\quad
   f_5 = \left( 1 + \frac{\alpha ' Q_5}{ r^2}\right) ~.
\end{align}
There is also non-zero 3-form NSNS-field strength.

Its near horizon limit is also $AdS_3 \times S^3 \times T^4$
\begin{equation}
 ds^2 = Q_5 \frac{r^2}{\alpha'} (- d x_0^2 + d x_1^2) 
       +  \frac{Q_5 \alpha'}{r^2} d r^2 + Q_5 \alpha' d\Omega_3^2 
       +  dx^A dx^A ~,
\end{equation}
where we use the fact that the string coupling in this limit is 
$g^2=v Q_5/ Q_1$.
Superstring theory on curved space supported by NSNS-flux may be
described by a super WZW model, which is a solvable model,
and we can use a super WZW model as
superstring theory on $AdS_3 \times S^3$ with NSNS-flux. 
The isometry of $AdS_3 \times S^3 $ is given by 
\begin{equation}
SO(2,2) \times SO(4) \backsimeq SL(2,\br)_L \times SL(2,\br)_R \times
SU(2)_L \times SU(2)_R ~, 
\end{equation}
therefore we use $SL(2,\br) \times SU(2)$ WZW model.
The level of this model is related with the radius $R^2$ and given by $Q_5$.
We will concentrate on this case because this model is easier to deal
with than superstrings with RR-flux.

We have to apply the S-duality also to the effective field theory, 
however we do not know much
about the moduli space of non-linear sigma model on
$Sym^{Q_1Q_5} (T^4)$\footnote{
The discussions on the moduli space of D1/D5 system are given, for example, 
in \cite{Dijkgraaf,Larsen}}.
Because of this fact, we can only use objects independent of the
moduli parameter.
For example, BPS states are independent of the moduli parameters and 
they are identified as supergravity modes in superstrings on 
$AdS_3 \times S^3 \times T^4$ \cite{KLL,HS,HHS,AGS}.

As investigated in the previous section, we can compare non BPS states with
stringy modes by applying the Penrose limit to the $AdS_5/CFT_4$
correspondence.
Thus it is natural to expect that we can obtain the similar results by
applying the Penrose limit to the $AdS_3/CFT_2$ correspondence.
Since we concentrate on almost BPS states with very large R-charge, 
these states are not sensitive to the deformation of moduli parameter 
and can be compared with the stringy modes\footnote{
However, there is a subtlety on this discussion and we will comment on
the subtlety in subsection \ref{RR}.}.

\subsection{Penrose Limit of $AdS_3/CFT_2$ Correspondence}
\label{PenroseLimit}

In this subsection, we will see the ``Penrose limit'' of $AdS_3/CFT_2$
correspondence.
It is convenient to express the metric of $AdS_3 \times S^3 \times T^4$ as
($R^2=Q_5 l_s^2$)
\begin{align}
 ds^2 = &R^2 \left[
  - \cosh^2 \rho dt^2 + d \rho^2 + \sinh^2 \rho d \varphi_1^2 
  + \cos^2 \theta d \psi^2 + d \theta^2 + \sin^2 \theta d \varphi_2^2
 \right] \nn&+ d x^A d x^A~,
\end{align}
where $x^A$ $(A=6,7,8,9)$ are the coordinates of $T^4$.
As discussed in the previous section, 
the Penrose limit corresponds to concentrating
on the particles moving at the $\psi$ direction (one of the $S^3$
coordinates) very fast and staying at $\rho = \theta = 0$.
More precisely, we rewrite as
\begin{equation}
 \rho = \frac{r_1}{R} ~,~~\theta = \frac{r_2}{R} ~,~~
 t = - f u  + \frac{1}{4 f R^2} v ~,~~ \psi = f u + \frac{1}{4 f R^2} v ~,
\label{Penrose}
\end{equation}
and take the limit $R \to \infty$. Then we obtain the plane wave
metric as\footnote{
We can use arbitrary $f$ in front of $x^i x^i d u^2$ by rescaling $u$
and $v$.}
\begin{equation}
 ds^2 = du dv - f^2 x^i x^i d u^2 + d x^i d x^i + d x^A d x^A ~,
\label{ppmetric}
\end{equation}
which is supported by the NSNS-flux $H_{u12} = H_{u34} = -2f$.
Here we have introduced the coordinates $x^i$ ($i=1,2,3,4$) as 
\begin{equation}
 d x^i d x^i = 
  d r^2_1 + r^2_1 d \varphi^2_1 + d r^2_1 + r^2_2 d \varphi_2^2 ~.
\end{equation}

It is known that the superstring theory on this background supported 
by the NSNS-flux is described by a (generalized) Nappi-Witten model
\cite{NW}.
Contrary to the case with RR-flux, we can quantize this system in the
covariant gauge since we can use the technique of WZW models.
Some years ago, the Nappi-Witten model was investigated as a solvable
model for string theory on a non-trivially curved background
\cite{KK,Sfetsos,Sfetsos2,NW2,NW2-2,NW2-3,NW2-4,NW2-5,NW2-6,Sfetsos3,KK2,RT2}.
Recently, this model attracts some interests again in the context of 
PP-Wave/CFT duality \cite{RT,KP,HS',HS'2}. 
The sigma model approach was taken in \cite{RT2-2,RT2-3,RT}
and the current algebra method was developed  
in \cite{KK,KK2,KP} by using the free field realization.

Then, consider what this limiting procedure means in the context of the
dual CFT. Using the redefinition of the light-cone coordinates 
(\ref{Penrose}), we find 
\begin{align}
 p_{u} &=  - i \partial_u =  f(i \partial_t - i \partial_{\psi}) 
  = f ( - \Delta + Q)
  ~, \nn
 p_{v} &=  - i \partial_v = 
     \frac{ - i \partial_t -  i \partial_{\psi}}{4 f R^2} 
 =  \frac{\Delta + Q}{4 f R^2}  ~,
\label{Penrose2}
\end{align}
where we denote $\Delta$ as the conformal weight and $Q$ as the 
R-charge. 
The momenta of states in the superstring theory are finite, 
thus the corresponding states in the dual CFT are the states with  
$\Delta - Q \sim {\cal O}(1)$ and $ \Delta + Q \sim {\cal O}(R^2)$.
This is the exactly same as the case in subsection \ref{BMN}


\subsection{Nappi-Witten Group and Lie Algebra}
\label{NW}

Nappi-Witten model \cite{NW} is a WZW model associated with the
central extension of the two dimensional Euclidean group, 
which is called as Nappi-Witten group%
\footnote{A nice review is given in \cite{FS} and we follow their
discussion.}. 
Although we will use a generalization of Nappi-Witten model,
we first focus on the original Nappi-Witten group for simplicity.

We denote an element of the two dimensional Euclidean group as 
$h(\theta,w)$, which acts to two dimensional coordinate $z$ as
\begin{equation}
 h (\theta , w) \cdot z = e^{ - i \theta} z + w ~.
\end{equation}
In other words, the parameters $\theta$ and $w$ represent rotation
and translation, respectively, thus we can use $h(\theta,w) = T (w) R
(\theta)$ with 
\begin{equation}
 T(w) \cdot z = z + w ~,~~ R(\theta) \cdot z = e^{ - i \theta} z ~.
\end{equation}  
In order to define a group we have to determine a multiplication
law and in this case we can read as
\begin{equation}
 h (\theta_1,w_1) h(\theta_2,w_2) 
 = h (\theta_1 + \theta_2, w_1 + e^{ - i \theta_1}w_2 ) ~,
\end{equation}
or equivalently,
\begin{align}
 R(\theta_1) R(\theta_2) = R (\theta_1 + \theta_2) ~,~~
 R(\theta) T(w) = T(e^{ - i \theta} w) R (\theta) ~.
\end{align}

The Nappi-Witten group is given by the central extension of the two
dimensional Euclidean group and representation is characterized by a
group cocycle as
\begin{align}
 T (w_1) T(w_2) = T (w_1 + w_2) 
 \exp \left( \frac{1}{2} \im (w_1 \bar{w}_2) \right)~.
\end{align}
Introducing  $Z (\kappa)$, we can rewrite as
\begin{align}
 T (w_1) T(w_2) = 
 T (w_1 + w_2) Z \left(  \frac{1}{2} \im  (w_1 \bar{w}_2)\right)~.
\end{align}
Then, an element of Nappi-Witten group is defined as
\begin{equation}
g(\theta,w,\kappa) = T(w) R (\theta) Z(\kappa) ~,
\label{g}
\end{equation}
and the multiplication law is given by
\begin{equation}
 g(\theta_1,w_1,\kappa_1)g(\theta_2,w_2,\kappa_2) 
 = g \left( \theta_1 + \theta_2 , w_1 + e^{ - i \theta_1} w_2 , 
   \kappa_1 + \kappa_2 + \frac{1}{2}
    \im (w_1 e^{ i \theta_1} \bar w_2) \right ) ~.
\end{equation}
We can easily see that $g(0,0,0)$ is the identity and
$g(\theta,w,\kappa)^{-1} = g(-\theta, -e^{i \theta} w , -\kappa)$.

The generators of Nappi-Witten Lie algebra can be defined from the
Nappi-Witten group as
\begin{equation}
 T (w) = \exp \left( \frac{i}{\sqrt{2}}( w P^* + \bar{w} P ) \right) ~,~~
 R (\theta) = \exp ( i \theta J) ~,~~
 Z (\kappa) = \exp ( i \kappa F) ~,
\end{equation}
and their Lie brackets can be read from the multiplication laws as
\begin{equation}
 {[} J, P {]} = P ~,~~ {[} J, P^* {]} = - P^* ~,~~ {[} P , P^* {]} = F ~, 
\label{NWCR}
\end{equation}
which is nothing but a Heisenberg algebra.
Using the Lorentzian metric $\langle P , P^* \rangle = 1$ and 
$\langle J , K \rangle = 1$, we obtain the metric 
$d s^2 (g) = \langle g^{-1} d g , g ^{-1} d g \rangle $ as
\begin{align}
  d s^2 =   2 d\theta d\kappa
    + \frac{i}{2} ( w d \bar{w} - \bar{w} d w) d \theta 
    + d w d \bar{w} ~.
\label{ppmetric2}
\end{align}
Considering the WZW model associated with the Nappi-Witten group,
NSNS-flux is included from Wess-Zumino term as
$H_{\theta w \bar{w}} = - i/2$.
The generalization to higher dimensional one is obtained by
replacing $w \to w_i$, $P \to P_i$ and $P^* \to P^*_i$ ($i=1,\cdots,I$),
which correspond to the generators of Heisenberg algebra $H_{2I+2}$.
In this thesis, we concentrate on the case of $I=2$
since this case is obtained by the Penrose limit of superstring
theory on $AdS_3 \times S^3$.

By redefining the coordinates as
\begin{equation}
 w_1 \to e^{ - i \theta /2} y_1 ~,~~ w_2 \to e^{ - i \theta /2} y_2 ~ ,
\end{equation}
 we obtain the plane wave metric of the type (\ref{ppmetric})
\begin{equation}
 d s^2 = 2 d \theta d \kappa - \frac{1}{4} y^i \bar y^i d \theta^2 +
 d y^i d \bar y^i~.
\end{equation}
The relation to the metric \eqref{ppmetric} is given by 
$v \leftrightarrow 2 \kappa$, $u \leftrightarrow  \theta$ and $f = 1/2$.
The Killing vectors on this metric 
can be read from the group multiplication law as \cite{KP}%
\footnote{
Since the generator $F$ is central, $F_L$ and $F_R$ are not independent
of each other ($F=F_L=F_R$). 
}
\begin{align}
 F &= - i \partial_{\kappa} ~, \nn
 J_L &= - i\partial_{\theta} 
  + \frac{1}{2} (y^i \partial_{ y^i} - \bar y^i \partial_{\bar y^i} ) ~,~~
 J_R = - i\partial_{\theta} 
  - \frac{1}{2} (y^i \partial_{ y^i} - \bar y^i \partial_{\bar y^i} ) ~,\nn
 P_{iL} &= e^{i \theta /2} (\sqrt{2} i \partial_{\bar y^i} 
                   + \frac{1}{2 \sqrt{2}} y^i \partial_{\kappa} ) ~,~~
 P_{iR} = e^{i \theta /2} (\sqrt{2} i \partial_{y^i} 
                   + \frac{1}{2\sqrt{2}} \bar y^i \partial_{\kappa} ) ~,\nn
  P^*_{iL} &= e^{-i \theta /2} (\sqrt{2} i \partial_{y^i} 
                   - \frac{1}{2\sqrt{2}} \bar y^i \partial_{\kappa} ) ~,~~
  P^*_{iR} = e^{-i \theta /2} (\sqrt{2} i \partial_{\bar y^i} 
                   - \frac{1}{\sqrt{2}} y^i \partial_{\kappa} ) ~.
\end{align}
We can see that the commutation relation of these generators correctly 
reproduce (\ref{NWCR}).

Let us see classical scalar wave functions on these background, which
satisfy 
\begin{equation} 
(\Delta - m^2) \phi = 0 ~,~~
  \Delta = 2 \partial_{\theta} \partial_{\kappa}
         + \frac{1}{4} y^i \bar y^i   \partial_{\kappa} \partial_{\kappa}
         + 4 \partial_{y^i} \partial_{\bar y^i}  ~.
\label{KG}
\end{equation}
The ground state is given by
\begin{equation}
 \phi (\theta, \kappa , y^i) = \exp \left(
   i p_- \kappa  + i p_+ \theta  - \frac{1}{4} |p_-| y^i \bar y^i 
  \right) ~,
\end{equation}
and the arbitrary states are obtained by the action of $n_{iL}$ times of 
$P_{iL}$ and $n_{iR}$ times of $P_{iR}$ for $p_- > 0$\footnote{
For $p_- < 0$ we should replace $P_{iL}$ and $P_{iR}$ with $P^*_{iL}$ 
and $P^*_{iR}$, respectively.}.
For these states, the condition (\ref{KG}) becomes
\begin{equation}
 2 p_- p_+ + | p_- | \sum_{i=1}^2 (n_{iL} + n_{iR} + 1 ) + m^2 = 0 ~,
\label{disp}
\end{equation} 
and the helicity in the transverse space becomes
\begin{equation}
h = J_L - J_R = sgn(p_-) \sum_{i=1}^2 (n_{iL} - n_{iR}) ~.
\label{helicity}
\end{equation}
For $p_- = 0$ states, the wave function could have the transverse momenta but
there is no solution for general $m^2$.

The above results can be reproduced by the systematic analysis of
unitary irreducible representation of $H_6$ Lie algebra \cite{KK,KK2,KP}. 
There are two Casimir operators in the $H_6$ Lie algebra, which are
given by the generator $F$ and
\begin{equation}
 C = P_i P_i^* + P_i^* P_i + 2 J F ~.
 \end{equation}
We define the eigenstates of these operators as
\begin{align}
 J \ket{j,\eta} = j \ket{j,\eta} ~,~~
 F \ket{j,\eta} = \eta \ket{j,\eta} ~,~~
 C \ket{j,\eta} = c \ket{j,\eta} ~.
\end{align}
The eigenvalues of $F$ and $C$ are not changed in the same irreducible
representation.
By using the Hermitian property $(P_i)^{\dagger} = P^*_i$, we find
\begin{align}
\bra{j,\eta} P_i  P^*_i \ket{j,\eta} &= 
 \frac{c}{2}  - \eta (j -1) \geq 0 ~, \nn
\bra{j,\eta} P^*_i  P_i \ket{j,\eta} &= 
  \frac{c}{2} - \eta (j + 1) \geq 0~.
\end{align}
Thus, the following representations are obtained.

\begin{description}
\item[Type I] 
           There are no highest weight state and lowest weight
	   state. The labels take the values of 
           $\eta = 0$, $j \in \br$ and $c > 0$. 
           The eigenvalue of $J$ becomes $\cdots, j-1, j, j+1, \cdots$, 
           therefore the fractional part of $j$ labels the irreducible
	   representation.  
\item[Type II] 
            There are the highest weight states 
            $ P_i \ket{j,\eta} = 0 $ with $c = 2 \eta (j+1)$.
            This representation is characterized by $ \eta > 0$ and 
           $j \in \br$ and the eigenvalue of $J$ is $j, j-1, \cdots$.
\item[Type III] 
            There are the lowest weight states 
            $ P^*_i \ket{j,\eta} = 0 $ with $c = 2 \eta (j-1)$.
            This representation is characterized by $ \eta < 0$ and 
            $j \in \br$  and the eigenvalue of $J$ is $j, j+1, \cdots$.   
\item[Type IV]
            There is the state which is both the lowest weight and the
	   highest weight. The labels take the values of 
           $\eta = 0$, $j \in \br$ and $c = 0$.
           This is the one dimensional representation and we will not
	   consider this representation.
\end{description}

Now, we can compare with the analysis of classical scalar wave function.
The eigenvalues can be identified as $\eta_R = \eta_L = p_-$, 
$j_L = p_+ + h/2$ and $j_R = p_+ - h/2$.
Thus, the Type I representation corresponds to the case of $p_- = 0$.
The Type II representation corresponds to the case of $p_- > 0$ 
and the ground state can be identified as the lowest weight state.
Then, we can show that the Casimir operator $C$ and the operator 
$h = j_L - j_R$ correctly reproduce the mass square (\ref{disp}) 
and the helicity (\ref{helicity}), respectively.
For the Type III representation, we can make the similar analysis with
$p_- < 0$.


\subsection{Superstring on NSNS PP-Wave in the Light-Cone Gauge}

\label{LC}
In this subsection, we quantize the superstring on the pp-wave with
NSNS-flux by using the sigma model approach \cite{RT2-2,RT2-3,RT}.
The metric of the type  (\ref{ppmetric2}) is considered first and the
metric of the type (\ref{ppmetric}) is analyzed by using the coordinate
transformation.
In the light-cone gauge, we can explicitly show that physical
spectrum does not include negative norm states.
In the next section, we compare it with the spectrum from the BRST
quantization.

Let us first consider the following sigma model action.
The bosonic part of the action is given by
\begin{equation}
 S_B = \frac{1}{2\pi \alpha '} \int d^2 \sigma [
  \partial_+ u \partial _- v + \partial _+ z^i \partial_- \bar{z}^i
    +  i f \partial _+ u 
  ( z^i \partial_- \bar{z}^i - \bar{z}^i \partial_- z^i ) 
  + \partial _+ x^A \partial _- x^A] ~,
\label{sigmab}
\end{equation}
where $i =1,2$ and $\bar{z}^i$ is the complex conjugate of $z^i$.
The coordinates of $T^4$ sector are given by $x^A$ $(A = 6,7,8,9)$ 
and we neglect these coordinates and the corresponding fermionic
parts for a while.
We use the notation of world-sheet coordinates as
$\sigma_{\pm} = \tau \pm \sigma$ and $\partial_{\pm} = \frac{1}{2}
(\partial_0 \pm \partial_1)$.
The fermionic part of the action is given by
\begin{align}
 S_F = \frac{1}{2\pi \alpha '} \int d^2 \sigma [
 i \psi^{\hat{u}}_L \partial_-  \psi^{\hat{v}}_L + 
 i \bar{\psi}^i_L \partial_- \psi^i_L 
 & - f (\partial_- \bar{z}^i \psi^{\hat{u}}_L \psi^i_L
 -  \partial_- z^i \psi^{\hat{u}}_L \bar{\psi}^i_L ) \nn
  &+  i \psi^{\hat{u}}_R \partial_+ \psi^{\hat{v}}_R + 
 i \bar{\psi}^i_R \partial_+ \psi^i_R 
  - 2  f \partial_+ u \bar{\psi}^i_R \psi^i_R
 ] ~,
\label{sigmaf}
\end{align}
where the hatted indices are the ones of tangent space. 
The equations of motion can be calculated as
\begin{align}
 &\partial_- \partial_+ u = 0 ~,~~
  \partial_+ [ \partial_- v 
  +  i f ( z^i \partial_- \bar{z}^i - \bar{z}^i \partial_- z^i) 
   - 2  f \bar{\psi}^i_R \psi^i_R] ~, \nn
 &\partial_+ \partial_- z^i + 2 i f \partial_+ u \partial_- z^i -
   f \partial_-(\psi^{\hat{u}}_L \psi^i_L ) ~,~~ 
 \partial_+ \partial_- \bar{z}^i - 2 i f \partial_+ u \partial_- \bar{z}^i +
   f \partial_- (\psi^{\hat{u}}_L \bar{\psi}^i_L ) ~,
\end{align}
for the bosonic part and 
\begin{align}
 &\partial_- \psi^{\hat{u}}_L = 0 ~,~~
  \partial_- \psi^{\hat{v}}_L 
 - f (\partial_- \bar{z}^i \psi^i_L 
  - \partial_- z^i \bar{\psi}^i_L ) = 0 ~, \nn
   &\partial_- \psi^i_L  
     - f \partial_- z^i \psi^{\hat{u}}_L = 0 ~,~~
   \partial_- \bar{\psi}^i_L  
  + f \partial_- \bar{z}^i \psi^{\hat{u}}_L = 0 ~, \nn
 &\partial_+ \psi^{\hat{u}}_R = 0 ~,~~ \partial_+ \psi^{\hat{v}}_R = 0 ~,
 ~~ \partial_+ \psi^i_R + 2 i f \partial_+ u \psi^i_R = 0 ~,~~
  \partial_+ \bar{\psi}^i_R - 2 i f \partial_+ u \bar{\psi}^i_R = 0 ~,
\end{align}
for the fermionic part.

Since the light-cone coordinate $u$ and the corresponding fermions
$\psi^{\hat{u}}_L$ and $\psi^{\hat{u}}_R$ 
satisfy the equations of motion for free fields, we can
adopt the light-cone gauge as $u=u_0 + k_+ \sigma_+ + k_- \sigma_-$ and
$\psi^{\hat{u}}_L=\psi^{\hat{u}}_R=0$. 
Using the super Virasoro constraints $T_{++} = T_{--} = G_+ = G_- = 0$, 
the coordinate $v$ and the corresponding fermions $\psi^{\hat{v}}_L$ and 
$\psi^{\hat{v}}_R$ can be written by the transverse coordinates and the
corresponding fermions as in the flat case.

The general solutions for the transverse coordinates and the
corresponding fermions are given by
\begin{align}
 z^i = e^{- 2 i f k_+ \sigma_+} X^i ~,~~ \bar{z}^i = e^{ 2 i f k_+ \sigma_+}
 \bar{X}^i ~, ~~
 \psi^i_R = e^{- 2 i f k_+ \sigma_+} \eta^i_- ~,~~
  \bar{\psi}^i_R = e^{2 i f k_+ \sigma_+ } \bar{\eta}^i_- ~,
\end{align}
where $ X^i = X_+^i (\sigma_+) + X_-^i (\sigma_-)$ and 
$\eta^i_-$ are free complex bosons and fermions, 
respectively. 
Since the equations of motion for the fermions $\psi^i_L$ and
$\bar{\psi}^i_L$ are the ones for the free fermions in the light-cone gauge, 
they can be expanded as
\begin{align}
  &\psi^i_L = \sqrt{2 \alpha'} \sum_{n \in \bz} 
 \tilde d_n e^{-in\sigma_+}~~(\mbox{R sector}) ~,\nn
 &\psi^i_L = \sqrt{2 \alpha'} \sum_{r \in \bz + 1/2} 
 \tilde c_r e^{-ir\sigma_+}~~(\mbox{NS sector}) ~.
\end{align} 
In order to recover periodicity conditions $z^i ( \sigma + 2 \pi ,
\tau) = z^i (\sigma,\tau)$ and 
$\psi^i_R (\sigma + 2 \pi, \tau) = \pm \psi^i_R$\footnote{
The ($+$) sign and the ($-$) sign correspond to R-sector and
NS sector, respectively.}, it is convenient to redefine the fields as
\begin{align}
 X^i_+ = e^{ 2 i f k_+ \sigma_+} {\cal X}^i_+ ~,~~
 X^i_- = e^{ - 2 i f k_+ \sigma_-} {\cal X}^i_- ~,~~
 \eta^i_- = e^{ - 2 i f k_+ \sigma_- } \chi^i_- ~,
\end{align}
with periodicity conditions ${\cal X}^i_{\pm} (\sigma + 2 \pi , \tau) =
{\cal X}^i_{\pm} (\sigma ,\tau) $ and $\chi_- (\sigma + 2 \pi , \tau) =
\pm \chi_- (\sigma , \tau)$. 
The mode expansions of these fields are given by
\begin{align}
 &{\cal X}^i_+ = i \sqrt{\frac{1}{2}\alpha '} 
   \sum_n \tilde{a}^i_n e^{-  i n \sigma_+} ~,~~
 {\cal X}^i_- = i \sqrt{\frac{1}{2} \alpha '} \sum_n a^i_n e^{-  i n \sigma_-}
 ~,\nn
 &\chi^i_- = \sqrt{2 \alpha'} \sum_{n \in \bz} d^i_n e^{-in\sigma_-}
 ~~(\mbox{R sector}) ~,\nn
 &\chi^i_- = \sqrt{2 \alpha'} \sum_{r \in \bz + 1/2} c^i_r e^{-ir\sigma_-}
 ~~(\mbox{NS sector}) ~.
\end{align}

We can calculate stress tensors in this system as
\begin{align}
 &T_{--} = \partial_- u \partial_- v + \partial_- z^i \partial_-
 \bar{z}^i +  i f \partial_- u (z^i \partial_- \bar{z}^i -
 \bar{z}^i \partial_- z^i) + i \bar{\psi}^i_R \partial_- \psi^i_R - 2 f
 \partial_- u \bar{\psi}^i_R \psi^i_R ~, \nn 
 &T_{++} = \partial_+ u \partial_+ v + \partial_+ z^i \partial_+
 \bar{z}^i +  i f \partial_+ u (z^i \partial_+ \bar{z}^i -
 \bar{z}^i \partial_+ z^i) + i \bar{\psi}^i_L \partial_+ \psi^i_L ~,
\end{align}
and zero mode parts are obtained as
\begin{align}
 &L_0 = \frac{1}{4 \pi \alpha '} \int^{2\pi}_0 d \sigma T_{--}
      = \frac{k_- q_-}{2 \alpha '} 
      + \frac{1}{4} \sum_n (n +  2 f k_+ ) n \bar{a}^i_n a^i_n 
      + \sum_n n \bar{d}^i_n d^i_n ~, \nn
 &\tilde{L}_0 = \frac{1}{4 \pi \alpha '} \int^{2 \pi}_0 d \sigma T_{++}
      = \frac{k_+ q_+}{2 \alpha '} 
      + \frac{1}{4} \sum_n (n - 2 f k_+ ) n \bar{\tilde{a}}^i_n
     \tilde{a}^i_n + \sum_n n \bar{\tilde d}^i_n \tilde d^i_n ~,
\label{Vira}
\end{align}
for the R sector\footnote{
Virasoro generators for the NS sector can be obtained in the similar way.}.
We have defined as
\begin{align}
y = \frac{1}{2}(u + v) = y_0 + k \tau + \cdots ~,\quad
t = \frac{1}{2} (-u + v) = t_0 + l \tau + \cdots  ~,
\end{align}
and hence
\begin{align}
 &u = u_0 + k_+ \sigma_+ + k_- \sigma_- ~,~~ 
  k_{\pm} = \frac{1}{2} (k-l) ~, \nn
 &v = v_o + q_+ \sigma_+ + q_- \sigma_- + \cdots ~,~~
  q_{\pm} = \frac{1}{2} (k+l) ~, 
\end{align}
where we denote $\cdots$ as higher modes.
The momentum canonically conjugate to $y$ is given by
\begin{align}
 p_y = \frac{1}{4 \pi \alpha '} \int^{2\pi}_0 d \sigma [
 \partial_0 y +  i f (z^i \partial_- \bar{z}^i
   - \bar{z}^i \partial _- z ^i + 2 i \bar{\psi}^i_R \psi^i_R ) 
 ] =\frac{1}{2} \frac{k}{\alpha '} - 2 f J_R ~,
\label{p_y}
\end{align}
and the momentum canonically conjugate to $t$ corresponding to conserved
energy is obtained as
\begin{equation}
 E = - \frac{1}{4 \pi \alpha '} \int^{2\pi}_0 d \sigma [ \partial_0 t 
     +  i f (z^i \partial_- \bar{z}^i 
     - \bar{z}^i \partial_- z^i  + 2 i \bar{\psi}^i_R \psi^i_R ) ] 
   = - \frac{1}{2} \frac{l}{\alpha '} + 2 f J_R ~.
\label{E}
\end{equation}
We have defined $J_R$ as the operator corresponding to negative of 
right-mover part of angular momentum
\begin{align}
 &J_R =  \frac{1}{4} \sum_n (n + 2 f k_+ ) \bar{a}^i_n a^i_n + K
 ~, \nn
  &K =  \sum_n \bar{d}^i_n d^i_n ~(\mbox{R sector}) ~,~~ 
   K =  \sum_r \bar{c}^i_r c^i_r ~(\mbox{NS sector}) ~.
\label{J_R}
\end{align}

Let us turn to the quantization of the model.
We can quantize by replacing the coefficients $a^i_n$ 
$\tilde a^i_n$, $ d^i_n$, $\tilde d^i_n$, $c^i_r$ and $\tilde c^i_r$ 
with the corresponding operators whose (anti-)commutation relations are
\begin{align}
 &{[} a^i_n , \bar{a}^i_m {]} = \frac{4}{ n +  2 f k_+ } \delta_{nm}
  ~,~~
 {[} \tilde{a}^i_n , \bar{\tilde{a}}^i_m {]} = 
  \frac{4}{ n - 2 f k_+ } \delta_{nm}
  ~, \nn
 & \{ d^i_n , \bar{d}^i_m \} = \delta_{nm} ~,~~ \{ c^i_r , \bar c^i_s \} =
 \delta_{rs} ~, ~~ 
  \{ \tilde d^i_n , \bar{\tilde d}^i_m \} = \delta_{nm} ~,~~ 
  \{ \tilde c^i_r , \bar{\tilde c}^i_s \} =
 \delta_{rs} ~,
\end{align}
which are determined to be consistent with canonical 
(anti-)commutation relations.
For the bosonic modes, it is convenient to rewrite as ($n > 0$)
\begin{align}
 &{b}^{i\dagger}_{n,+} = a^i_{-n} \omega_{n,-} ~,~~
 b^i_{n,+} = \bar{a}^i_{-n} \omega_{n,-} ~,~~
 {b}^{i\dagger}_{n,-} = \bar a^i_{n} \omega_{n,+} ~,~~
 b^i_{n,-} = a^i_{n} \omega_{n,+} ~, \nn
 &{\tilde b}^{i\dagger}_{n,+} = \tilde a^i_{-n} \omega_{n,+} ~,~~
 {\tilde b^i}_{n,+} = \bar{\tilde a}^i_{-n} \omega_{n,+} ~,~~
 {\tilde b}^{i\dagger}_{n,-} =\bar{\tilde a}^i_{n} \omega_{n,-} ~,~~
 {\tilde b^i}_{n,-} =  {\tilde a}^i_{n} \omega_{n,-} ~,
\label{relabel}
\end{align}
with $\omega_{n,\pm} =\frac{1}{2}  \sqrt{(n \pm 2 f k_+)}$. 
We assume $0 < 2 f k_+ <1$ and we relabel the modes if $f k_+$ is not in
this range. This procedure corresponds to the enlargement of Hilbert
space by spectral flow symmetry
 (\ref{spectral flow}) as we will see below. For the zero modes, we use 
\begin{align}
 b^{i\dagger}_0 = \frac{1}{2} \sqrt{2 f k_+} \bar{a}^i_0 ~,~~ 
 b^i_0 = \frac{1}{2} \sqrt{2 f k_+} a^i_0 ~,~~ 
 \tilde  b^{i\dagger}_0 
  = \frac{1}{2} \sqrt{2 f k_+} { \tilde a}^i_0 ~,~~ 
 \tilde b^i_0 = \frac{1}{2} \sqrt{ 2 f k_+} \bar{\tilde a}^i_0 ~.
\end{align}
These new modes satisfy the standard commutation relations
\begin{align}
  {[} b^i_{n,\pm} , b^{i\dagger}_{m,\pm} {]} = \delta_{nm} ~,~~
  {[} \tilde b^i_{n,\pm} , \tilde b^{i\dagger}_{m,\pm} {]} = \delta_{nm}
 ~,~~ {[} b^i_0 , b^{i\dagger}_0 {]} = 1 ~,~~
  {[} \tilde b^i_0 , \tilde b^{i\dagger}_0 {]} = 1 ~.
\end{align}
Fock vacuum is define by 
$b^i_{n,\pm} \ket{0} = \bar{d}^i_{-n} \ket{0}= d^i_{n} \ket{0} = 0$ 
for $n > 0$ and $\bar{c}^i_{-r} \ket{0} = c^i_r \ket{0} = 0$ for $r > 0$. 
The Fock vacuum for the left-movers is defined in the similar way.

Now, we can obtain the spectrum from the quantum expressions of 
Virasoro constraints. 
In order to do this, we include the coordinates of $T^4$ sector $x^A$
$(A=6,7,8,9)$ and the corresponding fermions $\psi^A_L$ and $\psi^A_R$.
Mode expansions of these fields are defined in the way similar to the
other fields. 
It is convenient to use the quantum expression of $J_R$ (\ref{J_R})
\begin{align}
 &\hat{J}_R =  b^{i \dagger}_0 b^i_0 + \frac{1}{2} - 
  \sum_{n=1}^{\infty} (b^{i \dagger}_{n,+} b^i_{n,+} 
     - b^{i \dagger}_{n,-} b^i_{n,-}) + \hat{K} ~, \nn
  &\hat K =  \bar d^i_0 d^i_0 - \frac{1}{2} +
    \sum_{n=1}^{\infty} (\bar{d}^i_n d^i_n + d^i_{-n} \bar{d}^i_{-n} ) 
     ~~(\mbox{R sector}) ~, \nn
  & \hat K =  \sum_{r = 1/2}^{\infty} 
  (\bar{c}^i_r c^i_r + c^i_{-r} \bar{c}^i_{-r}) ~~(\mbox{NS sector}) ~,
\end{align}
and operators $N_L$ and $N_R$ (for the R-sector)
\begin{align}
&N_R=\sum_{n=1}^{\infty} 
   n (b^{i \dagger}_{n,+} b^{i}_{n,+} 
   + b^{i \dagger}_{n,-} b^{i}_{n,-} 
   + b^{A \dagger}_{n} b^{A}_{n} 
   + \bar d ^i_n d^i_n + d^i_{-n} \bar d^i_{-n}
   + d^A_{-n} d^A_{n})  ~, \nn
 &N_L=\sum_{n=1}^{\infty} 
 n (\tilde b^{i \dagger}_{n,+} \tilde b^{i}_{n,+} 
   + \tilde b^{i \dagger}_{n,-} \tilde b^{i}_{n,-}
   + \tilde b^{A \dagger}_{n} \tilde b^{A}_{n} 
   + \bar{\tilde d} ^i_n \tilde d^i_n + \tilde d^i_{-n} \bar{\tilde d}^i_{-n} 
   + \tilde d^A_{-n} \tilde d^A_{n})  ~.
\end{align}
The operators for the NS-sector can be defined in the similar way.
By using the usual $\zeta$-regularization, we can calculate zero
energy shift and obtain the quantum expressions of Virasoro operators
(\ref{Vira}) as
\begin{align}
 &\hat L_0 = \frac{1}{2} \alpha ' ( - E^2 + p_A^2 + p_y^2) + \hat N_R +
  2 \alpha ' (E + p_y)f \hat J_R ~, \nn
 &\hat{\tilde{L}}_0 
 = \frac{1}{2} \alpha ' ( - E^2 + p_A^2 + p_y^2) + \hat N_L +
  2 \alpha ' (E + p_y) f \hat J_R ~.
\end{align}
We use $\hat N_{R,L} = N_{R,L} - a$ with $a=0$ for the R-sector and
$a=1/2$ for the NS-sector.
Virasoro conditions $\hat{L}_0 = \hat{\tilde L}_0 = 0$ lead to
level matching condition $\hat N_L = \hat N_R$ and
\begin{equation}
 E^2 - p_A^2 = \frac{1}{\alpha '} (\hat N_L + \hat N_R) + p_y^2 
 + 4 f (p_y + E) \hat J_R ~. 
\end{equation}
{}From this condition, the light-cone energy can be read as
\begin{equation}
 H_{l,c} = - p_u = \frac{1}{2}(E - p_y) 
  = \frac{p^2_A}{4 p_v}  + \frac{1}{4 \alpha ' p_v} (\hat N _L + \hat N _R) 
   + 2 f \hat J_R ~,
\label{spectrum} 
\end{equation}
 with $p_v = \frac{1}{2} (p_y + E)$.

Next, let us consider the case with the metric (\ref{ppmetric}).
Although spectrum in this case can be calculated straightforwardly, 
it is simpler by making use of the coordinate transformation.
In the polar coordinates $z^1 = r_1 e^{i \varphi'_1}$ and 
$z^2 = r_2 e^{ i \varphi'_2}$, the Lagrangian (\ref{sigmab}) is
written as
\begin{align}
 L = & ~ \partial_+ u \partial_- v + 2 f (r_1^2 \partial_+ u \partial_-
 \varphi'_1 + r_2^2 \partial_+ u \partial_- \varphi'_2) \nn
 &+ \partial_+ r_1 \partial_- r_1 + r_1^2 \partial_+ \varphi'_1 \partial_-
 \varphi'_1 + 
  \partial_+ r_2 \partial_- r_2 + r_2^2 \partial_+ \varphi'_2 \partial_-
 \varphi'_2 ~. 
\end{align}
By rotating the frame as
\begin{equation}
 \varphi'_1 = \varphi_1 - f u ~, \qquad  \varphi'_2 = \varphi_2 - f u ~,
\label{rotate}
\end{equation}
the Lagrangian becomes that for sigma model on the metric
(\ref{ppmetric}) as
\begin{align}
  L  = &~ \partial_+ u \partial_- v 
     - f^2 (r_1^2 + r_2^2) \partial_+ u \partial_- u  \nn
     &+\partial_+ r_1 \partial_- r_1 
     + r_1^2 \partial_+ \varphi_1 \partial_- \varphi_1 
     +\partial_+ r_2 \partial_- r_2 
     + r_2^2 \partial_+ \varphi_2 \partial_- \varphi_2 \nn
 &+ f  r_1^2 (
  \partial_+ u \partial_- \varphi_1 - \partial_- u \partial_+ \varphi_1 )
  + f r_2^2 (
  \partial_+ u \partial_- \varphi_2 - \partial_- u \partial_+ \varphi_2 )
   ~. 
\end{align}
The explicit form of the spectrum can be obtained from (\ref{spectrum})
by using the coordinate transformations (\ref{rotate}) and 
\begin{align}
\hat{J}  
 =  - i \frac{\partial}{\partial \varphi_1} 
    - i \frac{\partial}{\partial \varphi_2} 
 \leftrightarrow  - i \frac{\partial}{\partial \varphi'_1} 
    - i \frac{\partial}{\partial \varphi'_2} 
 = \hat{J} \equiv \hat{J}_L - \hat{J}_R ~.
\end{align}
Under the coordinate transformations, we find
\begin{align}
 &E = - i \frac{\partial}{\partial t} \leftrightarrow
 - i \frac{\partial}{\partial t} 
 - i f  \frac{\partial}{\partial \varphi'_1}
 - i f  \frac{\partial}{\partial \varphi'_2} = E + f \hat{J} ~,\nn
 &p_y = - i  \frac{\partial}{\partial y} \leftrightarrow
 - i  \frac{\partial}{\partial y}  
 + i f  \frac{\partial}{\partial \varphi'_1}
 + i f  \frac{\partial}{\partial \varphi'_2} = p_y - f \hat{J} ~,
\end{align}  
and hence $p_u \leftrightarrow p_u - f \hat{J}$ and $p_v \leftrightarrow p_v$.
Therefore, the spectrum is given by
\begin{equation}
 H_{l,c} = - p_u = \frac{1}{2}(E - p_y) 
  = \frac{p^2_A}{4 p_v}  + \frac{1}{4 \alpha ' p_v} (\hat N _L + \hat N _R) 
   +  f (\hat J_L + \hat J_R )~.
\label{spectrum2}
\end{equation}
We will compare this spectrum with the one  obtained from the BRST
quantization in the next section.

\newpage

\section{Superstring Theory on NSNS PP-Wave}
\label{NSNSppwave}

In the previous section, we have shown that superstring theory on
the pp-wave background with NSNS-flux can be obtained as the Penrose limit
of superstring theory on $AdS_3 \times S^3 (\times M^4) $ 
($M^4 = T^4$ or $K3$) with NSNS-flux.
Moreover, we have reviewed the (generalized) Nappi-Witten model and
quantized this model by using the sigma model approach in the light-cone
gauge. 
The same analysis can be given by using the current algebra approach.
Superstring theory on $AdS_3 \times S^3 (\times M^4) $
with NSNS-flux can be described by the $SL(2;\br) \times SU(2)$ 
super WZW model and the Penrose limit is realized by contracting 
currents on this model \cite{Sfetsos,Sfetsos2,Sfetsos3}. 
The superstring theory in the Penrose limit is given by the super WZW model 
whose target space is the 6 dimensional Heisenberg group $H_6$ 
as we saw in subsection \ref{PenroseLimit}.
In this section, we analyze the $H_6$ super WZW model by using the
free field realization in the current algebra approach  \cite{KK,KK2,KP}. 
We use this approach because the similarity with
the analysis  developed in \cite{GKS} for the case of $AdS_3 \times S^3$
becomes more transparent.


\subsection{$H_6$ Super WZW Model as a Penrose Limit}

The superstring theory on $AdS_3 \times S^3$ can be described by the
$SL(2;\br) \times SU(2)$ super WZW model. We set the level of 
each current algebra to positive integer $k$, which corresponds to the
number of NS5-branes $Q_5$.  
(The bosonic parts of   $SL(2;\br)$ and $SU(2)$ current algebras
have the levels $k+2$ and $k-2$, respectively.)
It is convenient to use the superfield formalism with
supercoordinates $(z,\theta)$.
The supercurrents of this system are given by  
\begin{align}
&\cJ^A(z,\theta) = \sqrt{\frac{k}{2}}\psi^A(z)
+\theta J^A(z)~ (\mbox{for $SL(2;\br)$})~,\nn
&\cK^a(z,\theta) = \sqrt{\frac{k}{2}}\chi^a(z)
+\theta K^a(z)~ (\mbox{for $SU(2)$})~.
\label{super SL(2) SU(2)}
\end{align}
The total currents
$J^A(z)$ and $K^a(z)$ $(A,a = \pm , 3)$
satisfy the following operator product expansions (OPEs)
\begin{align}
& J^3 (z) J^3 (w) \sim - \frac{k}{2(z-w)^2} ~,~~
 J^3 (z) J^{\pm} (w) \sim \frac{\pm J ^{\pm}(w)}{z-w} ~, \nn
& J^+ (z) J^- (w) \sim \frac{k}{(z-w)^2} - \frac{2 J^3 (w)}{z-w} ~,
\end{align}
\begin{align}
& K^3 (z) K^3 (w) \sim \frac{k}{2(z-w)^2} ~,~~
 K^3 (z) K^{\pm} (w) \sim \frac{\pm K ^{\pm}(w)}{z-w} ~, \nn
& K^+ (z) K^- (w) \sim \frac{k}{(z-w)^2} + \frac{2 K^3 (w)}{z-w} ~.
\end{align}
The free fermions $\psi^A$ and $\chi^a$ are defined by the OPEs
\begin{align}
&\psi^3(z)\psi^3(w) \sim -\frac{1}{z}~,~~~ 
\psi^+(z)\psi^-(w) \sim \frac{2}{z-w}~,
\end{align}
\begin{align}
 \chi^3(z)\chi^3(w) \sim \frac{1}{z}~,~~~ 
\chi^+(z)\chi^-(w) \sim \frac{2}{z-w}~,
\end{align}
and transformed by the action
of the total currents as follows
\begin{align}
& J^3(z)\psi^{\pm}(w) \sim \psi^3(z)J^{\pm}(w) \sim 
\frac{\pm \psi^{\pm}(w)}{z-w}~, \nn
& J^{\pm}(z)\psi^{\mp}(w) \sim \mp \frac{2\psi^3(w)}{z-w}~,
\end{align}
\begin{align}
& K^3(z)\chi^{\pm}(w) \sim \chi^3(z)K^{\pm}(w) \sim 
\frac{\pm \chi^{\pm}(w)}{z-w}~, \nn
& K^{\pm}(z)\chi^{\mp}(w) \sim \pm \frac{2\chi^3(w)}{z-w}~.
\end{align}

The Penrose limit of this model is given by 
a noncompact super WZW model associated with
the 6 dimensional Heisenberg group $H_6$, which is a natural
generalization of the  Nappi-Witten model \cite{NW}.
According to \cite{Sfetsos,Sfetsos2,Sfetsos3}, 
we can obtain the supercurrents of this model by contracting
those of the $SL(2;\br)\times SU(2)$ WZW model.
We redefine the  supercurrents \eqn{super SL(2) SU(2)} as 
\begin{align}
 \cJ(z,\theta) &= \cK^3(z,\theta) + \cJ^3(z,\theta) ~,~~ 
 \cF(z,\theta) = \frac{1}{k} (\cK^3(z,\theta) - \cJ^3(z,\theta)) ~, \nn
 \cP_1(z,\theta)  &= \frac{1}{\sqrt{k}} \cJ^+(z,\theta) ~,~~
 \cP_1^*(z,\theta)  = \frac{1}{\sqrt{k}} \cJ^-(z,\theta) ~, \nn
 \cP_2(z,\theta)  &= \frac{1}{\sqrt{k}} \cK^+(z,\theta) ~,~~
 \cP_2^*(z,\theta)  = \frac{1}{\sqrt{k}} \cK^-(z,\theta) ~,
\label{contraction}
\end{align}
and take the limit $k \to \infty$ with keeping the eigenvalues of
$K^3_0-J^3_0$ order $\cO(k)$ but the eigenvalues of 
$K^3_0+J^3_0$ much smaller than $k$.
Then we obtain the supercurrent algebra of the  $H_6$ super WZW model as
\begin{align}
& {\cal J} (\theta , z) = \psi_J (z) + \theta J (z) ~,~~
 {\cal F} (\theta , z) = \psi_F (z) + \theta F (z) ~, \nn
& {\cal P}_i (\theta , z) = \psi_{P_i} (z) + \theta P_i (z) ~,~~
 {\cal P}_i^* (\theta , z) = \psi_{P_i^*} (z) + \theta P_i^* (z) ~,
\end{align}
where $i = 1,2$.
The total currents $J(z)$, $F(z)$, $P_i(z)$ and $P^*_i(z)$ 
satisfy the OPEs
\begin{align}
 J(z) P_i (w) &\sim \frac{P_i (w)}{z-w} ~,~~
 J(z) P^*_i (w) \sim - \frac{P^*_i (w)}{z-w} ~, \nn
 P_i (z) P_j^* (w) &\sim \delta_{ij} 
 \left( \frac{1}{(z-w)^2} + \frac{F(w)}{z-w}\right) ~,\nn
 J (z) F (w) &\sim \frac{1}{(z-w)^2}~.
\label{H6OPE}
\end{align}
Other OPEs have no singular terms. 
Their superpartners $\psi_J$, $\psi_F$, $\psi_{P_i}$ and $\psi_{P_i^*}$ 
are free fermions defined by
\begin{equation}
 \psi_{P_i} (z) \psi_{P_j^*} (w) \sim \frac{\delta_{ij}}{z-w} ~,~~
 \psi_J (z) \psi_F (w) \sim \frac{1}{z-w} ~.
\end{equation}
The total currents $J$, $F$, $ P_i$ and $P_i^*$ 
non-trivially act on these free fermions as
\begin{align}
& J (z) \psi_{P_i} (w) \sim \psi_J(z)P_i(w) \sim
\frac{\psi_{P_i}}{z-w} ~,\nn
& J (z) \psi_{P^*_i} (w) \sim \psi_J(z)P^*_i(w) \sim
- \frac{\psi_{P^*_i}}{z-w} ~,\nn
& P_i (z) \psi_{P_j^*} (w) \sim \psi_{P_i}(z)P_j^*(w) \sim
\delta_{ij} \frac{\psi_{F}}{z-w} ~.
\end{align}

In terms of the metric in subsection \ref{PenroseLimit}, 
$-J_0^3$ and $K^3_0$ 
corresponds to the space-time energy $- i \partial_t$ and
the angular momentum $-i\partial_{\psi}$, respectively. 
Therefore, the contraction of the currents is identified as 
 the transformation of the coordinates (\ref{Penrose}), 
see also (\ref{Penrose2}). 
In fact, this contraction can be regarded as a stringy extension of
Penrose limit since this contraction defines a transformation from
superstrings on $AdS_3 \times S^3$ to superstrings on the pp-wave.
In the dual CFT side,
the eigenvalue of $-J^3_0$ and $K^3_0$ are interpreted as 
conformal weight $\Delta$ and R-charge $Q$, respectively
and the contraction means focusing on almost BPS states with 
\begin{equation}
  \Delta+Q \sim k\gg 1~, ~~~ \Delta-Q \ll k ~,
\end{equation} 
as in subsection \ref{PenroseLimit}.


$\cN=1$ superconformal symmetry is realized as follows.
Because the total currents have non-trivial OPEs with fermions, it is
convenient to introduce bosonic currents, which can be treated
independently of fermions. They are defined as
\begin{align}
& \hat{J}  = J - \psi_{P_1} \psi_{P_1^*} - \psi_{P_2} \psi_{P_2^*} ~,~~
   \hat{F} =  F ~, \nn
& \hat{P}_i = P_i - \psi_F \psi_{P_i} ~,~~
   \hat{P}_i^* = P_i^* + \psi_F \psi_{P_i^*} ~,
\end{align} 
which again satisfy the same OPEs \eqn{H6OPE} but have no singular OPEs
with the free fermions.
$\cN=1$ supercurrent is now defined in the standard fashion
\begin{align}
 G &= \hat{J} \psi_F + \hat{F} \psi_J 
 + \sum_{i=1}^2 (\hat{P}_i \psi_{P_i^*} + \hat{P}_i^* \psi_i 
    + \psi_F \psi_{P_i} \psi_{P_i^*} )  \nonumber \\
  &=   J \psi_F + F \psi_J 
 + \sum_{i=1}^2 (\hat{P}_i \psi_{P_i^*} + \hat{P}_i^* \psi_i) ~,
\end{align}
and the total currents can be given by acting $G$ on the fermions
$\psi_J$, $\psi_F$, $\psi_{P_i}$ and $\psi_{P_i^*}$ as it should be.

We can continue the analysis by using the abstract current algebra
techniques in principle.   However, it is easier to make use of the 
following free field realizations as given in  \cite{KK,KK2}.
We introduce free bosons $X^+$, $X^-$, $Z_i$ and $Z_i^*$ $(i=1,2)$
defined by the OPEs
\begin{equation}
 X^+ (z) X^- (w) \sim  - \ln (z-w) ~,~~ 
 Z_i (z) Z_j^* (w) \sim - \delta_{ij} \ln (z-w) ~,
\end{equation}
and  rewrite the fermions $\psi_J$, $\psi_F$, $\psi_{P_i}$ and
$\psi_{P_i^*}$ as 
\begin{equation}
 \psi_F = \psi^+ ~,~~\psi_J = \psi^- ~,~~
 \psi_{P_i} = \psi_i e^{i X^+} ~,~~ \psi_{P_i^*} = \psi_i^* e^{-iX^+} ~,
\end{equation}
where the new fermions are defined by
\begin{equation}
 \psi^+ (z) \psi^- (w) \sim \frac{1}{z-w} ~,~~
 \psi_i (z) \psi_j^* (w) \sim  \frac{\delta_{ij}}{z-w} ~.
\end{equation}
The total currents can be now expressed as 
\begin{align}
& F = i \partial X^+ ~,~~ J = i \partial X^- ~, \nn
& P_i = e^{i X^+} ( i \partial Z_i + \psi^+ \psi_i) ~,~~
  P_i^* = e^{- i X^+} ( i \partial Z_i^* - \psi^+ \psi_i^*) ~,
\end{align}
and the bosonic currents are written as
\begin{align}
& \hat{F} = i \partial X^+ ~,~~ 
  \hat{J} = i \partial X^- - \psi_1 \psi_1^* - \psi_2 \psi_2^* 
   - 2 i \partial X^+~, \nn
& \hat{P}_i = e^{i X^+}  i \partial Z_i  ~,~~
 \hat{P}_i^* = e^{- i X^+} i \partial Z_i^*  ~.
\end{align}
In terms of these free fields
the superconformal current is rewritten as the standard form
of flat background
\begin{equation}
 G =  \psi^{+} i \partial X^- +  \psi^{-} i \partial X^+ 
     + \psi_i^* i \partial Z_i + \psi_i i \partial Z^*_i  ~.
\end{equation} 
We also introduce free fields  $Y^i$ and $\la^i$ $(i=1,2,3,4)$
to describe the remaining  $T^4$ sector as 
\begin{equation}
 \lambda^i (z) \lambda^j (w) \sim \frac{\delta^{ij}}{z-w} ~,~~  
 Y^i(z)Y^j(w) \sim -\delta^{ij}\ln (z-w)~.
\end{equation}

In order to analyze space-time fermions, we have to introduce spin
fields, which are defined by using bosonized fermions.
We bosonize the fermions as
\begin{align}
 \psi^{\pm 0} &= \psi^{\pm} =  e^{\pm i H_0}~, \nn
 \psi^{+ j} &= \psi_j = e^{+ i H_j} ~, \nn
 \psi^{- j} &=  \psi_j^* = e^{- i H_j} ~, \nn
 \psi^{\pm 3} &=
 \frac{1}{\sqrt{2}}(\lambda^1 \pm i \lambda^2 ) = e^{\pm i H_3}~,\nn 
 \psi^{\pm 4} &=
 \frac{1}{\sqrt{2}}(\lambda^3 \pm i \lambda^4 ) = e^{\pm i H_4}~,
\end{align}
and define the spin fields as
\begin{equation}
 S^{\epsilon_0 \epsilon_1 \epsilon_2 \epsilon_3 \epsilon_4} 
 = \exp \left(  \frac{i}{2} \sum_{j=0}^{4} \epsilon_j H_j \right) ~.
\label{spin field}
\end{equation}
The GSO condition imposes $ \prod_{j=0}^4\ep_j= + 1$ in the
convention of this thesis.
Precisely speaking, 
the OPEs including the spin fields are affected by cocycle factors
and they depend on the notation of gamma matrices, which is summarized
in appendix \ref{gamma}.


\subsection{Hilbert Space of $H_6$ Super WZW Model}

The irreducible representations of the current algebra
of generalized Nappi-Witten model (that is the $H_6$ WZW model)
are classified in \cite{KK,KK2}. 
We shall here focus on 
the Type II representation (corresponding to the highest weight 
representation of the zero-mode subalgebra in subsection
\ref{NW})
and later we discuss the other types of representations. 
The vacuum state (in the NS sector) is characterized by
\begin{align}
& J_0 \ket{j,\eta} = j \ket{j,\eta} ~,~~
 F_0 \ket{j,\eta} =  \eta  \ket{j,\eta} ~,\nn
& P_{i,n} \ket{j,\eta} = 0 ~,~({}^\forall n \geq 0) ~,~~
 P_{i,n}^* \ket{j,\eta} = 0 ~,~({}^\forall n > 0) ~, \nn
& \Psi_r \ket{j,\eta} = 0 ~,~({}^\forall r  > 0~, ~ r\in \frac{1}{2}+\bz)~ ,
\label{type II}
\end{align}
where $\Psi$ represents all the fermionic fields $\psi_J$, $\psi_F$,  
$\psi_{P_i}$ and $\psi_{P_i^*}$. We assume that $j\in \br$ and $0<\eta
<1$.

It is useful to rewrite this representation \eqn{type II} 
in terms of the free fields $X^{\pm}$, $Z_i$, $Z_i^*$, $\psi^{\pm}$,
$\psi_i$ and $\psi_i^*$. 
This is nothing but a Fock representation 
with the Fock vacuum defined by the vertex operator
\begin{equation}
 V = \exp \left( i j X^+ + i\eta X^-\right) \sigma_{\eta} ~,  
\end{equation}
where $\sigma_{\eta}$ is the (chiral) twist field. 
This field imposes the boundary conditions
\begin{align}
& i\partial Z_i(e^{2\pi i}z) = e^{-2\pi i \eta} i\partial Z_i(z)~,~~
\psi_i(e^{2\pi i}z) = e^{-2\pi i \eta} \psi_i(z)~, \nn
& i\partial Z^*_i(e^{2\pi i}z) = e^{2\pi i \eta} i\partial Z^*_i(z)~,~~
\psi^*_i(e^{2\pi i}z) = e^{2\pi i \eta} \psi^*_i(z)~,
\end{align}
which ensure the locality of $H_6$ supercurrents.
More precisely, $\sigma_{\eta}$ should have the following OPEs
\begin{align}
& i \partial Z_i (z) \sigma_\eta (w) \sim 
 (z-w)^{- \eta} {\tau^i}_\eta (w) ~,~~
 i \partial Z_i^* (z) \sigma_\eta (w) \sim 
 (z-w)^{\eta - 1} {{\tau '}^{i}}_\eta (w) ~, \nn
& \psi_i (z) \sigma_\eta (w) \sim 
 (z-w)^{- \eta} {t^i}_\eta (w) ~,~~
 \psi^*_i (z) \sigma_\eta (w) \sim 
 (z-w)^{\eta} {{t'}^{i}}_\eta (w) ~, 
\end{align}
where ${\tau^i}_\eta$, ${{\tau '}^{i}}_\eta$, ${t^i}_\eta$ and
${{t '}^{i}}_\eta$ are descendant twist fields.  
This twist operator $\sigma_{\eta}$ has the conformal weight
\begin{equation}
h(\sigma_{\eta})= 2 \times \frac{1}{2} \eta (1-\eta) 
 + 2 \times \frac{1}{2} \eta^2 = \eta ~.
\end{equation}

As already discussed in \cite{KK,KK2,KP}, there is the spectral flow symmetry 
\begin{align}
 J_n \to J_n ~,~~ F_n \to F_n + p \delta_{n,0} ~,~~ 
 P_{i,n} \to P_{i,n + p} ~,~~P_{i,n}^* \to P_{i,n-p}^* ~, \nn
 \psi_{J,\,r} \to \psi_{J, \, r} ~,~~ \psi_{F,\,r} \to \psi_{F,\,r} ~,~~ 
 \psi_{P_i,\,r} \to \psi_{P_i,\, r + p} ~,~~
 \psi_{P^*_i,\,r} \to \psi_{P^*_i,\,r-p}~,
\label{spectral flow}
\end{align}
and hence we also consider 
the flowed representation as the natural extensions
of \eqn{type II}. The vacuum states are given by (where we use $p\in \bz$ as
the spectral flow number)  
\begin{align}
& J_0 \ket{j,\eta, p} = j \ket{j,\eta, p} ~,~~
 F_0 \ket{j,\eta, p} =  (\eta+p)  \ket{j,\eta, p} ~,\nn
& P_{i,n} \ket{j,\eta,p} = 0 ~,~({}^\forall n \geq -p) ~,~~
 P_{i,n}^* \ket{j,\eta,p} = 0 ~,~({}^\forall n > p) ~, \nn
& \psi_{J,\,r} \ket{j,\eta,p} =
  0 ~,~({}^\forall r  > 0)~ ,~~
 \psi_{F,\,r} \ket{j,\eta,p} = 
  0 ~,~({}^\forall r  > 0)~ , \nn
& \psi_{P_i,\,r} \ket{j,\eta,p} = 
  0 ~,~({}^\forall r  > -p)~ ,~~
  \psi_{P^*_i,\,r} \ket{j,\eta,p} = 
  0 ~,~({}^\forall r  > p)~ ,
\label{flowed type II}
\end{align}
where $n \in \bz$ and $r \in 1/2 + \bz$.
This spectral flow symmetry \eqn{spectral flow} 
is actually the counterpart of that for $SL(2;\br)\times SU(2)$ WZW
model \cite{HHRS,MO} and it is not difficult to confirm it directly 
by taking the contraction \eqn{contraction}.  
We will later focus on the sectors with non-zero spectral flow number $p$
to realize the (almost) BPS states and discuss
the correspondence with the symmetric orbifold theory.
We can again realize the flowed representation \eqn{flowed type
II} by means of the Fock representation, in which the Fock 
vacuum corresponds to the vertex operator
\begin{equation}
 V = \exp \left( i j X^+ + i(\eta+p) X^-\right) \sigma_{\eta} ~.  
\end{equation}

We should comment on the other types of irreducible representations
of $H_6$ current algebra. (The irreducible representations of the 
zero-mode subalgebra are given in subsection \ref{NW}.)
The Type III representations have the lowest weights
and the eigenvalues of $F$ take $-1 < \eta < 0$.
The Type I representations have neither the highest
nor lowest weights and the eigenvalues of $F$ are $\eta=0$.
There are also their spectrally flowed representations. 
As in the case of $SL(2;\br)$ WZW model, (see, for example, \cite{MO}) 
the spectral flow symmetry interchanges the Type III representation with
the Type II representation. In fact, we can easily see that 
\begin{equation}
\cH^{(\msc{II})}_{j,\eta,p} \cong \cH^{(\msc{III})}_{j,\eta-1,p+1}~,
\label{II III}
\end{equation}
where $\cH^{\msc{(II)}}_{j,\eta,p}$ is the flowed Type II
representation defined by using the vacuum \eqn{flowed type II} 
and $\cH^{\msc{(III)}}_{j,\eta,p}$ is the flowed Type III representation
defined in the similar way.  
(We must also make the redefinitions of 
fermionic oscillators as $\psi'_{P_i,\,r}:= \psi_{P_i,\,r+1}$ and 
$\psi'_{P^*_i,\,r}:= \psi_{P^*_i,\,r-1}$ to equate the both sides of 
\eqn{II III}.) Therefore, we only have to consider the Type II representation 
if assuming the spectral flow symmetry.  We also note that 
$p\geq 0$ representations correspond to positive energy states and 
$p<0$ representations to negative energy ones in the original 
$AdS_3$ string theory.

On the other hand, the Type I representations seem to be
the counterparts of the principal continuous series in string theory
on $AdS_3$.  
Because there are no twisted fields in this sector, 
the vacua of Type I representations  correspond to the vertex
operators 
\begin{equation}
 V = \exp \left( ipX^- + 
i ( j + n ) X^+ + i p_i^* Z_i + i p_i Z_i^*\right) ~, ~~ 
 n \in \bz ~.
\end{equation}
If we do not consider the spectral flow $(p=0)$, there are only  
trivial massless states with zero energy, that do not propagate along
the transverse plane ($Z_i$, $Z_i^*$ directions) just as the classical
wave functions with $p_- = 0$ in subsection \ref{NW}.
For the cases of flowed Type I representations ($p\neq 0$),  many
physical states are possible. 
The spectra of light-cone energies in these sectors are continuous
and the strings freely propagate along the transverse plane. 
It is quite natural to identify these sectors with ``long string
sectors'' in string theory on $AdS_3$ \cite{MO}.
On the other hand, the Type II (and Type III) 
representation should correspond to ``short string sectors'', 
because they are the counterparts of the discrete series 
in string theory on $AdS_3$.   
The strings in these sectors cannot freely propagate 
along the transverse plane because the string coordinates $Z_i$, $Z_i^*$
are twisted.


\subsection{Physical Vertices of Superstring  on NSNS PP-Wave}

In this section we turn to physical vertex operators in
superstring theory on the pp-wave background with NSNS-flux.
We use the BRST quantization and physical states correspond to the 
elements of the BRST cohomology.
We shall concentrate on the Type II representation with $p\geq 0$ 
for the time being and we will discuss later the Type I representation. 
In order to construct the physical vertices, 
it is convenient to make use of 
the free field representation previously discussed.
We fix the Fock vacuum as 
\begin{equation}
\ket{j,\eta,p} = \sigma_{\eta} e^{ijX^+ + i(\eta+p)X^-}
\ket{0}~,~~ 0<\eta<1,~~~ p\in \bz_{\geq 0}~.
\end{equation}
The light-cone formalism is manifestly ghost free, however the
covariant formalism needs a no-ghost theorem.
In order to prove the no-ghost theorem, 
it is convenient to construct DDF operators
which are BRST invariant and generate the physical spectrum.
In our case, the spectrum in the light-cone gauge is obtained in 
(\ref{spectrum}) or (\ref{spectrum2}), and we construct DDF
operators which generate these spectrum in this subsection.
Here we introduce the superghosts $(\gamma, \beta)$ or the
bosonized ones $(\phi, \xi, \eta)$.
The BRST charge has the standard form of the free superstring theory as
\cite{FMS}
\begin{equation}
Q_{\msc{BRST}} = \oint \left\lb c\left(T-\frac{1}{2}(\partial \phi)^2
  -\partial^2\phi -\eta\partial\xi + \partial c b\right)
+\eta e^{\phi}G-b\eta\partial\eta e^{2\phi}\right\rb  ,  \\
\end{equation}
where $T$ and $G$ are the total stress tensor and the superconformal 
current constructed from the free fields $X^{\pm}$, $Z_i$, $Z_i^*$, $Y^i$,
$\psi^{\pm}$, $\psi_i$, $ \psi^*_i$ and $ \la^i$.

The most important vertex operators are the generators
of space-time supersymmetry algebra  \eqn{CR1}, \eqn{CR2} and \eqn{CR3}, 
which are defined in appendix \ref{STSS}.
The generators of bosonic part are nothing but the zero-modes of
world-sheet $H_6$ (total) currents 
\begin{align}
& \cJ = \oint \psi^- e^{-\phi} 
    =  \oint i \partial X^- =J_0  ~, \nn
& \cF = \oint \psi^+ e^{-\phi}   
     =   \oint i \partial X^+ =F_0 ~,\nn
& \cP_i = \oint \psi_i e^{i  X^+} e^{-\phi}  
= \oint \left( i \partial Z_i + \psi^+\psi_i \right)e^{i X^+}= P_{i,\,0}~,\nn
&  \cP_i^* = \oint \psi^*_i e^{- i  X^+} e^{-\phi} 
= \oint\left( i \partial Z^*_i 
- \psi^+\psi_i^* \right) e^{- i  X^+} = P_{i,\,0}^*  ~,
\label{PVF}
\end{align}
where we implicitly identify the operators by using the picture changing
operator\footnote{The picture means the number $n$ for $e^{-n \phi}$ in
vertices. We can change the picture of the vertices by making use of 
the picture changing operators without affecting the BRST cohomology.}. 
The generators of fermionic part are obtained by using the 
spin fields (\ref{spin field}) as (in the $(-1/2)$ picture)
\begin{align}
& \cQ^{++a} = \oint S^{+++aa} e^{i X^+} e^{- \frac{\phi}{2}} ~,~~ 
\cQ^{--a} =  \oint S^{+--aa} e^{-i X^+} e^{- \frac{\phi}{2}} ~,\nn 
& \cQ^{+-a} = \oint S^{-+-aa}  e^{- \frac{\phi}{2}} ~,~~
\cQ^{-+a} =  \oint S^{--+aa} e^{- \frac{\phi}{2}} ~.
\label{PVS}
\end{align}
These operators manifestly BRST invariant and 
they can locally act on the Fock space associated with
$\ket{j,\eta,p}$ (irrespective of the values of $j$, $\eta$ and $p$).
We can directly check that they generate the super pp-wave 
algebra (\ref{CR1}), (\ref{CR2}) and (\ref{CR3}). 
In particular, we note that
\begin{align}
& \{\cQ^{-+a}, \cQ^{+-b} \} = \ep^{ab}\cJ~, ~~~ 
\lb \cJ, \cQ^{\pm\mp a}\rb  =0~, 
\end{align}
which indicates that $\cQ^{-+a}$ and $\cQ^{+-a}$ play the role of 
supercharges with the ``Hamiltonian'' $\cJ$.
In terms of appendix \ref{KS}, these are the ``dynamical'' supercharges.

In order to analyze the spectrum of physical states
we further need to introduce DDF operators.
We propose the following DDF operators as the ``affine extension'' 
of \eqn{PVF} and \eqn{PVS}
\begin{align}
\cP_{i,\,n} &= \frac{1}{\sqrt{p+\eta}}
\oint \psi_i e^{i\frac{n+\eta}{p+\eta}X^+}e^{-\phi}~, \nn
\cP^*_{i,\,n} &=  \frac{1}{\sqrt{p+\eta}}
\oint \psi_i e^{i\frac{n-\eta}{p+\eta}X^+}e^{-\phi}~,\nn
\cQ^{++a}_n &=  \frac{1}{\sqrt{p+\eta}}
\oint S^{+++aa} e^{i\frac{n+\eta}{p+\eta}X^+}
 e^{- \frac{\phi}{2}} ~,\nn 
\cQ^{--a}_n &=   \frac{1}{\sqrt{p+\eta}}
\oint S^{+--aa}  e^{i\frac{n-\eta}{p+\eta}X^+}
 e^{- \frac{\phi}{2}}~.
\label{DDF1}
\end{align}
These operators are BRST invariant and locally act on the Fock space 
associated with $\ket{j,\eta,p}$ (of the fixed $\eta$ and $p$). 
Later we will compare the spectrum generated by these DDF operators
with the light-cone spectrum.
It is obvious that
\begin{align}
&\sqrt{p+\eta}\cP_{i,\,p}= \cP_i ~,~~ 
\sqrt{p+\eta}\cP^*_{i,\,-p}= \cP^*_i ~, \nn
&\sqrt{p+\eta}\cQ^{++a}_p = \cQ^{++a} ~,~~ 
\sqrt{p+\eta}\cQ^{--a}_{-p}=\cQ^{--a} ~.
\end{align}
Note that the supercharges $\cQ^{\pm \mp a}$
do not have such affine extensions because the BRST invariance cannot 
be preserved. The DDF operators \eqn{DDF1} satisfy  
the following (anti-)commutation relations 
(up to the picture changing and BRST exact terms)
\begin{align}
&\lb \cP_{i,\,m}, \cP^*_{j,\,n}\rb = \frac{m+\eta}{p+\eta}
\delta_{ij}\delta_{m+n,0}~,~~~
\{ \cQ_m^{--a}, \cQ_n^{++b} \} =  \ep^{ab}\delta_{m+n,0}~, \nn
&\lb \cJ , \cP_{i,\,n} \rb =  \frac{n+\eta}{p+\eta}\cP_{i,\,n}~, ~~~
\lb \cJ , \cP^*_{i,\,n} \rb =  \frac{n-\eta}{p+\eta}\cP^*_{i,\,n}~, \nn
&\lb \cJ , \cQ^{++a}_{n} \rb =  \frac{n+\eta}{p+\eta}\cQ^{++a}_{n}~, ~~~
\lb \cJ , \cQ^{--a}_{n} \rb =  \frac{n-\eta}{p+\eta}\cQ^{--a}_{n}~.
\label{DDF comm 1}
\end{align}
It is also useful to remark that $(\cP_{i,\,n}, \cQ^{++a}_n)$ and
$(\cP^*_{i,\,n}, \cQ^{--a}_n)$ are the supermultiplets 
with respect to the supercharges $\cQ^{+-a}$ and $\cQ^{-+a}$.
More precisely, we find the relations
\begin{eqnarray}
  \cP_{1,\,n}  &\stackrel{\cQ^{-+a}}{\longrightarrow}&
  \cQ_n^{++a}  ~\stackrel{\cQ^{-+(-a)}}{\longrightarrow}~
  \cP_{2,\,n} ~\stackrel{\cQ^{-+*}}{\longrightarrow}~  0 ~ \nn
  0~ \stackrel{\cQ^{+-*}}{\longleftarrow}~ 
  \cP_{1,\,n}  &\stackrel{\cQ^{+-(-a)}}{\longleftarrow}&
  \cQ_n^{++a} ~\stackrel{\cQ^{+-a}}{\longleftarrow}~
  \cP_{2,\,n} ~,
\label{CRPQ}
\end{eqnarray}
\begin{eqnarray}
  \cP^*_{1,\,n}  &\stackrel{\cQ^{+-a}}{\longrightarrow}&
  \cQ_n^{--a}  ~\stackrel{\cQ^{+-(-a)}}{\longrightarrow}~
  \cP^*_{2,\,n} ~\stackrel{\cQ^{+-*}}{\longrightarrow}~  0 ~ \nn
  0~ \stackrel{\cQ^{-+*}}{\longleftarrow}~ 
  \cP^*_{1,\,n}  &\stackrel{\cQ^{-+(-a)}}{\longleftarrow}&
   \cQ_n^{--a} ~\stackrel{\cQ^{-+a}}{\longleftarrow}~
  \cP^*_{2,\,n} ~,
\label{CRP*Q}
\end{eqnarray}
and the explicit forms of (anti-)commutation relations are
\begin{align}
& {[} \cQ^{-+a} , \cP_{1,n} {]} = 
  \frac{n + \eta}{p + \eta}  
   \cQ^{++a}_n ~,~~
 {[} \cQ^{+-a} , \cP^{*}_{1,n} {]} = 
  \frac{n - \eta}{p + \eta}  
   \cQ^{--a}_n ~, \nn
& {[} \cQ^{+-a} , \cP_{2,n} {]} = 
  \frac{n + \eta}{p + \eta}  
   \cQ^{++a}_n ~,~~
 {[} \cQ^{-+a} , \cP^{*}_{2,n} {]} = - 
  \frac{n - \eta}{p + \eta}  
   \cQ^{--a}_n ~, \nn
& \{  \cQ^{+-a} , \cQ^{++b}_n \} = 
 - \epsilon^{ab} \cP_{1,n} ~,~~
 \{  \cQ^{-+a} , \cQ^{--b}_n \} =
 \epsilon^{ab} \cP^{*}_{1,n} ~, \nn
& \{  \cQ^{-+a} , \cQ^{++b}_n \} = 
 \epsilon^{ab} \cP_{2,n} ~,~~
 \{  \cQ^{+-a} , \cQ^{--b}_n \} = 
 \epsilon^{ab} \cP^{*}_{2,n} ~.
\end{align}

In order to construct the remaining DDF operators for the $T^4$ 
directions, it is convenient to relabel the fermions $\la^i$ as 
\begin{align}
 \lambda^{+-} &= \frac{1}{\sqrt{2}} (\lambda^1 + i \lambda^2 ) ~,~~ 
 \lambda^{-+} = \frac{1}{\sqrt{2}} ( - \lambda^1 + i \lambda^2 ) ~,\nn
 \lambda^{++} &= \frac{1}{\sqrt{2}} ( - \lambda^3 - i \lambda^4 ) ~,~~ 
 \lambda^{--} = \frac{1}{\sqrt{2}} ( - \lambda^3 + i \lambda^4 ) ~,
\end{align}
and the free bosons $Y^{a\da}$ as in the similar way.
These fields satisfy the following OPEs 
\begin{equation}
 \lambda^{a \dot{a}} (z) \lambda^{b \dot{b}} (w) \sim 
  \frac{ \epsilon^{ab} \epsilon^{\dot{a} \dot{b}} }{z-w} ~, ~~~
 Y^{a\da}(z)Y^{b\db}(w) \sim - \ep^{ab}\ep^{\da\db} \ln (z-w)~. 
\end{equation}
Then the DDF operators can be given by
\begin{align}
 \cA^{a \dot{a}}_n &= \frac{1}{\sqrt{p+\eta}}
 \oint \lambda^{a \dot{a}} e^{i \frac{n}{p + \eta}  X^+} e^{-\phi} 
=  \frac{1}{\sqrt{p+\eta}}
 \oint i\partial Y^{a \dot{a}} e^{i \frac{n}{p + \eta}  X^+}  
~, \nn
 \cB^{+ \pm}_n &= - \frac{1}{\sqrt{p+\eta}} \oint S^{+ - + \mp \pm}  
 e^{i \frac{n}{p+\eta}  X^+} e^{-\frac{\phi}{2}} ~, \nn
 \cB^{- \pm}_n &=  \frac{1}{\sqrt{p+\eta}} \oint S^{+ + - \mp \pm}  
 e^{i \frac{n}{p + \eta}  X^+} e^{-\frac{\phi}{2}} ~,
\label{DDF2}
\end{align}
and they satisfy  
\begin{align}
& {[}  \cA^{a \dot{a}}_m , \cA^{b \dot{b}}_n {]}
  = \frac{m}{p+\eta} \delta_{m+n} \epsilon^{ab} \epsilon^{\dot{a} \dot{b}} ~,~~
 {[}  \cB^{\al \dot{a}}_m , \cB^{\beta \dot{b}}_n {]}
  =  \delta_{m+n} \epsilon^{ab} \epsilon^{\dot{a} \dot{b}} ~,\nn
&\lb \cJ,  \cA^{a \dot{a}}_n \rb = \frac{n}{p+\eta}\cA^{a \dot{a}}_n~,~~~
\lb \cJ,  \cB^{\al \dot{a}}_n \rb = \frac{n}{p+\eta}\cB^{\al \dot{a}}_n~.
\label{DDF comm 2}
\end{align}
The pair of $(\cA^{a\da}_n, \cB^{\al\da}_n)$ again become supermultiplets 
with respect to $\cQ^{\pm \mp a}$ 
\begin{eqnarray}
  \cA^{a\da}_n  &\stackrel{\cQ^{\mp \pm (-a)}}{\longrightarrow}&
  \cB^{\pm \da}_n  ~\stackrel{\cQ^{\pm \mp(-a)}}{\longrightarrow}~
  \cA^{(-a)\da}_n ~\stackrel{\cQ^{**(-a)}}{\longrightarrow}~  0 ~ \nn
  0~ \stackrel{\cQ^{**a}}{\longleftarrow}~ 
  \cA^{a\da}_n   &\stackrel{\cQ^{\pm\mp a}}{\longleftarrow}&
  \cB^{\pm \da}_n  ~\stackrel{\cQ^{\mp \pm a}}{\longleftarrow}~
  \cA^{(-a)\da}_n ~,
\label{CRAQ}
\end{eqnarray}
more precisely,
\begin{align}
& {[} \cQ^{+-a} , \cA^{b \dot{b}}_n {]} = 
  \frac{n}{p + \eta} \epsilon^{ab} 
   \cB^{- \dot{b}}_n ~,~~
 {[} \cQ^{-+a} , \cA^{b \dot{b}}_n {]} = 
  - \frac{n}{p + \eta} \epsilon^{ab}  
   \cB^{+ \dot{b}}_n ~,\nn
& \{  \cQ^{-+a} , \cB^{- \dot{b}}_n \} = 
 - \cA^{a \dot{b}}_n ~,~~
 \{  \cQ^{+-a} , \cB^{+ \dot{b}}_n \} = 
 - \cA^{a \dot{b}}_n ~.
\end{align}


\subsection{Spectrum of Physical States}

Let us consider the spectrum which is generated by the DDF operators
constructed in the previous subsection. 
More precisely speaking, we are interested in  the physical states 
corresponding to almost BPS states in the dual CFT
characterized by 
\begin{equation}
     \Delta + Q \sim k \gg 1~, ~~~ \Delta-Q \ll k ~.
\label{BPS condition}
\end{equation}
This condition is equivalent to\footnote
   {In our convention, the BPS inequality is equivalent to $\cJ \leq 0$.}   
\begin{equation}
\cF \gtrsim 1~, ~~~|\cJ| \ll k ~,
\end{equation}
and all string excitations satisfy this condition  
in the Penrose limit $k \to + \infty $. 
BPS states correspond to the ones with $\cJ=0$ and belong to  
the short multiplets of the superalgebra \eqn{PVF} and \eqn{PVS}.
The condition $\cF \gtrsim 1$ only leads to the restriction $p\geq 1$
with respect to the spectral flow number $p$.
(We only consider the positive energy states.)

We concentrate on  the sectors with no momenta along $T^4$ direction
for the time being. 
It is not difficult to write down the complete list of BPS states for
each of the fixed $p$ and $\eta$. 
In the NS sector, we obtain (where we focus on the left-movers only)
\begin{align}
& \ket{\omega^0 ; \eta, p} 
= \psi_{1,-\frac{1}{2} + \eta} \ket{0,\eta,p} 
    \otimes c e^{-\phi} \ket{0}_{\msc{gh}} ~, \nn
& \ket{\omega^2 ; \eta, p}  = \psi_{2,-\frac{1}{2} + \eta} \ket{0,\eta,p} 
    \otimes c e^{-\phi} \ket{0}_{\msc{gh}} ~, 
\label{BPS NS}
\end{align}
and we have two more BPS states in the R-sector as
\begin{align}
& \ket{\omega^{1 \pm} ; \eta, p} =
  (S^{-++\mp \pm})_{-\frac{5}{8} + \eta } \ket{0,\eta,p} 
 \otimes c e^{-\frac{\phi}{2}} \ket{0}_{\msc{gh}} ~.
\label{BPS R}
\end{align}
Here we point out the next relation, which is useful for our discussion
\begin{eqnarray}
  \ket{\omega^0;\eta,p}  &\stackrel{\cB_0^{+\dot{a}}}{\longrightarrow}&
  \ket{\omega^{1\dot{a}};\eta,p}  
 ~\stackrel{\cB_0^{+(-\dot{a})}}{\longrightarrow}~
  \ket{\omega^2;\eta,p} ~\stackrel{\cB_0^{+ *}}{\longrightarrow}~  0 ~ \nn
  0~ \stackrel{\cB_0^{- *}}{\longleftarrow}~ 
  \ket{\omega^0;\eta,p}  &\stackrel{\cB_0^{-(-\dot{a})}}{\longleftarrow}&
  \ket{\omega^{1\dot{a}};\eta,p}~\stackrel{\cB_0^{- \dot{a}}}{\longleftarrow}~
  \ket{\omega^2;\eta,p} ~.
\label{BPS relation}
\end{eqnarray}
There is an obvious correspondence between the BPS states and
 the ``chiral part'' of 
the cohomology ring of $T^4$ by identifying $\cB^{+\da}_0$ with 
holomorphic one-form $dZ^{\da}$ on $T^4$. 
Therefore, emphasizing the correspondence to $H^*(T^4)$,
the non-chiral BPS states can be explicitly written as
\begin{equation}
\ket{\om^{(q,\bar{q})}; \eta, p} = \ket{\om^q;\eta,p}\otimes 
  \overline{\ket{\om^{\bar{q}};\eta,p}} ~,~~~
({}^{\forall} \om^{(q,\bar{q})} \in H^{q,\bar{q}}(T^4)) ~. 
\end{equation}
These states have degenerate charge $\cF = p+\eta$ 
for each sector of $p$ and $\eta$.

We can construct the other types of physical states by making  
the DDF operators \eqn{DDF1} and \eqn{DDF2} act on these BPS states.
We first note that the BPS states are actually the Fock vacua 
with respect to \eqn{DDF1} and \eqn{DDF2};
\begin{align}
& \cP_{i,\,n}\ket{\om;\eta,p}=0~,~({}^{\forall}n\geq 0)~,~~~
\cP^*_{i,\,n}\ket{\om;\eta,p}=0~,~({}^{\forall}n>0)~,  \nn
& \cQ_n^{++a}\ket{\om;\eta,p}=0~,~({}^{\forall}n\geq 0)~, ~~~
\cQ_n^{--a}\ket{\om;\eta,p}=0~,~({}^{\forall}n> 0)~, \nn
& \cA^{a\da}_n\ket{\om;\eta,p}=0~,~({}^{\forall}n\geq 0)~,~~~
\cB^{\al\da}_n\ket{\om;\eta,p}=0~,~({}^{\forall}n>0)~.
\label{Fock vacua}
\end{align}
(For $\cB^{\al\da}_0$, see \eqn{BPS relation}.)
We  hence  obtain  
\begin{align}
&  \cB_{-n_1}^{\alpha_1 \dot{a}_1} \cdots 
  \cQ_{-m_1}^{++b_1} \cdots 
  \cQ_{-k_1}^{--c_1} \cdots
  \otimes
  \bar{\cB}_{-\bar{n}_1}^{\bar{\al}_1 \bar{\dot{a}}_1} \cdots 
  \bar{\cQ}_{-\bar{m}_1}^{++\bar{b}_1} \cdots 
  \bar{\cQ}_{-\bar{k}_1}^{--\bar{c}_1} \cdots
\ket{\om; \eta, p} ~,\nn
& \hspace{1.7cm} n_i, \bar{n}_i, m_i, \bar{m}_i >0~,~ k_i, \bar{k}_i \geq 0~,~
 {}^{\forall}\om \in H^*(T^4)~,
\label{almost BPS1}
\end{align} 
as typical physical states.
These states become almost BPS under the condition
\begin{align}
& |\cJ+\bar{\cJ}| \equiv  
  \frac{1}{p+\eta} \left\lb 
\left(\sum_i n_i + \sum_i (m_i-\eta) + \sum_i (k_i+\eta) \right) 
 \right.\nn
& \hspace{2.5cm} +  \left. \left(\sum_i \bar{n}_i + \sum_i (\bar{m}_i-\eta) 
+ \sum_i (\bar{k}_i+\eta)\right) \right\rb  \ll  k  ~,
\end{align}
which is always satisfied for sufficiently large $k$.
Other states can be obtained by multiplying the supercharges 
$\cQ^{\pm\mp a}$.
Recall that we should include the helicity in the transverse plane
$\cJ-\bar{\cJ} = h\in \bz$ as in (\ref{helicity}).
Then the level matching condition becomes
\begin{align}
&\left(\sum_i n_i + \sum_i (m_i-\eta) + \sum_i (k_i+\eta)\right)
 \nn & \hspace{1.7cm} 
-\left(\sum_i \bar{n}_i + \sum_i (\bar{m}_i-\eta) 
+ \sum_i (\bar{k}_i+\eta) \right)   \in (p+\eta)\bz~.
\label{lm string}
\end{align}

We should note that under the limiting procedure $k \to \infty$, a
huge number of stringy excitations of the original $AdS_3$ superstring
theory are included in our physical Hilbert space of pp-wave superstring
theory.
These states could correspond to very massive 
states, (which could possess very large energies $-J^3_0$) 
although they have the small $\cJ$-charges.  
This fact gives us a theoretical ground for making it possible 
to identify a lot  of stringy excitations with the objects 
in the dual theory as in \cite{BMN}.


In order to complete our discussion,
we also  consider the sectors with non-trivial momenta
along $T^4$ direction. 
Although there are no BPS states in these
sectors, it is possible to construct almost BPS states. 
We only consider a rectangular torus for simplicity and use $R_a$ 
$(a=1,2,3,4)$ as the radii. 
The momenta of $T^4$ sector can be written as 
\begin{equation}
p_a = \frac{n_a}{R_a}+\frac{w^a R_a}{2}~,~~
\bar{p}_a = \frac{n_a}{R_a}-\frac{w^a R_a}{2}~,~~ (n_a, w^a \in \bz)~,
\label{T4 momenta}
\end{equation}
where $n_a$ and $w^a$ are KK momenta and winding modes,
respectively. We have used the convention $\al'\equiv l_s^2=2$.
The simplest physical states (in the NSNS sector) have the next form
\begin{equation}
\psi_{i,-\frac{1}{2} + \eta}\bar{\psi}_{\bar{i},-\frac{1}{2} + \eta} 
\ket{j,\bar{j},\eta,p;n_a,w^a} 
    \otimes c\bar{c} e^{-\phi-\bar{\phi}} \ket{0}_{\msc{gh}} ~,
\label{phys T4 momenta}
\end{equation}
where $\ket{j,\bar{j},\eta,p;n_a,w^a}$ corresponds to the vertex
operator 
\begin{equation}
e^{i(jX^++(p+\eta)X^-+ p_aY^a)}\otimes 
e^{i(\bar{j}\bar{X}^++(p+\eta)\bar{X}^-+ \bar{p}_a\bar{Y}^a)} ~.
\end{equation}
The on-shell condition leads to 
\begin{equation}
j = -\frac{1}{2(p+\eta)}\sum_a p_a^2~, ~~~
\bar{j} = -\frac{1}{2(p+\eta)}\sum_a \bar{p}_a^2~,
\end{equation} 
and the level matching condition $j-\bar{j} \in \bz $ amounts to 
\begin{equation}
 \sum_a n_aw^a \in (p+\eta)\bz  ~.
\end{equation}
The general physical states are obtained by making the DDF operators
\eqn{DDF1} and \eqn{DDF2} and supercharges $\cQ^{\pm \mp a}$ act 
on the above states \eqn{phys T4 momenta}. 
The condition for the almost BPS states is again given by  
\begin{equation}
|\cJ+\bar{\cJ}| \ll k~,
\end{equation} 
and the level matching condition is 
\begin{equation}
 \cJ -\bar{\cJ} \in \bz~.
\end{equation}
We have again a huge number of stringy states
for sufficiently large $k$.

Now we obtain the complete spectrum generated by the DDF
operators, thus we can compare the spectrum with the one from the
light-cone analysis (\ref{spectrum2}).
In our case, the spectrum of light-cone energies is given by 
$H_{\msc{l.c.}} = -\frac{1}{2}(\cJ+\bar{\cJ})$.
In particular, 
in the case of $p=0$\footnote{We can extend this correspondence to the
case of general $p$ by relabeling the modes as we mentioned just below
(\ref{relabel}).}, we can obtain 
\begin{equation}
H_{\msc{l.c.}} = -\frac{1}{2}(\cJ+\bar{\cJ}) 
  = \frac{1}{2\eta}(\mbox{N}+\bar{\mbox{N}}) 
  +\frac{1}{2} (\mbox{J}+\bar{\mbox{J}} ) 
+ \frac{1}{4\eta}\left(\sum_ap_a^2+\sum_a\bar{p}_a^2\right)  ~,
\label{spectrum3}
\end{equation}
where $\mbox{N}$ and $\bar{\mbox{N}}$ are the mode counting operators
and $\mbox{J}$ and $\bar{\mbox{J}}$ are the ``angular momentum
operators''. These operators act on the DDF operators as 
\begin{align}
& \lb \mbox{N}, \cO_n\rb = -n \cO_n~,~~~
(\cO_n = \cP_{i,\,n}, \cP^*_{i,\,n}, \cQ^{++a}_n, \cQ^{--a}_n, 
\cA^{a\da}_n, \cB^{\al\da}_n) ~,\nn
& \lb \mbox{J}, \cO_n\rb = - \cO_n~,~~~ 
(\cO_n = \cP_{i,\,n},\cQ^{++a}_n)~,\nn
&\lb \mbox{J}, \cO_n\rb = \cO_n~,~~~ 
(\cO_n = \cP^*_{i,\,n},\cQ^{--a}_n)~,\nn
&\lb \mbox{J}, \cO_n\rb = 0~,~~~ 
(\cO_n = \cA^{a\da}_n, \cB^{\al\da}_n)~.
\end{align}
The level matching condition is expressed as 
$\cJ-\bar{\cJ} \equiv h \in \bz$, which leads 
to the conditions $\mbox{N}-\bar{\mbox{N}}=0$ and
$\mbox{J}-\bar{\mbox{J}}\equiv h \in \bz$ 
for generic value of $\eta$.

In order to compare the spectrum \eqref{spectrum2}, we identify 
$\mbox{N}$ and $\bar{\mbox{N}}$ as $\hat{N}_L$ and $\hat{N}_R$ 
and $\mbox{J}$ and $\bar{\mbox{J}}$ as $\hat{J}_L$ and
$\hat{J}_R$ without zero mode shifts. 
The zero mode shifts are included in the definition of the ``Fock
vacua'' (\ref{Fock vacua}).
We also identify $f=1/2$, $\eta = 2 \alpha' p_v$ and 
$p_A^2  = \alpha' p^2_a + \alpha' \bar p^2_a$.
Precisely speaking, the fermions used in subsection \ref{LC} is the ones
in the light-cone RNS formalism, on the other hand, the fermions used in
this section corresponding to the ones in the light-cone Green-Schwarz
formalism. We can exchange these two notations of fermions by using the
triality symmetry on $SO(8)$ Lie algebra or the bosonization of
fermions.
Using these relations, we can show that the  spectrum \eqref{spectrum3}
is consistent with the result given in (\ref{spectrum2}).


To conclude the analysis of Hilbert space of superstrings on
the NSNS pp-wave, we should also comment on
physical states in the sectors of spectrally flowed Type I representations. 
The analysis is quite simple because it can be described by the usual free
fields without twist operators.
As we have already mentioned, it is plausible to suppose that 
they correspond to long strings in string theory on $AdS_3$ \cite{MMS,SW,MO}. 
More precisely, the strings in these sectors possess a continuous 
spectrum of the light-cone energies
and can freely propagate along the transverse plane. 
The corresponding excitations do not seem to exist in the symmetric
orbifold theory, which will be analyzed in the next section,
because it only includes the discrete spectrum.
The D1/D5 system (and also the F1/NS5 system) 
has the singularity related to the moduli point where the
D1-branes are emitted from the D5-branes.
The existence of the long strings are related to this singularity
\cite{SW}, however the symmetric orbifold theory does not have the 
singularity. 
Therefore we expect that we obtain the corresponding excitations by
correctly deforming the symmetric orbifold theory. 
More detailed discussions will be given in subsection \ref{RR}.

\newpage

\section{Comparison with SCFT on  $Sym^{M}(T^4)$}
\label{compare}

Superstring theory on $AdS_3 \times S^3 \times M^4$ has been
proposed to be dual to $\cN=(4,4)$ 
non-linear sigma model on the symmetric orbifold space 
$Sym^M(T^4) \equiv (T^4)^M/S_M$ \cite{Maldacena,MS}.
In the case of F1/NS5 system with $Q_1$ fundamental
strings and $Q_5$ NS5-branes, $M$ is given by $M=Q_1 Q_5$.
By taking the near horizon limit of F1/NS5 system, we can identify $Q_5$
as the level $k$ of $SL(2;\br)$ and $SU(2)$ WZW models as we mentioned
above. 
Since the sum of the winding numbers of world-sheet is $Q_1$
(see, e.g., \cite{KS}), the number $Q_1$ should be the implicit upper
bound of winding number $p$, which corresponds to the spectral flow index. 
In our analysis, we can use any $p$ since we take $Q_1 \to \infty$ limit. 
In the next subsection, we review non-linear sigma model on the
symmetric orbifold $Sym^M(T^4)$.
In subsection  \ref{comparison}, we analyze the spectrum of BPS and
almost BPS states with large R-charges 
and compare the spectrum of short string sectors with positive energies.
We have found many missing states in the superstring side and we discuss
this point by comparing the case with RR-flux in subsection \ref{RR}.


\subsection{Review of SCFT on $Sym^{M}(T^4)$}

$\cN=(4,4)$ superconformal field theory defined by
supersymmetric sigma model on the symmetric orbifold 
$Sym^{M}(T^4)\equiv (T^4)^M/S_M$  is described as follows.  
We use  $4M$  free bosons 
$X^{a\da}_{(A)}$ $(A=0,1,2,\ldots, M-1, a, \da = \pm)$ and free fermions
$\Psi^{\al\da}_{(A)}$ and $\bar{\Psi}^{\dal\da}_{(A)}$ 
as fundamental fields.
The superconformal symmetry is realized by the following currents
(where we only write the left-mover)\footnote{
We should emphasize that $(z, \bar z)$ are the coordinates of
space-time superconformal field theory and not related to that of the string
world-sheet used in the previous sections.}
\begin{align}
&  T(z) = -\frac{1}{2}\sum_{A}\, \ep_{ab}\, \ep_{\da\db} \, 
           \partial X_{(A)}^{a\da} \, \partial X_{(A)}^{b\db} 
          -\frac{1}{2}\sum_{A} \,  \ep_{\al\beta}\, \ep_{\da\db}\,
             \Psi_{(A)}^{\al\da}\, \partial \Psi_{(A)}^{\beta\db} ~, \nn
& G^{\al a}(z) = i\sum_{A}\, \ep_{\da\db}\, 
    \Psi_{(A)}^{\al\da} \, \partial X_{(A)}^{a\db} ~, \nn
& K^{\al\beta}(z) = - \frac{1}{2}\, \sum_{A} \, \ep_{\da\db} \,
            \Psi_{(A)}^{\al\da} \, \Psi_{(A)}^{\beta\db} ~.
\label{SCA}
\end{align}
In our convention, we set $\ep^{+-}=\ep_{-+}=1$ and $\ep^{-+}=\ep_{+-}=-1$
and the OPEs of free fields are written as 
\begin{align}
X_{(A)}^{a\da}(z)X_{(B)}^{b\db}(0) \sim 
-\delta_{AB}\ep^{ab}\ep^{\da\db}\ln z  ~, ~~
\Psi^{\al\da}(z)\Psi^{\beta\db}(0) \sim 
\delta_{AB}   \frac{ \ep^{\al\beta}\ep^{\da\db} }{z} ~.
\end{align}
The usual convention of $SU(2)$ current is given by 
\begin{equation}
 K^3 = K^{+-} = K^{-+} ~,~~ K^+ = K^{++} ~,~~ K^- = - K^{--} ~.
\end{equation}  
These currents generate $N=4$ (small) superconformal algebra (SCA)
with central charge $c=6M$ and their OPEs are
\begin{align}
 T (z) T (w) &\sim \frac{ 3 M}{(z-w)^4} + 
   \frac{ 2 T(w)} { (z-w)^2} + \frac{\partial T(w)}{z-w} ~, \nn
 T (z) K^{\alpha \beta} (w) &\sim \frac{K^{\alpha \beta} (w)}{(z-w)^2} 
   + \frac{\partial K^{\alpha \beta} (w)}{z-w} ~, \nn
 K^{\alpha \beta} (z) K^{\gamma \delta} (w) &\sim - \frac{M
 (\epsilon^{\alpha \gamma} \epsilon^{\beta \delta} + 
 \epsilon^{\alpha \delta} \epsilon^{\beta \gamma} )}{2 (z-w)^2} \nn
  &\quad - \frac{
   \epsilon^{\alpha \gamma} K^{\beta \delta}(w)
  + \epsilon^{\alpha \delta} K^{\beta \gamma}(w)
  + \epsilon^{\beta \gamma} K^{\alpha \delta}(w)
  + \epsilon^{\beta \delta} K^{\alpha \gamma}(w)}{2(z-w)}  ~, \nn
 T (z) G^{\alpha a} (w) 
 &\sim \frac{3  G^{\alpha a} (w)}{2(z-w)^2} 
   + \frac{\partial G^{\alpha a } (w)}{z-w} ~, \nn
 G^{\alpha a} (z) K^{\beta \gamma} (w) &\sim - 
 \frac{\epsilon^{\alpha \gamma} G^{\beta a} (w) 
     + \epsilon^{\alpha \beta} G^{\gamma a} (w) }{2(z-w)} ~,\nn
 G^{\alpha a} (z) G^{\beta b} (w) &\sim
 \epsilon^{ab} \left(
  \frac{ 2 M\epsilon^{\alpha \beta}}{(z-w)^3} 
 + \frac{2 K^{\alpha \beta} (w)}{(z-w)^2}
 + \frac{ \epsilon^{\alpha \beta} T (w) + \partial K^{\alpha \beta} (w)}{z-w}
\right) ~.
\end{align}
The subalgebra of ``zero-modes'' $\{ L_{\pm 1},\, L_0, \, 
G_{\pm 1/2}^{\al a}, \, K_0^{\al\beta} \}$ and the counterpart of the 
right-movers compose super Lie algebra
$PSU(1,1|2)_L \times PSU(1,1|2)_R$.
This subalgebra  corresponds to the
supersymmetric algebra on $AdS_3\times S^3$ geometry
discussed in appendix \ref{STSS}.

According to the general approach to orbifold conformal field theories
\cite{orbifoldCFT,orbifoldCFT2,orbifoldCFT3}, 
we have various twisted sectors.
The Hilbert space of each twisted sector is defined 
with the following boundary condition $(z\equiv e^{\tau +i\sigma})$ as
\begin{equation}
\Phi_{(A)}(\tau, \sigma +2\pi ) = \Phi_{g(A)}(\tau, \sigma) ~, 
\label{twisted bc}
\end{equation}   
where $\Phi_{(A)}(\tau, \sigma)$ represents $X^{a\da}_{(A)}$,
$\Psi^{\al\da}_{(A)}$ and $\bar{\Psi}^{\dal \da}_{(A)}$.
The Hilbert space of the symmetric orbifold theory can be decomposed as
(see, for example, \cite{DMVV})
\begin{equation}
 \cH (Sym^M (T^4)) = \bigoplus_{\gamma} \cH_{\gamma}^{C_{\gamma}}~,
\end{equation}
where we denote $\gamma$ as a conjugacy class, which labels a twisted
sector. 
The Hilbert space of the twisted sector has to be invariant under 
centralizer subgroup $C_{\gamma}$ whose element $g \in S_M$ satisfies
$g h g^{-1} =h$ for $h \in \gamma$.

\begin{figure}
\begin{center}
\includegraphics{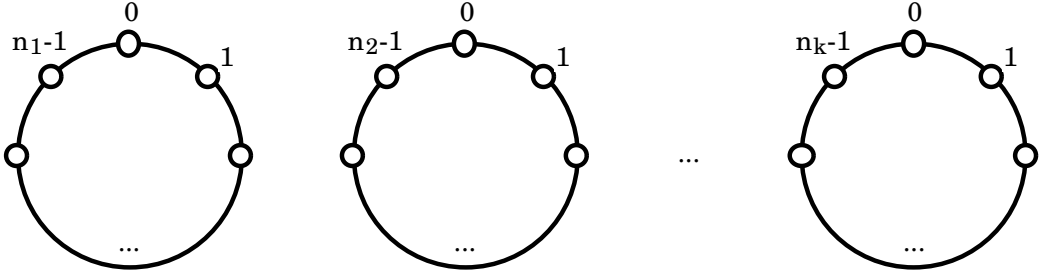}
\caption{The configuration corresponding to a twisted sector of the
 symmetric orbifold theory. 
 There are connected world-sheets with length $2 \pi n_i$
 in the $\bz_{n_i}$-twisted sector. In
 the notation in this figure, $\sum_{i=1}^k n_i = M$ and $n_i$ can take
 the same value as $n_j$ for $i \neq j$.}
\label{FIG:SymOrb}
\end{center}
\end{figure}

The conjugacy class of symmetric permutation can be written as the form
\begin{equation}
 \gamma = (1)^{N_1} (2)^{N_2} \cdots (l)^{N_l} ~,
\end{equation}
with $\sum_n n N_n = M$.
Each $(n)$ denotes the cyclic permutation of $n$ elements and $N_n$
denotes the multiplicity of the cycle $(n)$.
The centralizer group $C_{\gamma}$ can be given by the permutation of the
$N_n$ cycles $(n)$ and the rotation of the cycle $(n)$.
The Hilbert space of the twisted sector is then given by
\begin{equation}
 \cH_{\gamma}^{C_{\gamma}} = \bigotimes_{n>0} S^{N_n} \cH_{(n)}^{\bz_n} ~, 
\end{equation}
where we denote $S^{N_n} \cH$ as the (graded) symmetrization of 
tensor products  of $N_n$ Hilbert spaces $\cH$.
Because of the twisted boundary condition (\ref{twisted bc}),
there are connected world-sheets with length $2 \pi n$ in the Hilbert
space $\cH_{(n)}^{\bz_n}$ (see figure \ref{FIG:SymOrb}). 
The states in the Hilbert space
$\cH_{(n)}^{\bz_n}$ are invariant under the $\bz_n$ action and hence
the sector of these states is called as $\bz_n$-twisted sector.
States in the $\bz_n$-twisted sector correspond to
single-particle states and states in the total Hilbert space
generically correspond to multi-particle states.
In order to compare with the physical spectrum 
of first quantized superstring theory, it is enough to focus 
on the single-particle Hilbert space.
Hence, we shall concentrate on the $\bz_{N}$-twisted sector 
$(N\leq M)$ from now on. More precisely speaking, we consider the
conjugacy class
\begin{equation}
 \gamma = (1)^{M-N}(N) ~,  
\end{equation}
and only employ identity states in the Hilbert space $S^{N_{M-N}}
\cH_{(1)}$.

We label the objects in the $\bz_{N}$-twisted sector by the index 
$\left[A_0,A_1,\cdots,A_{N-1} \right] =   \left[0,1,\cdots,N-1 \right]$.
The coordinates of this sector are defined on the world-sheet
$0\leq \sigma \leq 2\pi $, which is rescaled from the
$N$-times one $0\leq \sigma \leq 2\pi N $.
These coordinates are given by
\begin{align}
\Phi (\tau, \sigma) = \Phi_{(n)}(\tau, N\sigma - 2\pi n) \quad
 \mbox{for} \quad
\frac{2\pi n}{N} \leq \sigma \leq \frac{2\pi (n+1)}{N} ~,
\label{long string}
\end{align} 
with $n=0,1,\ldots, N-1$.
These variables 
$X^{a\da}$, $\Psi^{\al\da}$ and 
$\bar{\Psi}^{\dal\da}$ can be used to construct
$\cN =4$ superconformal currents 
$\{L_{n},G_{r}^{\al a}, K_{n}^{\al\beta}\}$ 
with central charge $c=6$ in the manner similar to  \eqn{SCA}.
The $\bz_n$ action to the Hilbert space $\cH_{(N)}^{\bz_N}$ is given
by the action of $\Phi_{(n)} \to \Phi_{(n+1)}$.
The invariance under this action amounts to imposing
\begin{equation}
L_{0}- \bar{L}_{0} \in N\bz~,
\label{ZN}
\end{equation}
on the physical Hilbert space.

The superconformal currents compatible with the condition \eqn{ZN},
which can act on the physical Hilbert space independently 
of the right-movers, consist of only the modes of $n \in N\bz$.
More precisely, the superconformal currents properly
describing the $\bz_N$-twisted sector 
$\{\hat{L}_n,\hat{G}_r^{\al a}, \hat{K}^{\al\beta}_n\}$ 
should be defined as follows \cite{FKN,BHS} 
(from now on, we shall only present the NS sector)
\begin{align}
& \hat{L}_n = \frac{1}{N} L_{nN} + \frac{N^2-1}{4N}\delta_{n0} ~,   \nn
& \hat{G}_r^{\al a} =\left\{\begin{array}{ll}
               \frac{1}{\sqrt{N}} G_{Nr}^{\al a \,(NS)}~, &  (N=2q+1) ~, \\
	       \frac{1}{\sqrt{N}} G_{Nr}^{\al a \,(R)}~, &  (N=2q) ~,
  	\end{array} \right.  \nn
& \hat{K}^{\al\beta}_n = K^{\al\beta}_{nN} ~.
\label{hat}
\end{align}
One can check directly that 
these operators  generate $\cN=4$ superconformal algebra with $c=6N$.
The anomaly term in the expression of $\hat{L}_n$ 
corresponds essentially to the Schwarzian derivative of the conformal 
mapping $z \to z^N$. 
We should note that the modes of hatted current are counted by
$\hat{L}_0$.
{}From now on, we also count the modes of objects in this sector by 
$\hat{L}_0$ and hence the fractional modes are allowed.

{}From these definitions \eqn{hat}, we can find that 
the vacuum $\ket{0;N}$ of $\bz_N$-twisted sector  
possesses the following properties as
\begin{itemize}
 \item $N=2q+1$
\begin{align}
& \hat{L}_0\ket{0;N}= \frac{N^2-1}{4N} \ket{0;N} ~,  \nn
& \hat{K}^3_0\ket{0;N} = 0 ~.
\end{align}
 \item $N=2q$
\begin{align}
& \hat{L}_0\ket{0;N}= \left(\frac{N^2-1}{4N}
+\frac{1}{4N} \right)\ket{0;N} 
\equiv \frac{N}{4}\ket{0;N} ~,  \nn
& \hat{K}^3_0\ket{0;N}= -\frac{1}{2} \ket{0;N} ~.
\end{align}
\end{itemize}
When $N$ is even, the supercurrent 
$\hat{G}_r^{\al a}$ in the  NS sector is made of the one in the R sector
before imposing the $\bz_N$-invariance. 
The extra vacuum energy ``$ \frac{1}{4N}$''
and the extra R-charge  ``$-\frac{1}{2}$'' 
in the case of $N=2q$ originate from this fact.

Now, we focus on BPS states (chiral primary states)
in the $\bz_N$-twisted sector, which are defined by the next conditions 
\begin{align}
& \hat{L}_n \ket{\al}=\bar{\hat{L}}_n \ket{\al}=0 ~,~~
({}^\forall n \geq 1) ~, \nn
&  \hat{G}^{+ a}_r \ket{\al} = \bar{\hat{G}}^{+ a}_r \ket{\al}=0 ~,~~ 
 ({}^\forall r \geq -\frac{1}{2}) ~, \nn
&  \hat{G}^{- a}_r \ket{\al} = \bar{\hat{G}}^{- a}_r \ket{\al}=0 ~,~~ 
({}^\forall r \geq \frac{1}{2}) ~.
\label{hat CP}
\end{align}
These conditions lead inevitably to
\begin{equation}
 (\hat{L}_0-\hat{K}^3_0)\ket{\al}=0 ~.
\end{equation}

At this point it is not difficult to present the explicit forms 
of all the possible BPS states in the $\bz_N$-twisted sector.
(See, for example, \cite{Sugawara}.)
They are written as
\begin{equation}
\ket{\om^{(q,\bar{q})};N}= \ket{\om^q; N }\otimes
\overline{\ket{\om^{\bar{q}};N}} ~, ~~{}^{\forall}\om^{(q,\bar{q})} \in 
H^{q,\bar{q}} (T^4) ~, ~~(q,\bar{q}=0,1,2) ~,
\end{equation}
where the left(right)-moving parts are defined by
\begin{itemize}
\item $N=2q+1$
\begin{align}
&  \ket{\om^0;N}= \prod_{i=0}^{q-1}\,
\Psi^{++}_{-\left(\frac{1}{2N}+\frac{i}{N} \right)}
\Psi^{+-}_{-\left(\frac{1}{2N}+\frac{i}{N} \right)} \ket{0;N} ~, \nn
&  \ket{\om^{1\da};N}= \Psi^{+\da}_{-\frac{1}{2}}
\ket{\om^0;N} ~, \nn
&  \ket{\om^{2};N}= \Psi^{++}_{-\frac{1}{2}}
\Psi^{+-}_{-\frac{1}{2}}
\ket{\om^0;N} ~. 
\label{cp 1}
\end{align}
\item $N=2q$
\begin{align}
& \ket{\om^0;N}= \prod_{i=0}^{q-1}\,
\Psi^{++}_{-\frac{i}{N}}
\Psi^{+-}_{-\frac{i}{N}} \ket{0;N} ~, \nn
&  \ket{\om^{1\da};N}= \Psi^{+\da}_{-\frac{1}{2}}
\ket{\om^0;N} ~, \nn
&  \ket{\om^{2};N}= \Psi^{++}_{-\frac{1}{2}}
\Psi^{+-}_{-\frac{1}{2}}
\ket{\om^0;N} ~. 
\label{cp 2}
\end{align}
\end{itemize}
It is easy to check that
\begin{align}
& \hat{L}_0 \ket{\om^{(q,\bar{q})};N} = \hat{K}^3_0 
\ket{\om^{(q,\bar{q})};N} = 
\frac{q+N-1}{2} \ket{\om^{(q,\bar{q})};N} ~, \nn
&  \bar{\hat{L}}_0 \ket{\om^{(q,\bar{q})};N} = \bar{\hat{K}}^3_0 
\ket{\om^{(q,\bar{q})};N} =
 \frac{\bar{q}+N-1}{2} \ket{\om^{(q,\bar{q})};N}  ~.
\label{cp spectrum}
\end{align}


\subsection{Comparison with Superstring Spectrum}
\label{comparison}

We have obtained the spectrum of superstrings on pp-wave in section 
\ref{NSNSppwave} and in the previous subsection we have reviewed the
symmetric orbifold theory.
Therefore, it is time to compare the both Hilbert spaces.
In this subsection we compare the (almost) BPS states with large
R-charges in the $\bz_N$-twisted sector with the states in
superstring theory on the pp-wave. 
In the context of $AdS_3/CFT_2$ correspondence, $\hat{L}_0$ and 
$\hat{K}^3_0$ are identified with $-J^3_0$  and  $K^3_0$, respectively,
which are the zero-modes of the total currents in $SL(2;\br)$ and 
$SU(2)$ super WZW models.
Recalling the relations \eqn{contraction},
we must take the identification
\begin{equation}
 \cJ \leftrightarrow \hat{K}^3_0-\hat{L}_0~,~~~
 \cF \leftrightarrow \frac{1}{k} (\hat{K}^3_0+\hat{L}_0)~.
\label{JandF}
\end{equation}
The condition of almost BPS states are (\ref{BPS condition})
\begin{equation}
 \hat{L}_0 - \hat K^3_0 \ll k ~,~~~ \hat{L}_0 + \hat K^3_0 \sim k ~.
\label{BPScond} 
\end{equation}

Since the subalgebra of ``zero-modes'' $\{ \hat{L}_{\pm 1},\, \hat{L}_0, \, 
\hat{G}_{\pm 1/2}^{\al a}, \, \hat{K}_0^{\al\beta} \}$ 
corresponds to superalgebra on $AdS_3\times S^3$,
we can define the Penrose limit on this subalgebra just as
(\ref{bosonicPL}) and (\ref{fermionicPL}).
In order to be consistent with the relation (\ref{JandF}),
it is natural to redefine as
\begin{align}
& \cJ' = \hat{K}^3_0 - \hat{L}_0 ~,~~
  \cF' = \frac{1}{k} (\hat{K}^3_0+\hat{L}_0) ~,\nn
& \cP'_1 = - \frac{1}{\sqrt{k}} \hat{L}_{+1} ~,~~  
  \cP^{*'}_1 = -  \frac{1}{\sqrt{k}} \hat{L}_{-1} ~, \nn
& \cP'_2 = \frac{1}{\sqrt{k}} \hat{K}^+_0 ~,~~  
  \cP^{*'}_2 = \frac{1}{\sqrt{k}} \hat{K}^-_0 ~, \nn
& \cQ^{++\pm'} = \pm  \frac{1}{\sqrt{k}} \hat G^{+\pm}_{\frac{1}{2}} ~,~~
  \cQ^{--\pm'} = \pm  \frac{1}{\sqrt{k}} \hat G^{-\pm}_{-\frac{1}{2}} ~,\nn
& \cQ^{-+\pm'} = \pm  \hat G^{+\pm}_{-\frac{1}{2}} ~,~~
  \cQ^{+-\pm'} = \pm  \hat G^{-\pm}_{\frac{1}{2}} ~.
\label{zeromode}
\end{align}  
By taking $k \to \infty$ limit, these operators satisfy the commutation
relations of super pp-wave algebra (\ref{CR1}), (\ref{CR2}) and
(\ref{CR3}).

First, let us consider the BPS states.
As we observed in (\ref{cp spectrum}), the BPS states in the $\bz_N$-twisted
sector satisfy 
\begin{equation}
 \hat{L}_0 - \hat{K}^3_0 = 0 ~,~~ \hat{L}_0 + \hat{K}^3_0 \approxeq N ~.
\end{equation}
The BPS states in the string theory side have the eigenvalues  
$\cF = p+\eta$ in the sector with fixed $p,\eta$, the above relations
and (\ref{JandF}) imply the following identification
\begin{equation}
 \eta = \frac{l}{k} ~,\quad N=pk+l  ~,
\label{l/k}
\end{equation}
with $l=1,2,\cdots,k-1$\footnote
   {The threshold values  $l=0$ and $k$ cannot be included 
    because there is a restriction $0<\eta<1$ in the type II
    representations.  Moreover, the states with
    $l=k-1$ corresponds to the missing states in the original   
    $AdS_3\times S^3$ string theory for each $p$.
    (See, for example, \cite{HS2,HHS}.)
    One might worry that $\eta$ should be continuous in principle. 
    These subtleties are, however, harmless for sufficiently large $k$.}.
Under this identification, we can show the correspondence between the BPS
states (\ref{BPS NS}) and (\ref{BPS R}) in the string theory side and the BPS
states (\ref{cp 1}) or (\ref{cp 2}) in the dual CFT side.

At least with respect to the BPS states, 
we expect that the equal number of physical states exist  
in both of superstring theory on the $AdS_3 \times S^3 \times T^4$ 
and the pp-wave background. This statement is valid  
as long as superstring theory on the pp-wave is defined by the contraction
(\ref{contraction}). 
Let us fix the spectral flow number $p~( \geq 1)$.  
It is known \cite{HHS,AGS} that there are about
$k\times \dim H^*(T^4)$ BPS states with the R-charges
$\frac{1}{2}kp \lesssim Q  \lesssim \frac{1}{2}k(p+1)$ in
superstring theory on $AdS_3 \times S^3 \times T^4$.
This values of R-charges amounts to
$\cF = p + l/k$ ($0 \leq l \leq k$)
for each of the spectrally flowed sector.
This is consistent with the assumption $\eta = l/k$ $(l=1,2,\ldots,k-1)$
 (\ref{l/k}) and the following physical Hilbert space
\begin{equation}
 \cH_{\msc{pp-wave}}^{(p)} = 
 \bigoplus_l  \cH_{\msc{pp-wave}}(p, \eta=l/k)
\label{hilbert pp}
\end{equation}
includes the equal number of BPS states as those of superstring theory on  
$AdS_3\times S^3 \times T^4$.

Turning to the symmetric orbifold theory, we consider the following direct
sum of the single-particle Hilbert spaces of the 
$\bz_{N(l)}$-twisted sectors, 
(where we set $\dsp N(l)\equiv kp+l$ $(0< l < k)$)
\begin{equation}
 \cH_{\msc{symm}}^{(p)} = 
 \bigoplus_l  \cH_{\msc{symm}}(N(l)= kp+l) ~.
\label{hilbert symm}
\end{equation}
It is obvious that $\cH_{\msc{symm}}^{(p)}$ and 
$\cH_{\msc{pp-wave}}^{(p)}$ have the equivalent 
spectrum of BPS states. 
We should also point out that the identification
$\cB^{+\da'}_0 (\equiv \Psi^{+\da}_{-\frac{1}{2}}) = \cB^{+\da}_0$ 
is consistent with the relation between 
\eqn{BPS relation} and \eqn{cp 1}, \eqn{cp 2}.

Next, let us consider the almost BPS states.
For a moment, we again neglect the sectors with momenta along $T^4$
sector.
The almost BPS states in the dual CFT side are obtained by acting
the free oscillators $\Psi^{\pm\da}_{\mp\frac{1}{2}+\frac{n}{N}}$ and
$i\partial X^{a\da}_{\frac{n}{N}}$\footnote{
Strictly speaking, the mode expansions of $\Psi^{\al\da}$ and
$i\partial X^{a\da}$ depend on whether $N$ is even or
odd. However, we can safely neglect the difference $1/2N$ under the
assumption of large $N$.} 
on the BPS states under the condition
(\ref{BPScond}).
Since we have confirmed the correspondence of the BPS states, the
task is to look for the operators corresponding to the DDF operators.
For the DDF operators of $T^4$ sector, we can find the corresponding
operators as
\begin{align}
\cB^{\pm\da \,'}_n = \Psi^{\pm\da}_{\mp\frac{1}{2}+\frac{nk}{N}}~,~~
\cA^{a\da\, '}_n = \frac{1}{\sqrt{N}}\,i\partial X^{a\da}_{\frac{nk}{N}}~,
\label{DDF symmetric 2}
\end{align}
which satisfy the same (anti-)commutation relations as those of the 
DDF operators (\ref{DDF comm 2}).
For the transverse coordinates of pp-wave background, we also construct the
following operators
\begin{align}
& \cP^{(l)\,'}_{1,\,n} = -\frac{1}{\sqrt{N}}
\left\{\frac{N}{nk+l}\hat{L}_{\frac{nk+l}{N}}
-\left(\frac{N}{nk+l}-1\right)\hat{K}^3_{\frac{nk+l}{N}}\right\}~,\nn
& \cP^{*(l)\,'}_{1,\,n} = \frac{1}{\sqrt{N}}
\left\{\frac{N}{nk-l}\hat{L}_{\frac{nk-l}{N}}
-\left(\frac{N}{nk-l}+1\right)\hat{K}^3_{\frac{nk-l}{N}}\right\}~,\nn
& \cP^{(l)\,'}_{2,\,n} = \frac{1}{\sqrt{N}} \hat{K}^+_{-1+\frac{nk+l}{N}}~,
~~~
\cP^{*(l)\,'}_{2,\,n} = \frac{1}{\sqrt{N}} \hat{K}^-_{1+\frac{nk-l}{N}}~, 
\nn
& \cQ^{++\pm(l)\,'}_n =\pm \frac{\sqrt{N}}{nk+l}
\hat{G}^{+\pm}_{-\frac{1}{2}+\frac{nk+l}{N}}~, ~~~
\cQ^{--\pm(l)\,'}_n =\mp \frac{\sqrt{N}}{nk-l}
\hat{G}^{-\pm}_{\frac{1}{2}+\frac{nk-l}{N}}~.
\label{DDF symmetric}
\end{align}
Under the identification of (\ref{l/k}) and the large $N$ limit, 
we can show that these operators satisfy the same (anti-)commutation
relations as those of the corresponding DDF operators
 \eqn{DDF comm 1} and \eqn{DDF comm 2}.

It is quite important to note that there are the equal number of degrees 
of freedom after taking such large $N$ limit.  
The Hilbert space of $\bz_N$-twisted sector is spanned by 
the free oscillators $\Psi^{\pm\da}_{\mp\frac{1}{2}+\frac{n}{N}}$ and
$i\partial X^{a\da}_{\frac{n}{N}}$ and
for each energy level there are the equal number of 
bosonic and fermionic oscillators as those 
defined in (\ref{DDF symmetric 2}) and (\ref{DDF symmetric}). 
Thus, the almost BPS states in the symmetric orbifold theory can be
written in terms of the operators (\ref{DDF symmetric 2}) and
(\ref{DDF symmetric}) as
\begin{align}
&  \cB_{-n_1}^{\alpha_1 \dot{a}_1\,'} \cdots 
  \cQ_{-m_1}^{++b_1(l_{i})\,'} \cdots 
  \cQ_{-k_1}^{--c_1(l_{j})\,'} \cdots
  \otimes
  \bar{\cB}_{-\bar{n}_1}^{\bar{\al}_1 \bar{\dot{a}}_1\,'} \cdots 
  \bar{\cQ}_{-\bar{m}_1}^{++\bar{b}_1(\bar l_{i})\,'} \cdots 
  \bar{\cQ}_{-\bar{k}_1}^{--\bar{c}_1(\bar l_{j}) \,'} \cdots
\ket{\om; N(l)} ~,\nn
& \hspace{1.7cm} n_i, \bar{n}_i, m_i, \bar{m}_i >0~,~ k_i, \bar{k}_i \geq 0~,~
 {}^{\forall}\om \in H^*(T^4)~, ~~N(l)=kp+l ~,
\end{align} 
and the ones generated by acting (\ref{zeromode}).
By considering the case with $l_i = \bar l_i = l$, we obtain the states
corresponding to the string states (\ref{almost BPS1}).
The non-trivial consistency check is only the level matching condition. 
In the symmetric orbifold theory side, the level matching condition 
is given by
\begin{equation}
\hat{L}_0-\bar{\hat{L}}_0 \in \bz~.
\end{equation} 
This condition is consistent with 
the result of string theory side \eqn{lm string} under the above 
correspondence of DDF operators.

However, we should note here that the string Hilbert space 
$\cH_{\msc{pp-wave}}^{(p)}$ is strictly smaller than that of 
the symmetric orbifold theory;
\begin{equation}
\cH_{\msc{pp-wave}}^{(p)} \subsetneqq
\cH_{\msc{symm}}^{(p)}~. 
\end{equation} 
In fact, the general states (i.e., $l_i \neq \bar l_i \neq l$) 
have no counterparts in the string side since we cannot define the
corresponding DDF operators as local operators on the string Hilbert space.
These missing states in the string spectrum may be compensated by 
non-perturbative excitations. 
Because we are now assuming the small string coupling, 
the non-perturbative excitations usually become very massive.
Under the assumption of large $k$,
the space of almost BPS states can include such very massive 
excitations in principle. However, our world-sheet analysis 
as a perturbative string theory cannot include such excitations.

Finally we consider the sectors with non-vanishing momenta along $T^4$.
As in the spectrum of superstring theory, 
there are no BPS states in these sectors
but there are many almost BPS states. 
Recall that $\cJ'= \hat{K}^3_0 -\hat{L}_0$ and
$\hat{L}_0 = \frac{1}{N} L_{0} + \frac{N^2-1}{4N}$.
The operator $L_0$ includes the contribution of momenta with the standard 
normalization. Hence, we find that the contributions of momenta
to $\cJ'$ and $\bar{\cJ}'$ are given as follows (for the sector 
$\cH_{\msc{symm}}(N(l))$ with $N(l)\equiv kp +l$)
\begin{align}
& \Delta \cJ' = -\frac{1}{2N(l)} \sum_a
\left(\frac{n'_a}{R_a}+\frac{w'^aR_a}{2}\right)^2~,~~
\Delta \bar{\cJ}' = -\frac{1}{2N(l)} \sum_a
\left(\frac{n'_a}{R_a}-\frac{w'^aR_a}{2}\right)^2~,
\end{align}
where $n_a'$ and $w'^a$ are KK momenta and winding modes as before.
The level matching condition for the vacuum state becomes
\begin{equation}
\sum_a n'_aw'^{a} \in N(l)\bz~.
\end{equation}
By comparing the analysis given in the last section (under the  
identification $\eta = l/k$),
we obtain the following correspondence between the spectrum of
superstring theory and that of the symmetric orbifold theory as
\begin{equation}
   n'_a= \sqrt{k}n_a~,~~~ w'^{a} = \sqrt{k}w^a~, 
\end{equation}
where $n_a$ and $w^a$ are KK momenta and winding modes (\ref{T4
momenta}) in the string
theory side\footnote{We can treat $\sqrt{k}$ as an integer number since we
assume that $k$ is very large.}.

We have observed that there are again many missing states in the string
theory side. 
The essentially same aspect was already pointed out in the context of 
string theory on $AdS_3\times S^3 \times T^4$ \cite{KLL}.
It may be worthwhile to comment on how such discrepancy is removed  
if assuming fractional string excitations. 
These excitations do not exist in the perturbative string spectrum and
may be at least explained in the S-dual picture as discussed in section
\ref{D1D5}.  
The existence of $k\equiv Q_5$ NS5
leads to fractional strings with tension $ \tilde{T} = T / k$, 
where $T$ represents the tension of 
fundamental string. If we measure the radii of $T^4$ by the unit of
string length $l_s = 1/\sqrt{T}$, namely, $R_a = r_a l_s$,
the momenta \eqn{T4 momenta} becomes
\begin{equation}
p_a = \frac{1}{l_s}\left(\frac{n_a}{r_a}+w^ar_a\right)  ~,~~~
\bar{p}_a =\frac{1}{l_s}\left(\frac{n_a}{r_a}-w^ar_a\right) ~.
\label{T4 momenta 2}
\end{equation} 
For the fractional strings, we should replace the string length $l_s$
with $\tilde{l}_s \equiv \sqrt{k} l_s$. Thus, we obtain
\begin{equation}
p_a = \frac{1}{\sqrt{k}}\times
\frac{1}{l_s}\left(\frac{n_a}{r_a}+w^ar_a\right)  ~,~~~
\bar{p}_a = \frac{1}{\sqrt{k}}\times
\frac{1}{l_s}\left(\frac{n_a}{r_a}-w^ar_a\right) ~,
\label{T4 momenta 3}
\end{equation}
and the extra factor $1/\sqrt{k}$ completely compensates  the spectrum of 
missing states.


\subsection{Comments on Case with RR-Flux}
\label{RR}

As we saw in the last subsection, there are many missing states in
the string side.
This is because the superstring theory is compared with the
supersymmetric sigma model at the orbifold point, which is a different
moduli point from the one of the dual CFT.
The dual CFT can be obtain by deforming from the orbifold point 
and we expect that almost BPS states with large R-charge are not
sensitive to the deformation.
However, the dual CFT is at a singular moduli point, so this expectation
is too naive. In fact, there are some missing states in the superstring side
even for the BPS states and they are related to the divergence due to the
singularity as we said before.

One way to remove this difference is to deform the superstring theory in
order to include some non-perturbative excitations as mentioned before.
The other way is to deform the symmetric orbifold theory by using the
corresponding marginal operator.
In our case, we do not know how to deform the theory
from the orbifold limit, but
in the case with RR-flux, the corresponding deformation is conjectured
in \cite{GMS} and the 
spectra of two theories are compared in \cite{Gava} by making use of the
deformation. 
In order to clarify the point, 
we study the correspondence between superstrings on 6
dimensional pp-waves with RR-flux and non-linear sigma model on a
resolution of $Sym^{Q_1Q_5}(T^4)$ \cite{Lunin,GMS,Gava} in this
subsection.

The Penrose limit of $AdS_3 \times S^3$ with RR-flux is given
by 6 dimensional pp-wave background with RR-flux.
The metric of pp-wave is the same as that with NSNS-flux \eqn{ppmetric},
but now NSNS-flux is replaced by RR-flux.
Superstrings on the pp-wave with RR-flux can be quantized by
Green-Schwarz formalism in the light-cone gauge \cite{BMN,RT}.
This is the same situation as superstrings on the maximally
supersymmetric pp-wave with RR-flux and spectrum can be obtained in
the way similar to \eqn{ppspectrum} as
\begin{align}
 2 p^- = - p_+ = H_{\rm l.c.} = 
 \sum_{n} N_n \sqrt{1 + \frac{n^2}{(\alpha' p^+)^2}} 
  + \frac{L_0^{T^4} + \bar L_0^{T^4}}{\alpha' p^+ }~.
\end{align}

We have the relation between the light-cone
momenta in the superstring theory and conformal weights $\Delta$ and
R-charges $Q$ of operators in the space-time SCFT. 
Using that relation, the spectrum of the dual CFT is given by
\begin{align}
 \Delta -Q  = \sum_{n} N_n \sqrt{1 + \left(\frac{n g_s Q_5}{Q}\right)^2} 
  + g_s Q_5 \left(\frac{L_0^{T^4} + \bar L_0^{T^4}}{Q }\right)~,
\label{6ppspectrum}
\end{align}
where we use $R^2=\alpha' g_s Q_5$ \eqn{R2QQ}.
Kaluza-Klein modes in $T^4$ sector can be explained by assuming the
fractional D-strings as before.
For the other modes, 
it was conjectured in \cite{Lunin,GMS} that the general modes of 
four bosons and four fermions in the light-cone spectrum correspond to
\begin{equation}
 \hat L_{-1-\frac{n}{Q}} ~, \quad \bar{\hat L}_{-1-\frac{n}{Q}} ~,\quad
 \hat K^-_{-\frac{n}{Q}} ~, \quad \bar{\hat K}^-_{-\frac{n}{Q}} ~,\quad
 \hat G^{-\pm}_{-\frac12 -\frac{n}{Q}} ~, 
 \quad \bar{\hat G}^{-\pm}_{-\frac12-\frac{n}{Q}}~.
\end{equation}
They reproduces the first order of the spectrum \eqn{6ppspectrum}.
We should notice that these operators correspond to the states
missing in our analysis since now we take the limit of $n \ll Q$.
For this reason, we cannot compare this result with the previous one.

In order to see the next order, we have to deform the theory
from the orbifold limit. 
The moduli space of ${\cN=(4,4)}$ supersymmetric sigma model on 
$Sym^{M}(T^4)$ $(M=Q_1Q_5)$ is twenty dimensional.
The sixteen moduli parameters are the metric $G_{ij}$ and the
anti-symmetric tensor $B_{ij}$ on $T^4$. 
The other four moduli parameters are the blowing up modes which resolve
singularities.
The singularities appear at the fixed points
since we divide $M$ products of $T^4$ by $M$-th symmetric group.
Near the orbifold point, operators corresponding to the blowing up
modes are given by
\begin{equation}
 m^{ab}  = \sum_{0 \leq i \neq j \leq M-1} 
   G_{-\frac12}^{\alpha a} \bar G_{-\frac12}^{\beta b} 
 \sigma_{ij}^{\gamma \delta} 
\epsilon_{\alpha \gamma} \epsilon_{\beta \delta} ~,
\end{equation}
where $\sigma_{ij}^{\alpha \beta}$ is $\bz_2$ twist operators
acting on the covering space $(T^4)^M$.
The $\bz_2$ twist operators $\sigma_{ij}^{\alpha \beta}$ exchange the
$i$-th and $j$-th $T^4$ and their $SU(2)_R \times SU(2)_L$ R-charges are
$(\alpha/2 ,\beta/2)$.
The label $(a,b)$ is related to the global $SU(2)_I$ 
symmetry and $G^{\alpha a}_{-1/2}$ and $\bar G^{\beta b}_{-1/2}$ 
are doublets under the symmetry.
By taking the tensor products, we obtain $m_{\rm triplet}$ and 
$m_{\rm singlet}$. 
Here we choose $m_{\rm singlet}$ because $m_{\rm singlet}$ is
related to the $D1$ and $D5$ charges and $m_{\rm triplet}$ is related to
the charges of $D3$-branes wrapped on some two-cycles of $T^4$.
In summary, we use the following marginal deformation
\begin{equation}
 \delta S = \lambda \int dz^2 m_{\rm singlet} (z, \bar z) ~.
\end{equation} 
In \cite{GMS} it was proposed that $\lambda \sim g_6$ and a heuristic
explanation was given.
In fact, it was shown in \cite{Gava}
that we can reproduce the first non-trivial order
of the spectrum \eqn{6ppspectrum} by using the marginal deformation.

\newpage


\section{Superstring Theory on $H_6 \times T^4/\bz_2$
and SCFT on $Sym^M (T^4/\bz_2)$} 
\label{T^4/Z_2}

In this section, we extend our previous analysis to the correspondence
between superstring theory on $H_6  \times T^4/\bz_2$ background and 
superconformal field theory on the symmetric orbifold $Sym^{M}(T^4/\bz_2)$. 
The general K3 surface can be obtained by resolving the singularity of
$T^4/\bz_2$ and we investigate the case of the orbifold $T^4/\bz_2$ as a
solvable example of $K3$ surface.
We again find the good correspondence for the (almost) BPS 
spectrum.

\subsection{Spectrum of Superstring on $H_6  \times T^4/\bz_2$}

In this subsection, we investigate the superstrings on 
$H_6  \times T^4/\bz_2$ background.
The $\bz_2$-orbifold action acts on the coordinates of $T^4$ sector as 
\begin{equation}
\la^{a\da} ~\longrightarrow~ -\la^{a\da}~, ~~~
Y^{a\da}~ \longrightarrow~ -Y^{a\da}~.
\label{Z2 orbifold 1}
\end{equation}
Moreover, we assume the action on the free bosons $H_3$ and $H_4$,
which are the bosonizations of $\la^{a\da}$
\begin{equation}
 H_3 ~\longrightarrow~H_3+\pi~,~~~ H_4~\longrightarrow~H_4-\pi~,
\label{Z2 orbifold 2}
\end{equation}
so that the space-time supersymmetry is preserved. 
In fact, we can directly check that all the generators of
super pp-wave algebra $\{\cJ, \cF, \cP_i, \cP_i^*, \cQ^{\al\beta a}\}$ 
defined in \eqn{PVF} and \eqn{PVS} are invariant 
under the $\bz_2$-orbifold action \eqn{Z2 orbifold 1} and \eqn{Z2
orbifold 2}. 
As for the DDF operators \eqn{DDF1} and \eqn{DDF2},
the orbifold action is given by
\begin{equation}
\cA^{a\da}_n ~ \longrightarrow~ - \cA^{a\da}_n~, ~~~
\cB^{\al\da}_n ~ \longrightarrow~ - \cB^{\al\da}_n~,
\end{equation}
and the other DDF operators are invariant under this orbifold
action.

Now, we can write down the spectrum of (almost) BPS states.
First, let us consider untwisted sector and start with BPS states.
All the task we have to do is to single out $\bz_2$-invariant BPS states. 
Only the NS-NS and R-R BPS states are left and the NS-R and 
R-NS BPS states are projected out.
Thus, we obtain the 8 BPS states for each $p$ and $\eta$
and they are identified with the even cohomology of $T^4$.
As for almost BPS states, we first consider the following Fock vacua as
\begin{equation}
\ket{i,\bar{i};j,\bar{j};\eta,p;n_a,w^a; (\pm)} =
\ket{i,\bar{i};j,\bar{j};\eta,p;n_a,w^a} \pm 
\ket{i,\bar{i};j,\bar{j};\eta,p;-n_a,-w^a} ~,
\end{equation}
where $\ket{i,\bar{i};j,\bar{j};\eta,p;n_a,w^a}$ represents physical 
state with non-vanishing momenta along $T^4$, which was defined in
\eqn{phys T4 momenta}.   NS-NS BPS states 
are realized as the special cases of such physical states as
$\ket{i,\bar{i};0,0;\eta,p;0,0; (+)}$ and R-R BPS states can be obtained 
by multiplying $\cB^{+\da}_0$ and $\bar{\cB}^{+\da}_0$.
General physical states 
are constructed  by multiplying the DDF operators and supercharges 
$\cQ^{\pm\mp a}$ as in the case of $T^4$. 
We only need the following additional constraint as
\begin{align}
& \sharp \{\cB_{-n}^{\al\da},~ \cA_{-m}^{a\da}\} + 
   \sharp \{\bar{\cB}_{-n}^{\al\da},~ \bar{\cA}_{-m}^{a\da}\} = \mbox{even}~,
   ~\left( \mbox{for the Fock vacua $\ket{i,\bar{i};\cdots ; (+)}$} 
       \right)~, \nn
& \sharp \{\cB_{-n}^{\al\da},~ \cA_{-m}^{a\da}\} + 
   \sharp \{\bar{\cB}_{-n}^{\al\da},~ \bar{\cA}_{-m}^{a\da}\} = \mbox{odd}~,
   ~\left( \mbox{for the Fock vacua $\ket{i,\bar{i};\cdots ; (-)}$}
         \right) ~,
\end{align}
from the condition of $\bz_2$ invariance.

Next, we consider twisted sectors.
There are 16 twisted sectors including stringy excitations around 
the each fixed point of orbifold action. 
For the each twisted sector,  
we need to consider the following boundary conditions as
\begin{align}
& Y^{a\da}(e^{2\pi i}z ) = - Y^{a\da}(z ) ~,\nn
& \la^{a\da}(e^{2\pi i}z ) = - \la^{a\da}(z )  ~,~
                                      (\mbox{for  NS sector}) ~,\nn
& \la^{a\da}(e^{2\pi i}z ) = \la^{a\da}(z )  ~,~
                                      (\mbox{for R sector}) ~, 
\end{align}
and moreover,
\begin{equation}
H_3(e^{2\pi i}z) = H_3(z) + \pi~, ~~~ H_4(e^{2\pi i}z) = H_4(z) - \pi~.
\label{twisted H34}
\end{equation}

BPS states are given as follows.
In the NS vacua, there are  both of 
bosonic and fermionic twist fields; $\sigma_{\bsz_2}^{b}$, 
$\sigma_{\bsz_2}^{f}$, whose conformal weights are equal to
\begin{equation}
h(\sigma_{\bsz_2}^{b})= \bar{h}(\sigma_{\bsz_2}^{b})
= 4\times \frac{1}{16}=\frac{1}{4}~,~~~
h(\sigma_{\bsz_2}^{f})= \bar{h}(\sigma_{\bsz_2}^{f})
= 4\times \frac{1}{16}=\frac{1}{4}~.
\end{equation}
Based on this fact we can observe that there are no BPS
states in the NS-NS, NS-R and R-NS sectors. 
On the other hand, the R vacua can include only the bosonic twist field
$\sigma_{\bsz_2}^b$ and only the spin fields along  the $H_6$
direction, which we express as $S^{\ep_0\ep_1\ep_2}$.
This fact leads us to a unique R-R BPS state (per each twisted sector),
which is explicitly written as 
\begin{equation}
S^{-++}_{-\frac{3}{8}+\eta}\bar{S}^{-++}_{-\frac{3}{8}+\eta}
\sigma_{\bsz_2}^b\ket{0,\eta,p}\otimes 
c\bar{c}e^{-\frac{\phi}{2}-\frac{\bar{\phi}}{2}}\ket{0}_{\msc{gh}}~. 
\label{twisted BPS}
\end{equation}
In this way, we have found the 16 R-R BPS states in the twisted sectors
for each $p$ and $\eta$.
They correspond to the blow-up modes of $T^4/\bz_2$ orbifold
and reproduce the cohomology ring of $K3$ together with the
contributions from the untwisted sector.

The other physical states are also constructed straightforwardly. 
Contrary to the untwisted sector, the states with non-trivial
momenta along $T^4$ are not allowed. Thus, we only have to
consider the states created by the actions of DDF operators over the 
BPS states \eqn{twisted BPS}. The only non-trivial point is that 
the modes of DDF operators $\cB^{\al\da}_r$ should be half integers
$ r\in \frac{1}{2}+\bz$ in this case. This fact originates from 
the boundary condition \eqn{twisted H34}. We again need the constraint
\begin{equation}
\sharp \{\cB_{-r}^{\al\da},~ \cA_{-m}^{a\da}\} + 
   \sharp \{\bar{\cB}_{-r}^{\al\da},~ \bar{\cA}_{-m}^{a\da}\} = \mbox{even}~,
\end{equation}
to preserve the $\bz_2$-invariance.

\subsection{Comparison with SCFT on $Sym^M(T^4/\bz_2)$}

In the last subsection, we have obtained the spectrum of superstring
theory. In this subsection, we study superconformal field theory on
the symmetric orbifold $Sym^M(T^4/\bz_2)$ and compare with the spectrum of
the superstring theory.
The $\bz_2$-orbifoldization of non-linear sigma model on $Sym^M(T^4)$ is  
defined by the action
\begin{equation}
 X_{(A)}^{a\da}  \rightarrow  - X_{(A)}^{a\da}~, ~~~
 \Psi^{\al\da}_{(A)}  \rightarrow  - \Psi^{\al\da}_{(A)}~.
\end{equation} 
This action preserves the $\cN=(4,4)$ superconformal symmetry.
We again study the single-particle Hilbert space of $\bz_N$-twisted
sector with $N =pk+l$ ($l=1,2,\cdots,k-1$) as in the previous section.

We first consider the spectrum of BPS states.
In the untwisted sector of $\bz_2$ orbifoldization, 
only 8 $\bz_2$-invariant BPS states survive. 
They correspond to the even cohomology of $T^4$. 

The analysis of twisted sectors is more complicated. 
Focusing on one of the twisted sectors corresponding to 16 fixed
points, we can observe the following aspects:
\begin{itemize}
 \item $N=2q+1$

We have the mode expansions 
$i\partial X^{a\da}_{\frac{n}{N}+\frac{1}{2N}}$ and
$\Psi^{\al\da}_{\frac{n}{N}}$, ($n\in \bz$) and 
we obtain for the NS vacuum $\ket{0;N}^{\msc{(t)}}$
\begin{align}
& \hat{L}_0\ket{0;N}^{\msc{(t)}} 
= \left(\frac{N^2-1}{4N}+\frac{1}{2N}\right)\ket{0;N}^{\msc{(t)}}
\equiv \frac{N^2+1}{4N}\ket{0;N}^{\msc{(t)}} ~, \nn
& \hat{K}^3_0\ket{0;N}^{\msc{(t)}}
 = -\frac{1}{2}\ket{0;N}^{\msc{(t)}}~.
\end{align}
 \item $N=2q$

We have the mode expansions 
$i\partial X^{a\da}_{\frac{n}{N}+\frac{1}{2N}}$ and
$\Psi^{\al\da}_{\frac{n}{N}+\frac{1}{2N}}$, ($n\in \bz$) and we obtain
for the NS vacuum $\ket{0;N}^{\msc{(t)}}$
\begin{align}
& \hat{L}_0\ket{0;N}^{\msc{(t)}}
 = \left(\frac{N^2-1}{4N}+\frac{1}{4N}\right)\ket{0;N}^{\msc{(t)}}
\equiv \frac{N}{4}\ket{0;N}^{\msc{(t)}} ~, \nn
& \hat{K}^3_0\ket{0;N}^{\msc{(t)}} = 0~.
\end{align}
\end{itemize}
In these expressions the extra zero-point energies and the R-charges
assigned to the NS vacua are due to these twisted mode expansions.
Based on these aspects, we can find out the following BPS state that is 
unique for each of the twisted sectors as
\begin{align}
&\ket{\om^{(1,1)};N}^{\msc{(t)}}=
 \ket{\om^{1};N}^{\msc{(t)}} 
\otimes \overline{\ket{\om^{1};N}}^{\msc{(t)}} ~,\\
& \ket{\om^1;N}^{\msc{(t)}}  
= \prod_{i=0}^{q}\,\Psi^{++}_{-\frac{i}{N}}\Psi^{+-}_{-\frac{i}{N}}
  \ket{0;N}^{\msc{(t)}}~, ~~~(N=2q+1) ~, \nn
& \ket{\om^1;N}^{\msc{(t)}}  
= \prod_{i=0}^{q-1}\,\Psi^{++}_{-\frac{i}{N}-\frac{1}{2N}}
   \Psi^{+-}_{-\frac{i}{N}-\frac{1}{2N}}
  \ket{0;N}^{\msc{(t)}}~, ~~~(N=2q) ~,
\end{align}
and we obtain 
\begin{equation}
\hat{L}_0\ket{\om^1;N}^{\msc{(t)}}=
\hat{K}^3_0\ket{\om^1;N}^{\msc{(t)}}
= \frac{N}{2}\ket{\om^1;N}^{\msc{(t)}}~.
\end{equation}

In summary, we have obtained the ($8+16=24$) BPS states, which precisely
correspond to the cohomology of $K3$ for each of the $\bz_N$-twisted
sectors.  They have (approximately) degenerate charges
$  \cF' ~ (= (\hat{K}^3_0+\hat{L}_0) /k ) =
p+l/k$
and $\cJ' ~ (= \hat{K}^3_0-\hat{L}_0)=0$.
As in the case of $T^4$,  
we have the good correspondence between the superstring theory and symmetric
orbifold theory under the identifications $N=kp+l$ and $\eta=l/k$
(\ref{l/k}).

With respect to the almost BPS states, the discussion is almost parallel 
to the case of $T^4$.  
We can explicitly write down the almost BPS states in
the symmetric orbifold theory and compare them with the states in the
superstring  theory 
by using the identification of DDF operators as before.
However, in this case we must identify the DDF operators 
$\cB^{\al\da}_{n+\frac{1}{2}}$  (where $n \in \bz$) in the twisted sectors
with $\Psi^{\pm\da}_{\mp\frac{1}{2}+\frac{nk}{N}+\frac{k}{2N}}$
rather than $\Psi^{\pm\da}_{\mp\frac{1}{2}+\frac{nk}{N}}$.  
(We again neglect the small difference of mode $1/2N$.)
The short string spectrum is again completely 
embedded in the Hilbert space of symmetric orbifold theory and there are 
many missing states. 

\newpage

\section{Conclusion}
\label{conclusion}

In this thesis we have investigated the ``Penrose limit'' of $AdS_3/CFT_2$
correspondence.
The $AdS_3/CFT_2$ correspondence is the correspondence between 
superstring theory on $AdS_3 \times S^3 \times M^4$ 
($M^4 = T^4$ or $T^4/\bz_2$) and
superconformal field theory on the symmetric orbifold $Sym^{Q_1Q_5}(M^4)$.
The Penrose limit of superstring theory on 
$AdS_3 \times S^3 \times M^4$ with NSNS-flux is given by superstring theory
on NSNS pp-wave background.
This theory can be described by a generalization of Nappi-Witten model
\cite{NW} and it can be quantized 
by the sigma model approach \cite{RT2-2,RT2,RT2-3} in the light-cone gauge 
and by the current algebra approach in the covariant gauge \cite{KK,KK2,KP}.
By using the free field realization in the current algebra approach, 
we have constructed the complete set of DDF operators.
The spectrum of physical states is classified by short string
sectors and long string sectors, as in superstring 
theory on $AdS_3$ \cite{MO}.

In the dual CFT side, the Penrose limit corresponds to focusing on 
the subsector of almost BPS states with large R-charges.
We have compared the general short string excitations 
with single-particle states in the dual CFT. 
We have shown that all the short string states are successfully 
embedded into the Hilbert space of symmetric orbifold theory.
We have also found many missing states in the Hilbert space
of superstring theory.
We may interpret them as non-perturbative excitations or
remove the extra spectrum in the dual CFT 
by correctly deforming from the orbifold point.
In the case with RR-flux, spectra of the two theories are the same at only
the leading order at the orbifold point \cite{Lunin,GMS} 
as discussed in subsection \ref{RR}.
In that case, we can reproduce the spectrum at the first non-trivial
order by using a marginal deformation \cite{Gava}.
We want to investigate on this point in detail in near future.

The subjects we have investigated are some specific aspects of $AdS/CFT$
duality.
In general, a duality is very powerful tool since we may map from one
theory with strong coupling constant to its dual theory with weak
coupling constant.
Although we can only deal with a perturbation theory, we obtain
some information of a theory at non-perturbative region by using the
duality. 
As for $AdS_5/CFT_4$ correspondence, 
the CFT side is at strong coupling region and the
AdS side is at weak coupling region.
Therefore, it is expected that we can calculate, for example,
correlation functions by using the S-matrix of supergravity on $AdS_5$.
In this thesis we have concentrated on the $AdS_3/CFT_2$ correspondence.
This duality is related to the D1/D5 system and hence the black hole
physics. Superstrings on the background including black holes are very
difficult to deal with since the coupling constant is very strong inside
the black holes.
In future we may describe superstrings on black hole solutions by using its
dual CFT.

The duality is very strong tool but there was the limitation that
we can treat only BPS quantities protected by supersymmetry. Recently,
the authors of \cite{BMN} showed that if the quantities are very
near BPS, we can analyze beyond BPS quantities.  
We have followed their strategy and obtained some information about the
correspondence of non BPS spectrum.
The CFT side of $AdS_5/CFT_4$ correspondence is 4 dimensional $\cN=4$
super Yang-Mills theory and that of $AdS_3/CFT_2$ correspondence is 2
dimensional $\cN=(4,4)$ supersymmetric sigma model on $Sym^{Q_1Q_5}(T^4)$.
They are completely different, and hence our results are a non-trivial
evidence of the PP-Wave/CFT correspondence.

There are a lot of works left to investigate more on the $AdS/CFT$
correspondence. 
We can interpret the $AdS/CFT$ correspondence as an example which
realizes holographic principle \cite{Hooft,Susskind}.  
According to the holographic principle, degrees of freedom of gravity
modes can be projected into a low dimensional screen.
As mentioned in \ref{AdS_5/CFT_4}, we have a holographic map from
states in supergravity on $AdS_5 \times S^5$ to operators in super
Yang-Mills theory \cite{GKP,Witten}.
It is know that it is difficult to interpret the PP-Wave/CFT
correspondence in a holographic way\footnote{However, there are 
several works on this subject \cite{DGR,LOR,DSY}.},
so it is important to study on this aspect.
The other important direction is to investigate the correspondence of more
general spectrum.
By studying this further, we may be able to complete the dictionary of
$AdS/CFT$ correspondence.

The correspondence between superstring theory on pp-wave background 
and subsector of super Yang-Mills theory is very interesting 
since it provides an explicit example of the string/gauge theory duality. 
There have been subsequent works after \cite{BMN} and they gave deep
insights into both string theory side and gauge theory side.
In particular, if we understand the duality between superstring theory and
supersymmetric gauge theory more, we will obtain the information of 
superstring theory on non-trivial background which we do not know how to deal
with (for example, a background including black hole solutions)
by using a simpler supersymmetric gauge theory.

\newpage

\section*{Acknowledgement}

I am very grateful to Prof.~Kazuo Fujikawa for encouraging me
and reading the manuscript and Prof.~Tohru Eguchi and 
Prof.~Yutaka Matsuo for valuable discussions.
I would like to thank Dr.~Yuji Sugawara for the collaboration on the
work which this thesis is based on and careful reading of the manuscript.
I also thank  Dr.~Kazuo Hosomichi, Dr.~Masatoshi Nozaki, 
Dr.~Tadashi Takayanagi, Dr.~Satoshi Yamaguchi and all the
other people in high energy theory group at University of Tokyo for a
lot of useful discussions.

\newpage

\appendix
\section{Gamma Matrices}
\label{gamma}

The cocycle factors of spin fields are defined by using gamma matrices. 
In this thesis we use the following Gamma matrices 
\begin{align}
 \Gamma_{\pm 0} &= \sigma_{\pm} \otimes \bone 
      \otimes \bone \otimes \bone \otimes \bone ~,\nn
 \Gamma_{\pm 1} &= \sigma_3 \otimes \sigma_{\pm} 
      \otimes \bone \otimes \bone \otimes \bone ~,\nn
 \Gamma_{\pm 2} &= \sigma_3 \otimes \sigma_3 
      \otimes \sigma_{\pm} \otimes \bone \otimes \bone ~,\nn
 \Gamma_{\pm 3} &= \sigma_3 \otimes \sigma_3
     \otimes \sigma_3 \otimes \sigma_{\pm} \otimes \bone ~,\nn
 \Gamma_{\pm 4} &= \sigma_3 \otimes \sigma_3 
     \otimes \sigma_3 \otimes \sigma_3 \otimes \sigma_{\pm} ~,
\end{align}
with Pauli matrices
\begin{equation}
 \sigma_1 = \left(
\begin{array}{rcl}
 0 & 1 \\
 1 & 0 
\end{array}
\right)  ~,~
 \sigma_2 = \left(
\begin{array}{rcl}
 0 & -i \\
 i & 0 
\end{array}
\right) ~,~
\sigma_3 = \left(
\begin{array}{rcl}
 1 & 0 \\
 0 & -1 
\end{array}
\right) ~,
\end{equation} 
and $\sigma_{\pm} = \frac{1}{2}(\sigma_1 \pm i \sigma_2) $.
Charge conjugation matrix can be written as
\begin{equation}
 C =  \epsilon \otimes \sigma_1 
     \otimes \epsilon \otimes \sigma_1 \otimes \epsilon ~,~~
 \epsilon = i \sigma_2 ~,
\end{equation}
which has the property as 
\begin{equation}
 C \Gamma_{\mu} C^{-1} = - (\Gamma_{\mu})^{T} ~,~ C^{\dagger} = C^{-1} ~.
\end{equation}
OPEs including spin fields are given by
\begin{align}
 \psi^{\mu} (z) S^A (w) &\sim \frac{1}{(z-w)^{\frac{1}{2}}}  
   (\Gamma^{\mu})^A_{~B} S^B (w) ~, \nn 
 \psi^{\mu} \psi^{\nu} (z) S^A (w) &\sim 
 - \frac{1}{z-w} 
   (\Gamma^{\mu \nu})^A_{~B} S^B (w) ~, \nn 
 S^A (z) S^B (w) &\sim \frac{1}{(z-w)^{\frac{3}{4}}} 
  (\Gamma_{\mu} C)^{AB} \psi^{\mu} (w)~.
\end{align}

\newpage

\section{Super PP-Wave Algebra from Super $AdS_3 \times S^3$ Algebra}
\label{STSS}

In this appendix
we examine space-time supersymmetry algebra on pp-wave
background by contracting that of  $AdS_3 \times S^3$.
It is known that supersymmetry on $AdS_3 \times S^3$ 
is represented by super Lie group 
$PSU(1,1|2) \times PSU(1,1|2)$. 
Its even part corresponds to the isometry of this background.
The isometry of $AdS_3$ space is identified as 
$SU(1,1) \times SU(1,1) \backsimeq SO(2,2)$ and the isometry of $S^3$ 
is identified as $SU(2) \times SU(2) \backsimeq SO(4)$. 
We denote the generators of $SU(1,1)$ Lie algebra as
$m^{\alpha}_{~\beta}$ $(\alpha , \beta = 1 ,2 )$
and the generators of
$SU(2)$ Lie algebra as $m^i_{~j}$ $(i,j = \hat{1} ,\hat{2} )$
by following the notation of \cite{MT}. 
(We concentrate on holomorphic sector and
anti-holomorphic sector can be analyzed in the similar way.)
Their commutation relations are given by 
\begin{equation}
 {[} m^{\alpha}_{~\beta} , m^{\gamma}_{~\delta} {]}= 
 \delta^{\gamma}_{~\beta} m^{\alpha}_{~\delta} - 
 \delta^{\alpha}_{~\delta} m^{\gamma}_{~\beta} ~,~
 {[} m^i_{~j} , m^k_{~n} {]}= 
 \delta^k_{~j} m^i_{~n} - \delta^i_{~n} m^{k}_{~j} ~,
\end{equation}
and Hermitian conjugations are defined as
\begin{align}
& (m^1_{~1})^{\dagger} = m^1_{~1} ~,~
 (m^2_{~1})^{\dagger} = m^1_{~2} ~,~
 (m^1_{~2})^{\dagger} = m^2_{~1} ~, \nn
& (m^{\hat{1}}_{~\hat{1}})^{\dagger} = m^{\hat{1}}_{~\hat{1}} ~,~
 (m^{\hat{2}}_{~\hat{1}})^{\dagger} = - m^{\hat{1}}_{~\hat{2}} ~,~
 (m^{\hat{1}}_{~\hat{2}})^{\dagger} = - m^{\hat{2}}_{~\hat{1}} ~.
\end{align} 
The generators of odd sector $q^{\alpha}_{~i}$ and $q_{~\alpha}^{i}$ 
correspond to $8 (+ 8)$ supercharges and their commutation relations are
given by
\begin{align}
& {[} m^{\alpha}_{~\beta} , q^k_{~\gamma}  {]}= 
 - \delta^{\alpha}_{~\gamma} q^k_{~\beta} 
 + \frac{1}{2}  \delta^{\alpha}_{~\beta} q^k_{~\gamma} ~,~~
 {[} m^i_{~j} , q^k_{~\alpha} {]} =
  \delta^k_{~j} q^i_{~\alpha} 
 - \frac{1}{2}  \delta^i_{~j} q^k_{~\alpha} ~, \nn
& {[} m^i_{~j} , q^{\alpha}_{~k} {]} = 
 - \delta^i_{~k} q^{\alpha}_{~j} 
 + \frac{1}{2}  \delta^i_{~j} q^{\alpha}_{~k} ~,~~
 {[} m^{\alpha}_{~\beta} , q^{\gamma}_{~k} {]} = 
  \delta^{\gamma}_{~\beta} q^{\alpha}_{~k} 
 - \frac{1}{2}  \delta^{\alpha}_{~\beta} q_{~k}^{\gamma} ~, \nn
& \{ q^i_{~\alpha} , q^{\beta}_{~j}\} =
 i (\delta^i_{~j} m^{\beta}_{~\alpha} + \delta^{\beta}_{~\alpha} m^i_{~j})~.
\end{align}
Hermitian conjugations are defined as 
\begin{align}
& (q^{1}_{~\hat{1}})^{\dagger} = - i q^{\hat{1}}_{~1} ~,~
 (q^{2}_{~\hat{2}})^{\dagger} = - i q^{\hat{2}}_{~2} ~,~
 (q^{2}_{~\hat{1}})^{\dagger} = i q^{\hat{1}}_{~2} ~,~
 (q^{1}_{~\hat{2}})^{\dagger} = i q^{\hat{2}}_{~1} ~, \nn
& (q_{~1}^{\hat{1}})^{\dagger} = i q_{~\hat{1}}^{1} ~,~
 (q_{~2}^{\hat{2}})^{\dagger} = i q_{~\hat{2}}^{2} ~,~
 (q_{~1}^{\hat{2}})^{\dagger} = - i q_{~\hat{2}}^{1} ~,~
 (q_{~2}^{\hat{1}})^{\dagger} = - i q_{~\hat{1}}^{2} ~.
\end{align}

Now, we take a light-cone basis and a Penrose limit. 
First, we consider the even sector. 
We redefine the generators as 
\begin{align}
 J &= - m^1_{~1} + m^{\hat{1}}_{~\hat{1}} ~,~~
 F = - \frac{1}{R} \left( m^1_{~1} + m^{\hat{1}}_{~\hat{1}} \right) ~, \nn
 P_1 &=  \frac{i}{\sqrt{R}} m^2_{~1} ~,~~ 
 P_1^*  = - \frac{i}{\sqrt{R}} m^1_{~2} ~, \nn
 P_2 &= -  \frac{i}{\sqrt{R}} m^{\hat{1}}_{~\hat{2}} ~,~~ 
 P_2^*  =  - \frac{i}{\sqrt{R}} m^{\hat{2}}_{~\hat{1}} ~,
\label{bosonicPL}
\end{align}
and take the limit of $R \to \infty$. 
Then we obtain the commutation relations as
\begin{equation}
 {[} J , P_i {]} = P_i ~,~~ 
 {[} J , P_i^* {]} = - P_i^* ~,~~ 
 {[} P_i , P_j^* {]} = \delta_{ij} F ~,
\label{CR1}
\end{equation}
which is the same as the ones of $H_6$ Lie algebra. 
Hermitian conjugations are given by
\begin{align}
 (J)^{\dagger} = J ~,~
 (F)^{\dagger} = F ~,~
 (P_i)^{\dagger} = P_i^* ~,~
 (P_i^*)^{\dagger} = P_i ~.
\end{align} 

The analysis of odd part can be done just like the even part.
We redefine the generates in the odd sector as 
\begin{align}
& Q^{--+} = - \frac{i}{\sqrt{R}} q^1_{~\hat{1}} ~,~~ 
   Q^{+++} = -  \frac{i}{\sqrt{R}} q^2_{~\hat{2}} ~, \nn
& Q^{++-} = - \frac{1}{\sqrt{R}} q_{~1}^{\hat{1}} ~,~~ 
   Q^{---} =  - \frac{1}{\sqrt{R}} q_{~2}^{\hat{2}} ~, \nn
& Q^{-++} = q^1_{~\hat{2}} ~,~~ 
   Q^{+-+} =  q^2_{~\hat{1}} ~, \nn
& Q^{-+-} = i q_{~2}^{\hat{1}} ~,~~ 
   Q^{+--} = i q_{~1}^{\hat{2}} ~,
\label{fermionicPL}
\end{align}
and take the limit of $R \to \infty$.
Their commutation relations with the generators in the even sector becomes 
\begin{align}
& {[} J , Q^{+ + a} {]} =  Q^{+ + a}  ~,~~
 {[} J , Q^{- - a} {]} = -  Q^{- - a}  ~, \nn
& {[} P_1 , Q^{- + a} {]} = - Q^{+ + a}  ~,~~
 {[} P_1^* , Q^{+ - a} {]} =  Q^{- - a}  ~, \nn
& {[} P_2 , Q^{+ - a} {]} = - Q^{+ + a}  ~,~~
 {[} P_2^* , Q^{- + a} {]} = - Q^{- - a}  ~, 
\label{CR2}
\end{align}
and the other commutation relations vanish.
The anti-commutation relations of the generators in 
the odd sector are obtained as
\begin{align}
& \{ Q^{--a} ,Q^{++b} \} = \epsilon^{ab} F ~,~~
   \{ Q^{-+a} ,Q^{+-b} \} = \epsilon^{ab}  J ~, \nn
& \{ Q^{++a} ,Q^{+-b} \} =  \epsilon^{ab}  P_1 ~,~~
   \{ Q^{-+a} ,Q^{--b} \} =   \epsilon^{ab} P_1^* ~,\nn
& \{ Q^{-+a} ,Q^{++b}\} = \epsilon^{ab} P_2 ~,~~
   \{ Q^{+-a} ,Q^{--b} \} = \epsilon^{ab} P_2^* ~,
\label{CR3}
\end{align}
and Hermitian conjugations are given by
$ (Q^{\epsilon_1, \epsilon_2, \epsilon_3})^{\dagger} = 
 Q^{-\epsilon_1, -\epsilon_2, -\epsilon_3}$.
The authors of \cite{Hatsuda,Hatsuda2} obtained superalgebras on the
pp-wave backgrounds by contracting
superalgebras on $AdS_5 \times S^5$,  $AdS_4 \times S^7$ and $AdS_7
\times S^4$. 
Our result is the counterpart of the case of $AdS_3 \times S^3$ and
a natural supersymmetric extension of $H_6$ Lie algebra is obtained.

\newpage

\section{Supersymmetries on NSNS PP-Waves Based on Killing Spinors}
\label{KS}

We consider Killing spinors in Type IIB supergravity on
pp-wave backgrounds with NSNS-flux\footnote{ 
We follow the notation of \cite{BKO}.}.
RR-fields are not considered since we only introduce non-trivial
NSNS-flux. 
The relevant part of Type IIB supergravity action is given by
\begin{equation}
 S =\frac{1}{2} \int d^{10} x \sqrt{-g} e^{-2 \phi} 
     ( - R + 4 (\partial \phi)^2 - \frac{1}{3} H^2) ~,
\end{equation} 
where $\phi$ is dilaton and $H$ is field strength
$ H_{\mu \nu \rho} = \partial_{[ \mu} B_{\nu \rho ]} $.
We only consider backgrounds with constant dilaton $\phi$.
Unbroken supersymmetry can be seen from the bosonic part of
supersymmetry transformation of fermions, which are given by
\begin{align}
 \delta \psi_{\mu} &= (\partial_{\mu} - 
      \frac{1}{4} \Omega_{+ \mu \hat \nu \hat \rho}
    \Gamma^{\hat \nu \hat \rho}) \epsilon ~,\nn
 \delta \lambda &= (\Gamma^{\mu} \partial_{\mu} \phi 
         - \frac{1}{6} H_{\mu \nu \rho } \Gamma^{\mu \nu \rho}) \epsilon~,
\label{C2}
\end{align} 
where $\Omega$ are spin connections with torsion 
\begin{equation}
\Omega_{+ \mu \hat \nu \hat \rho} = \omega_{\mu\hat \nu \hat \rho } 
 + H_{\mu \hat \nu \hat \rho } ~. 
\end{equation}
Here  $\mu,\nu,\rho,\cdots$ are space-time indices and hatted ones
represent the indices of tangent space. 
Killing spinor conditions are given by the equations that 
the right hand sides of (\ref{C2}) are zero.

From now on we concentrate on a specific configuration.
Here we use the model of \cite{RT} and its bosonic part
is given by
\begin{equation}
 L = \partial u \bar{\partial} v + \cF_{ij} x^i \partial u \bar{\partial} x^j
   + \partial x^i \bar{\partial} x^i ~,
\end{equation}
where $\cF_{ij}$ is a constant and $i = 1,\cdots,8$. 
The nontrivial components of $B$-field are 
$B_{uj} = \frac{1}{2} \cF_{ij} x^i$ and the field strength is
$H_{uij} = - \frac{1}{2} \cF_{ij}$.
The spin connections can be calculated straightforwardly \cite{HT,RT}
and the nontrivial components are
\begin{equation}
 \Omega_{+ u \hat{i} \hat{j}} = - \cF_{ij} ~.
\end{equation}
Therefore, the Killing spinor conditions become 
\begin{equation}
 (\partial_u + \frac{1}{4} \cF_{ij} \Gamma^{\hat{i} \hat{j}}) \epsilon =0 ~,~
 \partial_v \epsilon =0 ~,~ \partial_i \epsilon = 0 ~, 
\label{killing1}
\end{equation}
and
\begin{equation}
 \cF_{ij} \Gamma^{\hat{u} \hat{i} \hat{j} } \epsilon = 0 ~.
\label{killing2}
\end{equation}
The condition (\ref{killing1}) can be always satisfied by the following
form of spinor as
\begin{equation}
 \epsilon (u) = 
 \exp ( - \frac{1}{4} \int^u  \cF_{ij} \Gamma^{\hat{i} \hat{j}} ) \epsilon_0 ~.
\end{equation}
with a constant spinor $\epsilon_0$.
However, the condition (\ref{killing2}) can not be always satisfied.
In general, this condition breaks a half of supersymmetries by
\begin{equation}
 \Gamma^{\hat{u}} \epsilon = 0 ~.
\end{equation}
It is known that pp-wave backgrounds always preserve
partial supersymmetries called as ``kinematical'' supersymmetries. 
However, there are special cases with enhanced supersymmetries
called as ``dynamical'' supersymmetries.
The well-known example is the Penrose limit of $AdS_5 \times S^5$ 
\cite{BFHP,BFHP2,BFP}
and this background preserves 16 dynamical supersymmetries in addition to
16 kinematical supersymmetries. 
The relatively less-known examples are given in M-theory, Type IIA and 
Type IIB theory \cite{extra,extra2,extra3,extra4,extra5,extra6,extra7}.

Let us see the example which is obtained by the Penrose limit of 
$AdS_3 \times S^3 (\times T^4)$. 
This case is given by $\cF_{ij} = 2 f \epsilon_{ij} (i,j = 1,2)$ and 
$\cF_{kl} = 2 f \epsilon_{kl} (k,l =3,4)$. (See (\ref{sigmab}).)
By using the notation of gamma matrices in Appendix A, 
the condition (\ref{killing2}) implies
\begin{equation}
  \Gamma^{+0} (\Gamma^{+1} \Gamma^{-1} - \Gamma^{-2} \Gamma^{+2} ) 
 \epsilon = 0 ~.
\label{killing3}
\end{equation}
Thus, besides the Killing spinors satisfying $\Gamma^{+0} \epsilon = 0$,
there are 8 Killing spinors satisfying 
$(\Gamma^{+1} \Gamma^{-1} - \Gamma^{-2} \Gamma^{+2} ) \epsilon = 0 $.
In the notation of supercharges in this thesis, the former Killing spinors
correspond to 16 kinematical supercharges (\ref{PVS}), (\ref{DDF2})
\begin{align}
& \cQ^{++a} \sim \oint S^{+++aa} e^{i X^+} ~,~
 \cQ^{--a} \sim \oint S^{+--aa} e^{- i X^+} ~, \nn
& \cB_0^{+a} \sim \oint S^{+-+(-a)a} ~,~
 \cB_0^{-a} \sim \oint S^{++-(-a)a} ~,
\end{align}
and the latter Killing spinors correspond to 8 dynamical supercharges 
(\ref{PVS})
\begin{equation}
  \cQ^{-+a} \sim \oint S^{--+aa} ~,~
 \cQ^{+-a} \sim \oint S^{-+-aa}  ~.
\label{extraSC}
\end{equation}
In appendix \ref{STSS}, we have analyzed the supersymmetry on the
pp-wave background from the contraction of super $AdS_3 \times S^3$
algebra and
we have also found 8 dynamical supersymmetries (\ref{extraSC}), which is
consistent with the analysis in this appendix.
The dynamical supersymmetries are very important for our analysis 
because they correspond to the linearly realized supersymmetries 
in the light-cone gauge. 
More precisely, the dynamical supercharges generate the super
transformations which preserve the light-cone Hamiltonian.

In \cite{HS'2} we have proposed the other superstring vacua on NSNS pp-waves
with enhanced supersymmetry.
The superstring theories on these backgrounds are defined by the
combination of 4 dimensional Nappi-Witten model and a general $\cN=2$
rational superconformal field theory (RCFT) with $c=9$.
The Nappi-Witten model and the RCFT are non-trivially
related to each other due to the GSO projection. 
In these configurations, it seems difficult to analyze in the similar way,
because there are no naive geometric interpretations. 
However, the existence of extra supercharges implies cancellation
between NSNS-fluxes like (\ref{killing3}).
On the other hand,
in the case of Penrose limit of the near horizon of NS5-branes \cite{Gomis}, 
there is only one type of NSNS-flux and hence there is no cancellation
as in (\ref{killing3}).
Therefore, the model admits only 16 kinematical supersymmetries 
and no dynamical supersymmetries.

\newpage


\end{document}